\documentclass [11pt]{article}
\pdfoutput=1
\usepackage{amsfonts}
\usepackage{graphicx}
\usepackage{amsmath,amssymb, dsfont}
\usepackage{latexsym}
\usepackage{mathrsfs}
\usepackage{bbold}
\usepackage{color}
\usepackage{slashed}
\usepackage{cancel}
\usepackage{soul} %
\usepackage{setspace} %
\usepackage{comment}
\usepackage{booktabs}
\usepackage{multirow}
\usepackage{cite}
\usepackage[small]{caption}

\definecolor{red}{rgb}{1,0,0}

\makeatletter
\def\section{\@startsection {section}{1}{\z@}{-3.5ex plus -1ex minus
 -.2ex}{2.3ex plus .2ex}{\large\bf}}
\def\subsection{\@startsection{subsection}{2}{\z@}{-3.25ex plus -1ex
minus -.2ex}{1.5ex plus .2ex}{\normalsize\bf}}
\makeatother
\makeatletter

\@addtoreset{equation}{section}

\makeatother

\newcommand{\bea}{\begin{equation} \begin{aligned}} \newcommand{\eea}{\end{aligned} \end{equation}}
\def\be{\begin{equation}} \def\ee{\end{equation}}

\usepackage[colorlinks,linkcolor=black,citecolor=blue,urlcolor=blue,linktocpage,pagebackref]{hyperref}

\allowdisplaybreaks

 \def\elmax{\ell_{{\rm max}}}

\begin{document}

\thispagestyle{empty}

\begin{center}

	\vspace*{-.6cm}

	\begin{center}

		\vspace*{1.1cm}

		{\centering \Large\textbf{A Monte Carlo approach to the conformal bootstrap}}

	\end{center}

	\vspace{0.8cm}
	{\bf Alessandro Laio$^{a,b}$, Uriel Luviano Valenzuela$^{a,c}$, and Marco Serone$^{a,c}$}

	\vspace{1.cm}

	${}^a\!\!$
	{\em SISSA, Via Bonomea 265, I-34136 Trieste, Italy}
	
		\vspace{.3cm}
	
	${}^b\!\!$
	{\em ICTP, Strada Costiera 11, I-34151 Trieste, Italy}
	
		\vspace{.3cm}
 
	${}^c\!\!$
	{\em INFN, Sezione di Trieste, Via Valerio 2, I-34127 Trieste, Italy}
	
		\vspace{.3cm}

	\vspace{.3cm}

\end{center}

\vspace{1cm}

\centerline{\bf Abstract}
\vspace{2 mm}
\begin{quote}

We introduce an approach to find approximate numerical solutions of  truncated bootstrap equations for Conformal Field Theories (CFTs) in arbitrary dimensions. The method is based on a stochastic search via a Metropolis algorithm guided by an action $S$ which is the logarithm of the truncated bootstrap equations for a single scalar field correlator. 
While numerical conformal bootstrap methods based on semi-definite programming put rigorous exclusion bounds on CFTs, this method looks for approximate solutions, which correspond to local minima of $S$, when present, and can be even far from the extremality region.  By this protocol we  find that if no constraint on the operator scaling dimensions is imposed,  $S$ has a single minimum, corresponding to the Free Theory. 
If we fix the external operator dimension, however, we encounter minima that can be studied with our approach. 
Imposing a conserved stress-tensor, a $\mathbf{Z}_2$ symmetry and one relevant scalar, we identify two  regions  where local minima of $S$ are present.  When projected in the $(\Delta_\sigma, \Delta_{\epsilon})$-plane, $\sigma$ and $\epsilon$ being the external and the lightest exchanged operators, one of these regions essentially coincides with the extremality line found in previous bootstrap studies. The other region is along the generalized free theories in $d = 2$ and below that in both $d = 3$ and $d = 4$. We empirically prove that some of the minima found are  associated to known theories, including the $2d$ and $3d$ Ising theories and the $2d$ Yang-Lee model.

\end{quote}

\newpage

\tableofcontents

\section{Introduction} 

Starting from the pioneering work \cite{Rattazzi:2008pe},  numerical bootstrap methods have become a widespread work-tool for studying  conformal field theories (CFTs) in $d>2$ space-time dimensions, see \cite{Poland:2018epd} for a review with an extensive list of references. The most established and developed methods make use of semidefinite programming algorithms applied to 4-point functions of local operators to rule in or rule out theories, the state of the art program being currently SDPB 2.0 \cite{Simmons-Duffin:2015qma,Landry:2019qug}. In this approach one makes assumptions on the CFT spectrum (typically on the low-lying scaling dimension of primary operators) and is able to rigorously establish if that assumption is allowed or not in the space of all possible CFTs. By imposing suitable assumptions to a set of 4-point functions one can restrict the allowed region to a small island in parameter space, so that precise and rigorous properties of theories living at the extremality region can be determined \cite{Kos:2014bka,Kos:2015mba,Kos:2016ysd}. Such CFTs at the edge of the extremality bounds can also be studied using the so called extremal functional method, first discussed in \cite{Poland:2010wg} and later developed in \cite{El-Showk:2012vjm,El-Showk:2016mxr}. Although not as rigorous as the previous method, the extremal functional allows a quite accurate and precise determination of the CFT data, see e.g.\cite{Simmons-Duffin:2016wlq} for an application to the $3d$ Ising model. These powerful methods generally require that the theory in question lies at the boundary of the allowed region in parameter space to start with. In contrast, the ``interior" of the space of allowed CFTs is harder to characterize using the current techniques. 

In fact, it could be useful to have some numerical tool to explore the structure of the space of allowed CFTs and, in principle, a way to construct approximate CFT data that are consistent with crossing symmetry of $4$-point correlation function of local operators.\footnote{See \cite{Reehorst:2021ykw} for recent progress in this direction.} Locating, even approximately, putative CFTs could be a useful first step to understand which assumptions to impose so that these theories become extremal and hence amenable to, e.g., a fully rigorous bootstrap analysis.
The aim of this paper is to propose a novel numerical method to analyze the bootstrap equations and possibly find a subset of its approximate solutions. More specifically, we will try to answer to the following question: Given a finite set of $n_i$ operators with spin $l_i$ below a given scaling dimension $\Delta^*$, exchanged in some 4-point correlation function, is there a set of CFT data (scaling dimensions and OPE coefficients) for the $n_i$ operators which best satisfy the truncated crossing equations? If so, where is the putative solution located in the allowed CFT parameter space?  
 
In one-dimensional systems, where there is no spin, accurate solutions can be  constructed \cite{Afkhami-Jeddi:2019zci}. For $d>1$ this problem is much harder because one has to look for solutions by choosing among the many possible partitions of the operators in the different spin channels. In addition to that, it is unclear how to choose the initial conditions for any search and which constraints should be imposed to possibly get a finite set of solutions. Finally, it is far from clear that the search would lead to an actual theory. In particular, deterministic methods such as gradient descent, would uniquely give an answer, but most likely this answer won't correspond to an approximate solution to a physical CFT.  This is easily understood as follows. Suppose we find a suitable set of constraints which guarantee the existence of a finite set of approximate solutions (we will see below that these are indeed necessary, but easy to identify). If we define a positive definite function $F$ as, say, the modulus square of the conformal bootstrap equations for a single 4-point correlator, we see that in the infinite dimensional space of all possible CFT data, scaling dimensions $\Delta_i$ and Operator Product Expansion (OPE) coefficients $\lambda_{ijk}$, generally $F\geq 0$. Putative physical theories can only be associated to points where $F=0$, where crossing symmetry is realized.\footnote{Of course this is only a necessary condition, since other constraints might arise from other correlators, etc.} Suppose now to use a  deterministic numerical method that allows us, starting from an arbitrary point in the (truncated) CFT data parameter space where $F>0$, to minimize $F$. The details of how to actually achieve this do not matter for the argument we want to make here. In a truncated numerical approach, of course, $F$ will never be zero, but we will nevertheless uniquely get the minimum value $F_0$, {\it given} the chosen initial conditions. It is clear that generally $F_0$ will be an approximation to a {\it local} minimum of $F$ (with $F_{\rm min}>0$) and not an approximation to one of the global minima of $F$ (where $F=0$). Thus, the solution found will not correspond to any CFT; it will be in the ``swampland".

In order to alleviate the problem of being trapped at fake CFT minima, we use a stochastic minimization via a Metropolis algorithm. The function $S$ we choose to minimze (the ``action") is the logarithm of the truncated bootstrap equations for a single scalar field correlator, $S\sim \log F$. Configurations with higher actions are from time to time accepted with a probability which depends on a parameter $T$ (``temperature"). In this way, thermal fluctations allows us to ``escape" from local minima in the quest for the global ones.\footnote{Very recently numerical methods based on reinforcement learning have been used to find approximate solutions to the bootstrap equations \cite{Kantor:2021kbx,Kantor:2021jpz}. Though this is an interesting line of research to pursue, it is not clear to us if and how such techniques can alleviate the problem of fake local minima.}  The problem of choosing among the many possible partitions of the operators in the different spin channels is solved by adding as many operators as possible to each spin channel and then allowing operators to possibly decouple along the minimization process. The complex dependence of $S$ on the truncated CFT data requires further non-trivial numerical manipulations, which will be discussed in section \ref{sec:method}, to finally determine the approximate CFT data of a putative theory.

We consider in this work the simplest bootstrap set-up, namely a single 4-point function of identical scalars in euclidean CFTs  and focus on the exploration of the CFT parameter space where the external scalar is close to the unitarity bound. We start in section \ref{sec:method} by presenting the method. In section \ref{subsec:DiscBotEq} we explain how the bootstrap equations are truncated and discretized, in section \ref{subsec:MMsm} how the stochastic minimization is defined and performed and finally in section \ref{sec:protocol} the various steps entering in the minimization process are reported and explained. 

The results of our numerical explorations, all based on severe truncations of the bootstrap equations with $\Delta^*\lesssim15$ and a total number of operators included $N_{\rm Ops}\lesssim20$, are reported in section \ref{sec:results}. The first general qualitative feature that emerged from our analysis is discussed in section \ref{subsec:sigmafree} and is the predominance of the free theory scalar theory. If no constraints are imposed
 -- aside unitarity, the presence of an energy momentum tensor and a $\mathbf{Z}_2$ global symmetry -- the minimization of $S$ leads only to a single global minimum, associated to the free theory.  It is enough to impose a single constraint, such as fixing the scaling dimension of the external operator, to get a finite set of minima.
At fixed $\Delta_\sigma$ we first study in section \ref{subsec:1opspin} the simplest spin partition with one operator per spin. We test the method by rediscovering the free theory and give indications that there are no other CFTs with one operator per spin, at least in their lowest spin channels and lowest lying operators.
Our most interesting results are contained in section \ref{subsec:Z2cft}, where we study the more realistic set-up of more operators per spin.
Now the situation is reversed and we get generally too many minima, unless extra assumptions are imposed. Demanding the presence of only one $\mathbf{Z}_2$-even relevant scalar drastically reduces the number of minima and makes the analysis feasible. A general qualitative feature that emerged from our analysis is the presence of two regions in the space of CFT data where the bootstrap equations are more easily satisfied for CFTs admitting a conserved energy-momentum tensor,
a $\mathbf{Z}_2$ global symmetry and one $\mathbf{Z}_2$-even relevant scalar. When projected in the $(\Delta_\sigma, \Delta_{\epsilon})$-plane, $\sigma$ and $\epsilon$ being the external and the lightest exchanged operator, respectively, one of these regions essentially coincides with the extremality line found in previous bootstrap studies. The other region is along the generalized free theories (GFTs) in $d=2$ and below the GFT line in both $d=3$ and $d=4$. 
The minima found in the $(\Delta_\sigma,\Delta_\epsilon)$-plane are reported in figures \ref{fig:d2-overview}, \ref{fig:d3-overview} and \ref{fig:d4-overview}  for the $d=2,3,4$ cases. The extremal minima in the $d=2$ case are identified with the generalized minimal models of \cite{Liendo:2012hy}, one extremal minimum in $d=3$ corresponds to the $3d$ Ising model, while all the other minima have not been identified. Unfortunately the numerical accuracy we have does not allow to reach higher values of $\Delta^*$ to establish more firmly if (and which of) these minima are numerical artifact or not.
CFTs without an energy momentum tensor are considered in section \ref{sec:DT}. The minima now cluster around the GFT line and GFTs themselves are well reproduced, especially in $d=2$ where our numerical accuracy is higher.
We finally briefly show in section \ref{subsec:2dYL} how our algorithm can be applied to non-unitary theories by rediscovering the $2d$ Yang-Lee model. 

We discuss in the final outlook the limitations of our algorithm and possible ways to improve it in the future. Three appendices contain some other details of our numerical procedure and further more technical results.

\section{Method}
\label{sec:method}
  
    We present in this section the numerical method we used to find approximate solutions of the bootstrap equations for a 4-point function of identical scalar primary operators in a generic CFT in $d$ dimensions. 

  \subsection{Discretization of the Bootstrap Equations}
  \label{subsec:DiscBotEq}
     
  The crossing equations for 4 identical scalar fields $\sigma$ with scaling dimension $\Delta_\sigma$ can be written as (notation as in \cite{Poland:2018epd})
      \begin{equation}
        \sum_{\Delta>0,\ell}
        \lambda_{\sigma\sigma{\cal O}_{\Delta_\ell}}^2\mathcal{F}_{\Delta,\ell}(z)
        =I(z)\,.
        \label{eq:bootstrap}
      \end{equation}
       Here $\lambda_{\sigma\sigma{\cal O}_{\Delta_\ell}}$ are OPE coefficients, $\Delta$ and $\ell$ are the scaling dimensions and spin of the exchanged primary operator ${\cal O}_{\Delta_\ell}$ (even spin traceless symmetric tensors), $z$ and $\bar z$ are the usual Dolan-Osborn conformal invariant cross ratios \cite{Dolan:2000ut,Dolan:2003hv}, $I(z,\bar z)=|z|^{2\Delta_\sigma}-|1-z|^{2\Delta_\sigma}$ is the identity contribution to the OPE expansion, and finally 
        \begin{equation}
     \mathcal{F}_{\Delta,\ell}(z) =  
        |1-z|^{2\Delta_\sigma}
        g_{\Delta,\ell}(z,\bar z)
        -
        |z|^{2\Delta_\sigma}
        g_{\Delta,\ell}(1-z,1-\bar z) \,,
         \end{equation}
      with $g_{\Delta,\ell}(z,\bar z)$ the conformal blocks, normalized as in \cite{Rattazzi:2008pe} with an extra $2^\ell$ factor.
      
As discussed in the introduction, we do not look for rigorous disallowed regions in the space of CFT data, but rather for approximate solutions of the above equations, namely for approximate possibly allowed CFT data.
To this aim,  we severely truncate \eqref{eq:bootstrap} to the finite set of primary operators with scaling dimension up to a given $\Delta^*$ (chosen to be integer), and assign a given number of operators $n_\ell$ to each spin channel $\ell$ up to some value $\ell_{{\rm max}}$. If unitarity is imposed, $\Delta^*$ and $\ell_{{\rm max}}$ are related. We take
\begin{equation}
    \ell_{{\rm max}} = \Delta^* - d  -1\,,
\end{equation}
so that the scaling dimension of the operators with spin $\ell_{{\rm max}}$ have a large enough range where they can be varied, compatibly with unitarity: $\Delta^* - 3 \leq \Delta_{\ell_{{\rm max}}}\leq \Delta^*$.
The total number of exchanged operators ${\cal O}_a$ that we consider is 
\begin{equation}
  \sum_{\ell=0,2,\ldots, \ell_{{\rm max}}} n_\ell = N_{{\rm Ops}}.
\end{equation}
 We then choose a finite sample of $N_z$ points in the complex $z$-plane in the neighborhood of $z=\bar z = 1/2$ and define the $N_z\times N_{{\rm Ops}}$ matrix ${\mathcal M}$ and the $N_z$ vector $I_i$
       \begin{equation}
      \mathcal{M}_{i,a} \equiv \mathcal{F}_{\Delta_a,\ell}(z_i), \quad  I_i\equiv I( z_i),\quad i=1,\ldots, N_z\,, \; a=1,\ldots,N_{{\rm Ops}}\,.
      \label{eq:matrix}
  \end{equation}  
     We define an ``action" as (spin dependence omitted for simplicity)
   \begin{equation}
      \exp(S(\Delta_a,\rho_a))\equiv  \frac{1}{N_z} \sum_{i=1}^{N_z}\sigma_i^{-2} \bigg|\sum_{a=1}^{N_{{\rm Ops}}} \rho_a {\cal M}_{i,a} - I_i  \bigg|^2\,,
      \label{eq:DefAction}
\end{equation}
 where $\rho_a\equiv \lambda_{\sigma \sigma {\cal O}_a}^2$, by taking a weighted sum of the absolute square of the truncated
  crossing equations \eqref{eq:bootstrap} at the $N_z$ points, with
    \begin{equation}
   \sigma_i = |\mathcal{F}_{\Delta^*,0}(z_i)|.
      \label{eq:error}
\end{equation}
The factor \eqref{eq:error} is introduced to take into account that the points with the best convergent conformal block expansion are around $z=1/2$, point where on the other hand the function ${\cal F}$ vanishes. 
For any finite truncation, our aim would be then to minimize $S$
in the space of CFT data.

   \subsection{Metropolis Montecarlo and stochastic minimization}
    \label{subsec:MMsm}
    
 Finding the $2N_{{\rm Ops}}$ CFT data $(\rho_a,\Delta_a)$ that minimize $S$ is
 a challenging task. First of all, in $d>1$ dimensions, where we have spin, one
 should in principle look for solutions in all possible partitions of the
 $n_\ell$ operators in each spin channel.  In addition to that, deterministic
 methods, such as gradient descent, can be highly inefficient because they
 converge to a  minimum of $S$ 
  depending on the initial conditions. The
 actual dependence of $S$ on the CFT data is expectedly complicated and depends on the constraints imposed,
 and can be characterized by the presence of several local minima. One of the main motivations of this work
 is in fact to shed some light on the ``landscape/swampland'' space of CFT data. As we will explain in what follows, both problems (avoiding to land on a local minimum and the need to consider all spin partitions) are significantly alleviated by the use of stochastic searches based on Monte Carlo (MC) methods. We will in particular sample the parameter space via the Metropolis Monte Carlo method, whose structure we now briefly describe. 
    
Given a function $f(x): U \to \mathbb{R}$, where $U\subset \mathbb{R}^n$, the minima of $f$ in $U$ are sampled by taking an initial point $x^{(0)}\in U$ and proposing an update 
        \begin{equation}
          x^{(1)} = x^{(0)} + \delta x\,,
          \label{eq:proposed}
    \end{equation}
 which is accepted with a probability given by 
 \begin{equation}
 P(x^{(0)}\to x^{(1)}) = 
\min\left(1,\exp\left(-\frac{ f(x^{(1)}) - f(x^{(0)})}{T}\right)\right)\,.
\label{eq:Metro}
 \end{equation}
 The parameter $T$ can be interpreted as a temperature and the exponential factor in \eqref{eq:Metro} as a Boltzmann suppression factor. While a gradient descent method would uniquely find the local minimum of $f$ connected to the point $x_0$, the Metropolis algorithm, accepting from time to time moves where the function increases rather than decreases, is able to overcome  barriers between different minima and sample wider regions of $U$. The ``efficiency'' in overcoming barriers depends on $T$. At fixed $\delta x$, the higher $T$ the easier it is. Of course, if $T$ is chosen too large, we would explore the entire region $U$ but we might not detect the minima, because with such a diffusive dynamics the probability  density $P(x)$ will tend to be uniform in $x$. 
 
 In our MC implementation, the variables $x$ which are changed 
 at each iteration procedure are the scaling dimensions $\Delta_a$ (and possibly the external scaling dimension $\Delta_\sigma$), while the function $f$ is identified with the action $S$ in \eqref{eq:DefAction}.
 The procedure works as follows. Starting from some initial scaling dimensions $(\Delta_\sigma^{(0)}, \Delta_a^{(0)})$,  we  analytically minimize $S$ with respect to the OPE coefficients $\rho_a$, given that they enter  quadratically in the action. We get
\begin{equation}
  \rho_a^{(0)} =\sum_b  A^{-1}_{ab} \sum_i {\rm Re} ({\mathcal M}_{b,i})  \sigma_i^{-2} I_i, \quad   A_{ab} \equiv \sum_i ( \mathcal{M}_{ai}^*  \mathcal{M}_{bi} +\mathcal{M}_{ai}  \mathcal{M}_{bi}^* ) \sigma_i^{-2} \,,
  \label{eq:rhoLeastSquare}
\end{equation}
where all the terms in the right hand side of \eqref{eq:rhoLeastSquare} are evaluated at $(\Delta_\sigma^{(0)}, \Delta_a^{(0)})$.
We then plug \eqref{eq:rhoLeastSquare} into \eqref{eq:DefAction} to get an action  $S^{(0)}=S(\Delta_a^{(0)},\rho_a^{(0)})$ which depends on operator scaling dimensions only (spin and external operator dependence omitted for simplicity). At this stage, the scaling dimensions $(\Delta_\sigma^{(0)}, \Delta_a^{(0)})$ are changed as in \eqref{eq:proposed}, the OPE coefficients recomputed as above and then the move is accepted or not according to \eqref{eq:Metro}. This procedure is then iterated $N_{{\rm steps}}$ times.  In this way we explore different the local minima of $S$,   
and then study the collective trends of these minima as a function of $\Delta^*$. Of course, for the CFT data $(\Delta_\sigma, \Delta_a,\rho_a)$ associated to a physical theory we should have
\begin{equation}
    \lim_{\Delta^*, N_{\rm Ops}\rightarrow \infty} S(\Delta_a, \rho_a) = - \infty\,.
    \label{eq:limitCFT}
\end{equation}
It is then important to establish that, in a finite truncation, the value of $S$ at the minima decreases as $\Delta^*$ increases.

This is in a nutshell how our protocol works. In the next subsection we will describe in detail the various steps involved.

  \subsection{The search protocol in detail}
\label{sec:protocol}

We choose to sample the $N_z$ points in the $z$-plane from a uniform grid with the constraint 
\begin{equation}
\lambda(z)<\lambda_0\,,    
\end{equation} 
with $\lambda(z)$ defined as in  (3.11) of \cite{CastedoEcheverri:2016fxt}. This condition defines a compact and convex region around $z=\bar z = 1/2$, the point of best convergence of the bootstrap equations. The parameter $\lambda_0$ is, together with $\Delta^*$ and $T$, the most relevant ``hyperparameter'' of our computational pipeline, and  has to be chosen carefully in order for the method to work. 
 
The number of MC iterations $N_{{\rm steps}}$ needed to get sensible results can be quite large ($\sim 10^8$). In order to be able to perform exhaustive searches in parameter space we developed a numerical approach which allows estimating \emph{efficiently} the action $S$ defined in \eqref{eq:DefAction}.  Indeed, conformal blocks are computationally-expensive functions of $\Delta$, $\ell$ and $z$, which would make an extensive sampling impossible with moderate computational resources without an appropriate numerical optimization.   We tabulate the blocks for a fixed $z-$point sample and each $\ell$, and then use cubic splines \cite{num_rec} to interpolate their values. In this way we get a speedup of $O(10^4)$. The interpolation grid is chosen in such a way that  the numerical error induced by this procedure is negligible compared to the truncation error (see figure  \ref{fig:hist_err} and the discussion in appendix \ref{sec:error}). 

It is important here to note that this speed up comes with the tradeoff of limiting our working precision to that of $\mathrm{double}$ type in Fortran (i.e. $64-$bit). This might come as a surprise to our more technical bootstrap readers, but an important result of this paper is indeed that approximate solutions 
to the bootstrap equations can be found without resorting to $200$ digits of precision. Nevertheless, limited precision is a bottleneck to go to larger values of $\Delta^*$ and to improve the quality of our approximate solutions.

We now describe each step of the protocol more thoroughly. 
Further details can be found in appendix \ref{sec:impDetails}.

          \paragraph{1. Determining the hyper parameters} 
 Once a physical system of interest (understood as a global symmetry structure and the relevant bounds on the operators' scaling dimensions) has been chosen, some exploratory work must be done in order to make our search protocol as efficient as possible. In appendix \ref{sec:impDetails} we list the choices for each search discussed in section \ref{sec:results}. 
We will here concentrate on motivating our choice for $T$, since we have found empirically that the other hyperparameters are less determinant to the success of the search.

 Since we will explore different truncations, the order of magnitude of $\delta S$, see eq.\eqref{eq:Metro},  can be very different from spectrum to spectrum\footnote{We have found empirically that the most important parameter in this case is $\elmax$, not necesarily $\Delta^*$ on its own.} and thus the temperature $T$ must be adjusted accordingly. Although we will use an adaptive step that guarantees that in the stationary state roughly half of the moves will be rejected, choosing a temperature that is too high will make $\delta\Delta$ too coarse and the MC will ``miss'' narrow features of the landscape. On the other hand, if $T$ is too small our method will move too slowly around the landscape or might even ``freeze'' in the basin of a local minimum.

 In the following we will empirically choose $T$ in such a way that the  projection of the trajectory to each coordinate axis (which in our case correspond to the operators' scaling dimensions) covers the whole allowed range at least 10 times, but still low enough so that the dynamics does not become diffusive, namely the local minima can still be detected as metastable states. 
   
    For further detail, we show in figure \ref{fig:trajectories} an example of archetypal trajectories. This figure and further considerations about the choice of $T$ are presented and analyzed in     \ref{sec:Temperature}. 
   
     \paragraph{2. Finite temperature MCs}
    For each ansatz 
    $\Delta_a^{(0)}$\footnote{For simplicity of notation, we do not explicitly report $\Delta_\sigma^{(0)}$ as input data.}
    we sample the CFT parameter space at the temperature $T$ chosen according to the criteria discussed in the previous step. 
    This is the most important and computationally expensive step, but one must bear in mind that it gives all the necessary information to understand the landscape of possible CFTs at several different truncations.

    The variation of the scaling dimensions $\delta \Delta_a$
    is taken randomly with uniform distribution over an hyperparallelepiped centered around the origin. This was observed to be more efficient than moving one operator at a time since the landscape shows highly correlated features and thus attempting moves in which all the operators change simultaneously improves the sampling efficiency. The user can choose the length of each side but the overall scale is adaptive and changes at each MC  step. This is done in order to guarantee a $50\%$ acceptance rate in the stationary state. In practice, when a move is rejected (resp. accepted) this overall scale is decreased (resp. increased) by a fixed factor. 
     Each operator scaling dimension is confined within a finite region within two boundary values decided by the user. 
     The MC then explores the landscape for $\sim 10^8$ steps. Our choice of the temperature guarantees that after this number of iterations the initial conditions are irrelevant.
For each of these spin partitions we store each $1000$th configuration visited. We will refer to these as frames.

    \paragraph{3. Separation by sectors}
     In a given MC run not all the operators in the ansatz will have sizable OPE coefficients at each iteration step. In this work we assume that if the OPE coefficient associated to a primary operator is smaller than a given tolerance (we found $10^{-8}$ to be a sensible choice for the $\elmax$ used in this paper)
     this operator is decoupled and can be neglected. If unitarity is imposed, OPE coefficients that turn out to be negative from \eqref{eq:rhoLeastSquare} are set to zero in that iteration.
     These two effects lead to a classification of the different points visited by the MC in terms of the effective spectrum  contributing to the crossing equations. In order to describe these different sectors we introduce the following shorthand notation to indicate each one of them:
     $ (m_0)\_(m_2)\_ \cdots \_ (m_{\elmax})$. For example, a point with 3 scalars, 2 spin-2 operators, 2 spin 4 and one spin-6 would belong to spin sector $3\_2\_2\_1$. Crucially, 
     \begin{equation}
      \sum_{\ell} m_{\ell} \leq N_{{\rm Ops}}\,.   
     \end{equation}
     
This implies that there is no need to consider different partitions of operators in each spin channel. All partitions with $m_\ell\leq n_\ell$ will be automatically searched for by  MC. 
Moreover, the same sector can be visited in two different MC runs with different $n_\ell$. Thus, this step effectively recombines all the information from the different spin partitions studied in the previous step.

     \paragraph{4. Identification of putative local minima}

Now that we have sampled the landscape we must identify the points that are more likely to be close to a local minimum of the action. For a low-dimensional parameter space we could visualize the action as a contour plot and identify the points by inspection. It is clear, though, that for more than $3$ scaling dimensions, this approach cannot be followed.

    Our proposal is thus the following. 
    For each set of frames obtained in the previous step 
    (namely, those corresponding to spin sectors with more than $120$k frames and those describing the trajectory of every search)\footnote{
It might seem redundant to analyze sector by sector frames that are already contained in the trajectories that result from the MC search described in point 2.
 Since exhaustivity is paramount and the definition of local minimum depends on the neighboring points, it makes sense to consider the same frame in two different settings in order not to miss any interesting features.}
we sort their elements according to the number of nearest neighbors with higher action, in the philosophy of Density Peak clustering\cite{rodriguez2014}. 
We consider as putative local minima  the  20  configurations with highest rank. We have observed empirically that in each
 sector the number of actual minima is always much smaller than this number, and that many of the smaller rank frames actually converge to other minima, so this is a conservative estimate.
  To clarify, the point with the highest rank will be the global minimum of that set of frames because every other point will have higher action.

    \paragraph{5. Local minimization}
    For each putative local minimum  we launch Newton - Rhapson (NR) minimizations. This deterministic algorithm is justified because the points found in the previous step are 
    likely contained in the basin of attraction of a local minimum.  More details about the implementation of this algorithm can be found in appendix \ref{sec:NR}.
    
All the endpoints from the NR minimizations are collected and those that violate the bounds imposed on the spectrum  are discarded.\footnote{Spectra where a whole spin sector is missing, such as $2\_1\_0\_1\_1$, have been ignored for simplicity.} 
We then assign to the corresponding sector the minima where one or more operators  decouple (in the sense described above). In the case of two or more minimizations converging to the same point (i.e. the relative difference for any scaling dimension is smaller than $20\%$) we keep only the point with the lowest action. At the end of this step we have a list of approximate solutions to crossing that are local minima of $S$.
   
     \paragraph{6. Convergence and  boundary check}
     Since we usually impose boundaries on the subleading operators in order to better control the type of CFTs that we might find, it is important to rule out the possibility of some of the local minima found in the previous step actually being induced by the presence of this boundary. 
     
     To this end,
    a final NR minimization is performed from all the minima found in the previous step, relaxing all the boundaries except strict unitarity (if imposed). We consider as true minima the ones that reach convergence according to our NR implementation and are contained within the region of interest: this means a minimum can be at the boundary of unitarity, but not at one of the boundaries introduced in step 1. 
  
   \paragraph{7. Identifying akin minima at different truncations}
  \label{sec:branch}

Minima belonging to diferent spin sectors but with  similar CFT data in their common operators should be identified.
In this case the minimum with more primaries should have a value of $S$ lower than the minimum with less primaries. This is a necessary (but by no means sufficient) condition to associate such minima to approximate solutions of the crossing equations of one and the same CFT. 
In order to find akin minima and quantify their similarity we organize all the minima in a  directed graph where the minima are the nodes and $A\rightarrow B$ if: (i) the sector to which
    $A$ belongs is contained in the one of $B$, namely if $B$ has at most
    $\mathtt{maxMismatch}$ more operators than $A$ and (ii) the operators which are in common between the two sectors differ in their scaling dimensions by at most $\mathtt{relTol}$  in relative error.
 The final result of our protocol are the leaves of this directed graph, namely the nodes with no outgoing links. 
 One can also study the branches (i.e. subsets of this directed graph) in order to understand the asymptotic behaviour of certain values, such as the action $S$ or the $\Delta$ of some leading operators.

    \section{Results}
    \label{sec:results}
    
We report in this section the results obtained with our method. We have considered a single 4-point function with four identical scalar operators in $d=2,3,4$ dimensions. 
Except for a short discussion of the Yang Lee CFT (see section \ref{subsec:2dYL}), we have considered CFTs where the fusion rules are compatible with a $\mathbf{Z}_2$ discrete symmetry under which the external operator is odd.\footnote{Note that in general such a $\mathbf{Z}_2$ symmetry is not required to be a genuine faithful symmetry of the whole CFT. We will encounter an instance of this phenomenon when discussing $2d$ minimal models in section \ref{subsec:Z2cft}.}
We will use a Ising-like notation and denote the external operator by $\sigma$ and the lowest dimensional exchanged scalar (besides the identity) by $\epsilon$.
  As will be later explained, demanding that $\epsilon$ be the only $\mathbf{Z}_2$-even relevant 
 (in the renormalization group sense) primary operator that can appear in the $\sigma \times \sigma$ OPE 
 is necessary to avoid the proliferation of minima.
 We consider both local and non-local theories, demanding or not the existence of a spin-two energy-momentum tensor operator at the unitarity bound $\Delta_T=d$. 
Given its relevance in our analysis, we have also considered the space of non-local theories where the scaling dimension of the first spin 2 operator is fixed and equals that of Generalized Free Theories (GFT), $\Delta_T= 2\Delta_\sigma+ 2$. We then get further insights by allowing $\Delta_T$ to vary.
We focus mostly on unitary theories, though our method does not rely on it. As a proof of concept, we study the non-unitary landscape around the 2d Yang-Lee model.

As already anticipated in the introduction, when $\Delta_\sigma$ is allowed to vary, the landscape of $S$ becomes very simple: the global minimum is at the lower edge of the unitarity region, which in $d=3,4$ coincides with the Free Theory (FT) of a scalar field. Moreover, for generic spectra we find that it is in fact the only reachable minimum.
For this reason in most of our analysis we will study ``slices'' of the landscape by keeping fixed $\Delta_\sigma$ in each MC run.

The section is structured as follows. We start in section \ref{subsec:sigmafree} by showing that the FT is the global (and in many cases the only) minimum of $S$. From section \ref{subsec:1opspin} we look for local minima by keeping the external operator fixed. We first consider the simplest possible scenario, namely very sparse CFTs where just one operator per spin appears below $\Delta^*$. We then discuss in section \ref{subsec:Z2cft} more general CFTs with multiple operators per spin. In section \ref{sec:DT} we discuss non-local theories with a special focus on the GFT. Finally, we briefly discuss the 2d Yang-Lee model as a proof of concept of the validity of the method for non-unitary theories in section \ref{subsec:2dYL}.

\subsection{The Free Theory is the global minimum of $S$}
\label{subsec:sigmafree}

Before attempting to identify the local minima of $S$ it is useful to have a qualitative idea of how $S$ behaves as a function of the CFT data. 
This analysis can be made by launching MC runs with a given temperature and spin partition of operators, and simply looking for the values of the $\Delta$'s corresponding to the lowest value of $S$ explored during the run. 

\begin{figure}[t!]
    \centering
    \includegraphics[width=0.9 \linewidth]{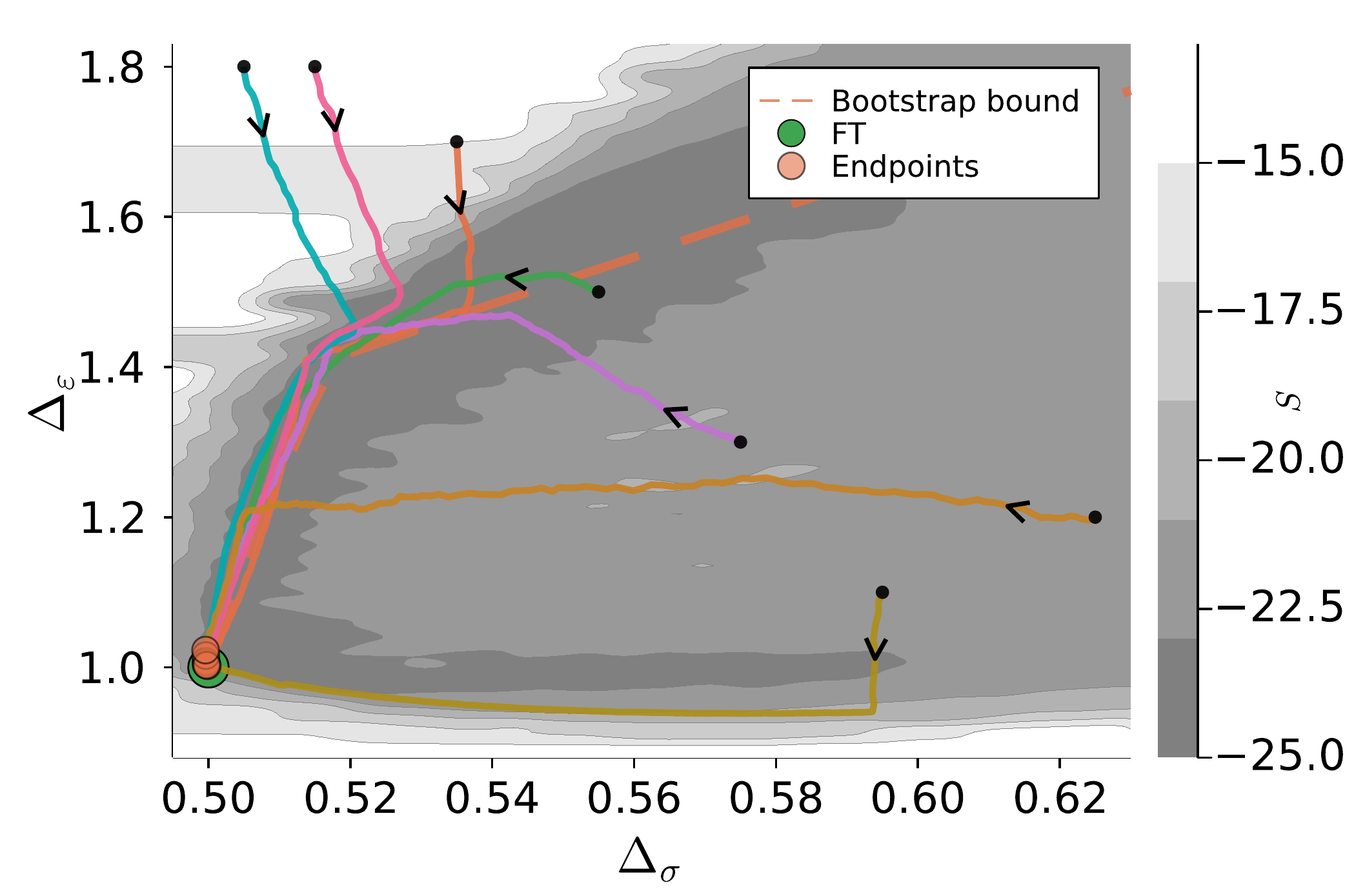}
\caption{Convergence of different trajectories to the $d=3$ free scalar CFT. The contour lines represent the minimum value of the action in each bin for a spectrum of the kind $3\_2\_2\_1$.
 The dashed line is the upper bound for $\Delta_\epsilon$  from \cite{El-Showk:2012cjh}. The black arrows on the colored lines indicate the direction of the MC evolution. 
}
    \label{fig:lowTMCevolution}
\end{figure}

 The most important property that comes out from this analysis, common to every scenario studied in this paper, is that the {\it global} minimum of $S$ corresponds to the FT (or the point $\Delta_\sigma = \Delta_\epsilon = 0$ for $d=2$). This has been observed in all our searches, and in particular for all the spin partition of operators which we considered.  
Of course, this does not imply the non-sensical result that the FT is the only consistent CFT. Rather, it implies that among all the physical CFTs allowed within the given assumptions at fixed truncation, within our numerical accuracy and in absence of further constraints, only the FT shows up as an isolated minimum of $S$.
   
In order to test the (non-)existence of other minima, we performed a very large number of MC minimizations at $T=10^{-4}$.
 This temperature, being of the order of the numerical error of the action,\footnote{See appendix \ref{sec:error}.} guarantees that the 
MC can only diffuse towards configurations of lower action, since the temperature is too low to cross any relevant barrier.
Independently of the choice of $n_{\ell}$ and of the  operator scaling dimensions $(\Delta_\sigma^{(0)}, \Delta_a^{(0)})$, all the  low temperature MC runs we performed  converge to the FT point 
within  $10^8$ steps, although longer times might be needed for higher $N_{{\rm Ops}}$.\footnote{There is an important exception described in detail in section \ref{subsec:1opspin}. In $d=3$ for spectra of the kind $1\_1\_1$ and $1\_1\_1\_1$ there are actually two local minima, though the FT still is the global one.}

\begin{figure}[t!]
    \centering
    \includegraphics[width=0.49 \linewidth]{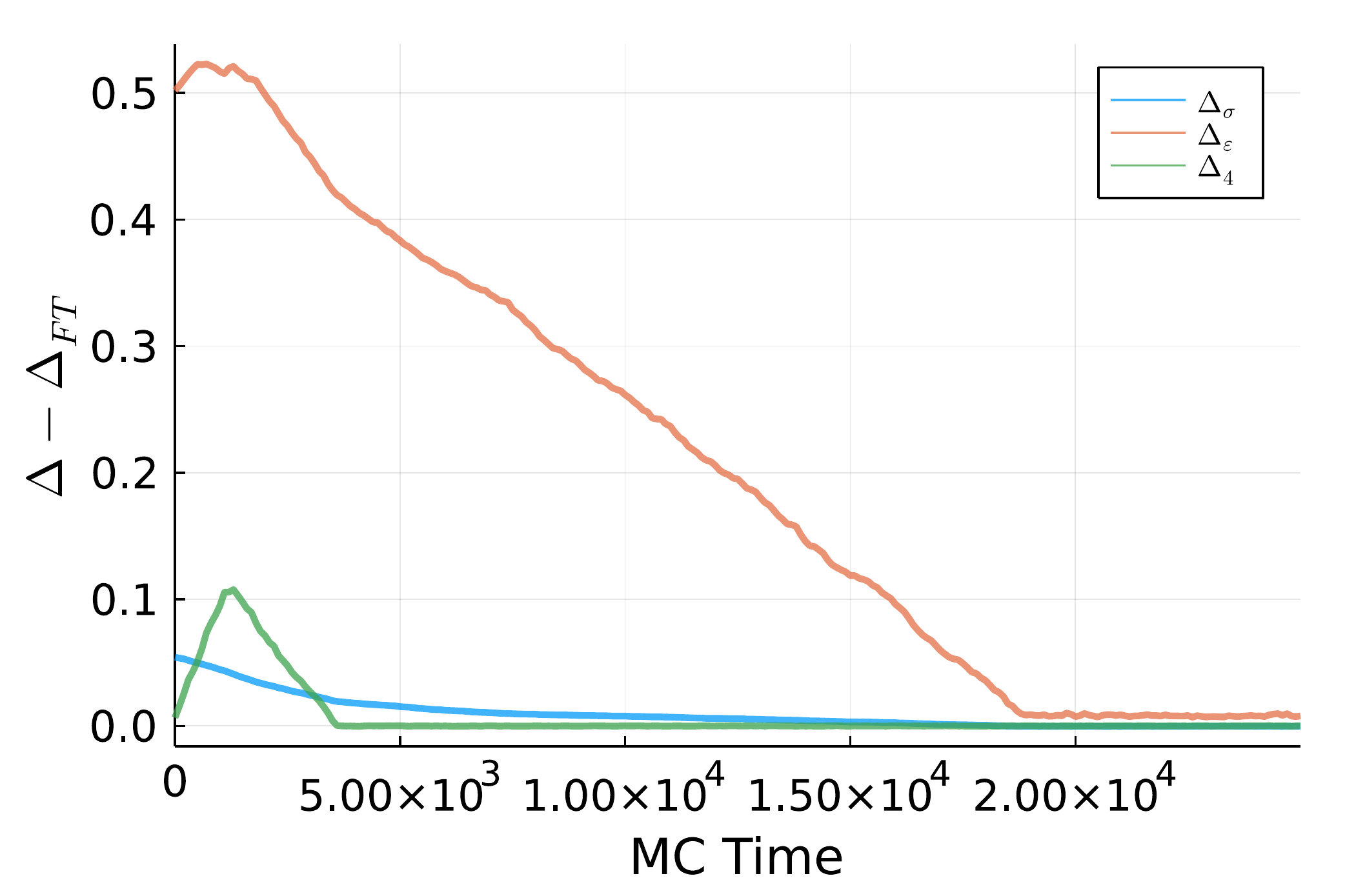}
    \includegraphics[width=0.49 \linewidth]{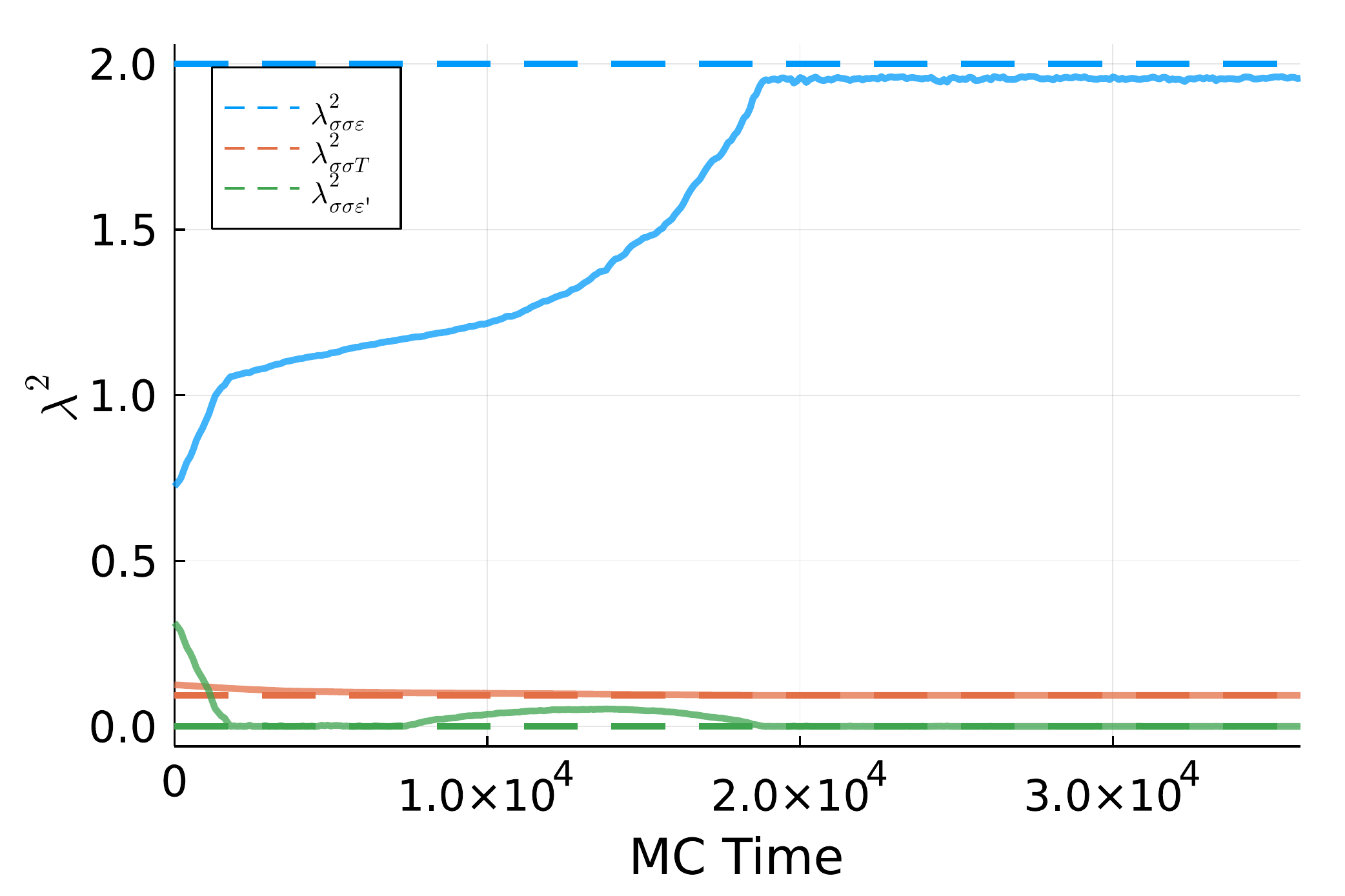}
\caption{
    Convergence in $ t_{\rm MC}$ for representative scaling dimensions (left) and squared OPE coefficients (right).
}
    \label{fig:MCconvergence}
\end{figure}

We report as an example in figure \ref{fig:lowTMCevolution} the low$-T$ evolution
for a $3d$ CFT spectrum with $\elmax=6$, $N_{{\rm Ops}} = 8$ and an initial spin partition $3\_2\_2\_1$.
Here the first spin two operator is fixed at $\Delta_T = 3$ and no gap on the number of relevant scalars has been imposed.\footnote{Note that in $d=3$ (the case shown in figure \ref{fig:lowTMCevolution}), if a scalar gap is imposed (for example, demanding that there can only be one exchanged relevant scalar), depending on the 
initial CFT data it can happen that the FT cannot be reached and a local minimum is induced.}
The starting point of each colored line (indicated with a black bullet) corresponds to a MC with initial values of $(\Delta_\sigma^{(0)}, \Delta_\epsilon^{(0)})$ 
at that point. The black arrows on the colored lines indicate the direction of the MC evolution. As can be seen, the final point of all the MC runs (highlighted by red circles) have
 $\Delta_\sigma\approx 1/2, \Delta_\epsilon\approx 1$. 
The contour plot shown in the background is indicative of the value of $S$ for the same spin partition but sampled at a much higher temperature (roughly 100 times larger). 
 A two-dimensional representation of $S$ is then obtained by binning the points visited on the $(\Delta_\sigma,\Delta_\epsilon)$-plane and plotting the minimum value of $S$ in each bin.\footnote{Understood as a cell of fixed width in $\Delta_\sigma$ and $\Delta_\epsilon$.} They are reported as a visual aid.
As can be seen from the figure, the initial values of $(\Delta_\sigma^{(0)}, \Delta_\epsilon^{(0)})$ can be taken in the excluded region, in which case they quickly approach
 the extremality line. The discrepancy between the rigorous bootstrap extremal line and the trajectories of our MC solutions is due to the severity of our truncation.
During the MC evolution the OPE coefficients of the extra scalars and higher spin operators decrease in such a way that at the final point we 
effectively have the FT spectrum with one operator per spin. We illustrate this phenomenon by plotting in figure \ref{fig:MCconvergence} the scaling dimensions of $\sigma$, $\epsilon$ and the spin$-4$ current as well as the squared OPE coefficients of $\epsilon$, $T$ and $\epsilon'$ (the second exchanged scalar) associated to the green line starting at $(0.55,1.5)$ in figure \ref{fig:lowTMCevolution}.  
A similar situation occurs for other choices of initial partitions and by keeping $\Delta_T$ unconstrained. The same also holds in $d=2$ and $d=4$ dimensions. This is the first important result of our analysis: for most of the spin partitions studied in this work, the landscape of $S$ has a single minimum, which can be reached by trivial gradient descent from (generically) any initial configuration. Note that most trajectories in figure \ref{fig:lowTMCevolution} reach the FT point following two broad paths in the ($\Delta_\sigma,\Delta_\epsilon)$-plane: i) the extremality line from above, ii) a curve, almost constant in $\epsilon$, from below. 
The presence of regions in the ($\Delta_\sigma,\Delta_\epsilon)$-plane where the bootstrap equations are more easily satisfied is going to be a common theme in our results. 

As we are going to show next, if one fixes at least another $\Delta$ one can find local minima along these slices. In the following subsections we will apply the protocol described in section \ref{sec:protocol} for fixed values of $\Delta_\sigma$ and we will show that 
some of the local minima that are present correspond to known theories (for example the Ising model). However, it is important to remark that none of the minima that we will describe are stable if $\Delta_\sigma$ is allowed to vary, not even the minimum corresponding to the Ising models in $d=2$ and $d=3$, in agreement with the results presented in this subsection. 
 \begin{figure}[t!]
    \centering
    \includegraphics[width=0.99 \linewidth]{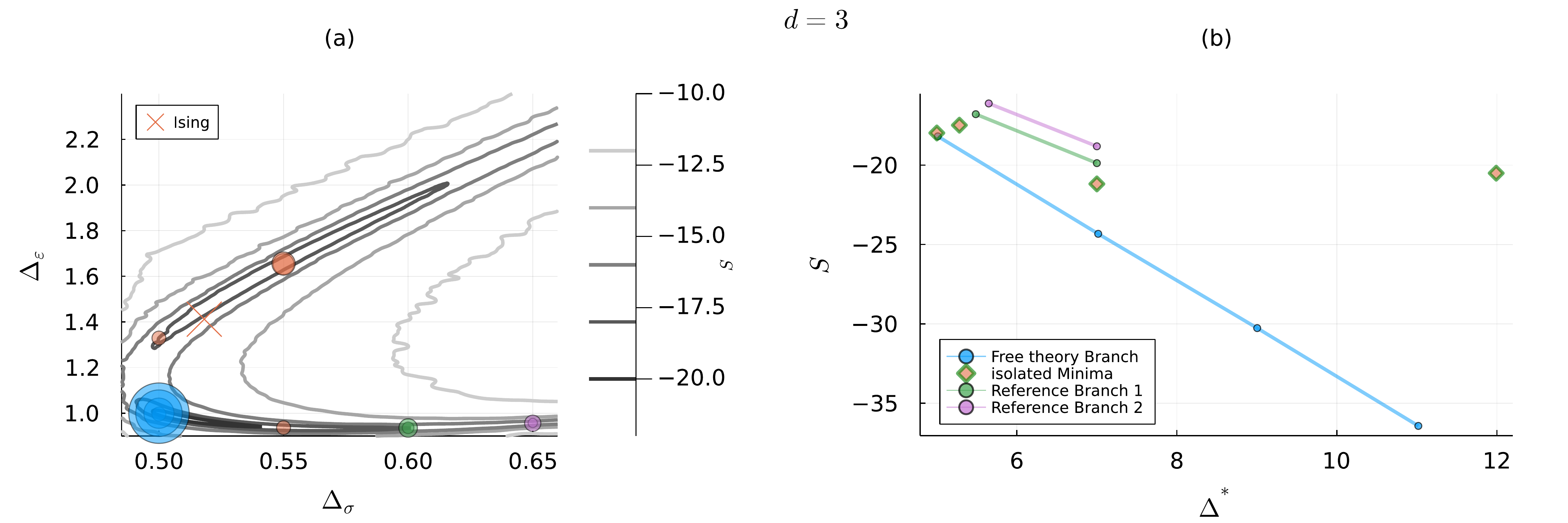}
    \includegraphics[width=0.99 \linewidth]{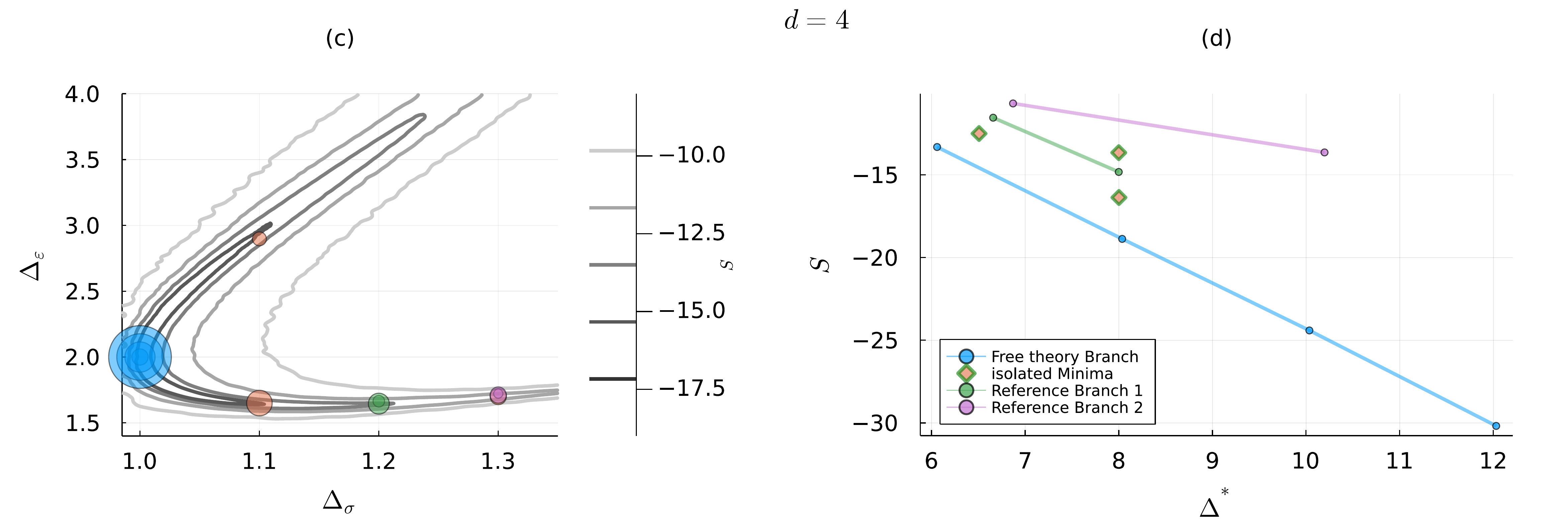}
\caption{Overview of the different minima found in $d=3$ (top) and $d=4$ (bottom) with spectra containing one operator per spin. The color code is the same across the panels.
    \emph{Left:} Minima at fixed $\Delta_\sigma$. 
The radii indicate the value of the action $S$. The larger the radius the smaller $S$. The contour lines are taken from an high$-T$ search with $\ell_{max} = 6 $.
    \emph{Right:} Value of $S$ for each point along the colored branches in the left panel, identified by its $\Delta^*$.}
    \label{fig:FT3d-dsigdeps-oop}
\end{figure}

\begin{figure}[t!]
    \centering
    \includegraphics[width=0.85
    \linewidth]{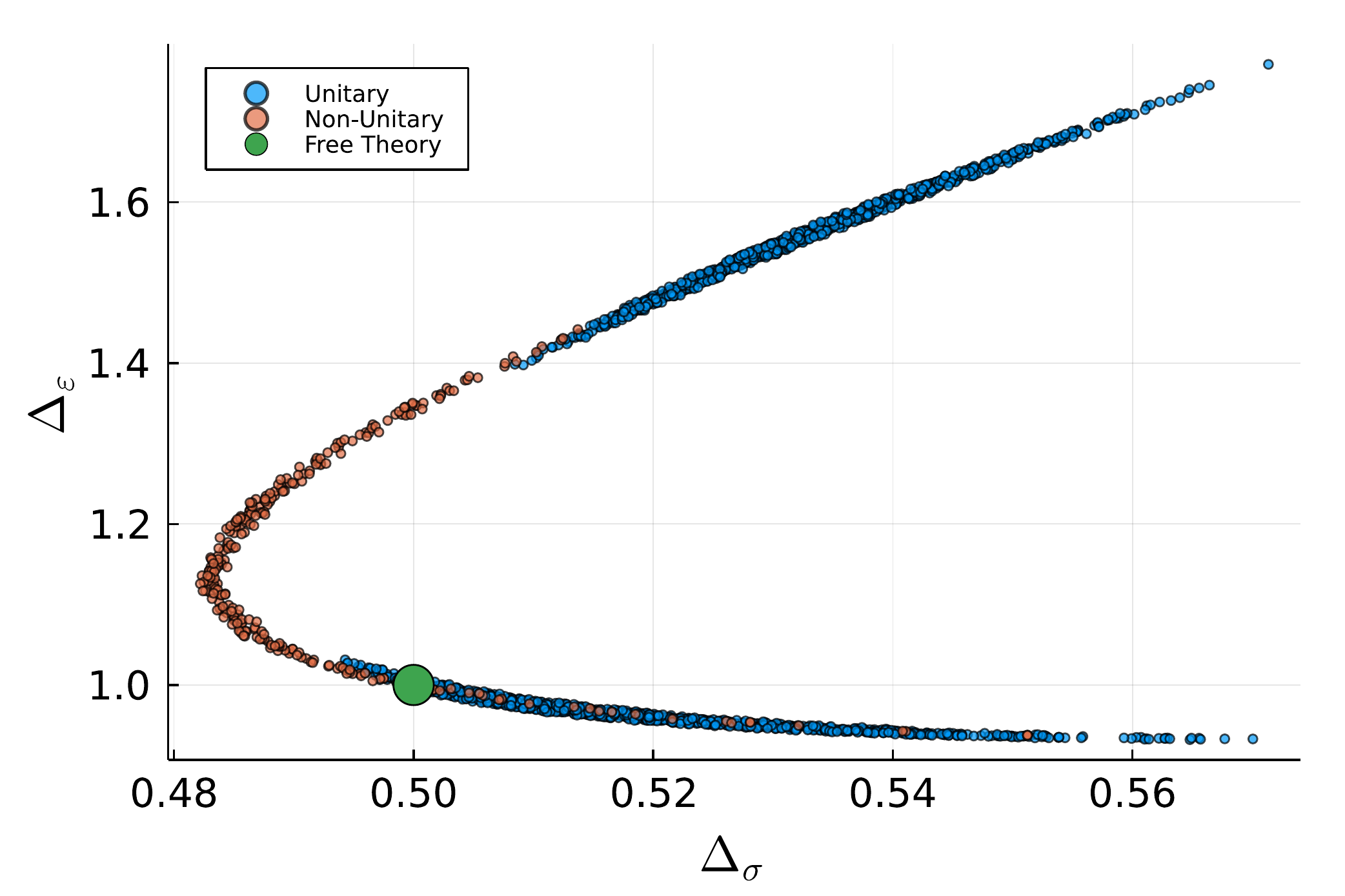}
\caption{Locus of the lowest-action points visited by a wide-search MC in $d=3$ with spectrum $1\_1\_1\_1$.
When unitarity is imposed we find two disjoint basins. Allowing for non-unitary configurations gives rise to a single connected basin.}
    \label{fig:d3-nonUnit-OOPS}
\end{figure}

\subsection{One operator per spin: $d=3,4$} 
\label{subsec:1opspin}

We now describe the results of a  systematic search of local minima at fixed $\Delta_\sigma$ by the protocol described in subsection \ref{sec:protocol}. It is instructive to begin by exploring the simplest possible scenario where
\begin{equation}
    n_\ell=1, \quad \ell=0,2,\cdots , \ell_{{\rm max}}\,.
    \label{eq:nl1}
\end{equation}
Although this assumption looks very restrictive, some basic features of the shape of the landscape of CFTs is already visible in this set-up.
Step $3$ of the protocol is clearly unnecessary when $n_{\ell}=1$, since all the configurations in each run will belong to just one spin sector. 
We focus on $d=3$ and $d=4$ CFTs and ask the following question: aside from the FT, are there other viable approximate solutions to the crossing equations in which \eqref{eq:nl1} is satisfied? This question might seem academic, given that interacting CFTs have an infinite number of Regge trajectories \cite{Caron-Huot:2017vep}. However, we are severely truncating the spectrum, so the question  we address here is if we can have CFTs where there is a substantial gap between the first Regge trajectory and the others.

An extensive high-$T$ search in $d=3$ with $\elmax = 6$ and free $\Delta_\sigma$ -- the contour lines in the background of figure \ref{fig:FT3d-dsigdeps-oop} (a) -- shows that there are two (very elliptic) disjoint basins of attraction.
The one at the bottom of the panel contains the FT. The upper one, elongated along the extremality bootstrap line, approximately contains the Ising point in the $(\Delta_\sigma,\Delta_\epsilon)$-plane. Note that the latter basin of attraction has a higher action with respect to the former one.

We searched local minima of $S$ at different fixed values of $\Delta_\sigma=1/2+n/20$, $n=0,1,2,3$, with input spin sectors $1\_1\_1$, $1\_1\_1\_1$, $1\_1\_1\_1\_1$ and $1\_1\_1\_1\_1\_1$. The minima that we found are represented in figure \ref{fig:FT3d-dsigdeps-oop} (a). 
In order to better appreciate the value of $S$ at the minima $i$, the latter are indicated with a circle whose radius $r_i$ is  given by
\begin{equation}
    r_i \propto S_{{\rm max}}-S_i + \alpha\,,
    \label{eq:riSmax}
\end{equation}
where $S_{{\rm max}}$ is the value of $S$ in the minimum with larger action, and $\alpha$ is an offset so that this minimum can still be visible in the figure.

For $n=0,1$ two classes of minima are found, one for each of the two basins of attraction. For $n=2,3$, instead, we find only  minima along the lower basin, with $\Delta_\epsilon<1$. It should be clear  that there is nothing special to the above chosen values of $\Delta_\sigma$. For each fixed value of $\Delta_\sigma$ we expect to find two local minima if $\Delta_\sigma\lesssim 0.6$ and one for higher values. 

We then use the protocol described in section \ref{sec:branch} with \texttt{relTol}$=0.1$ and \texttt{maxMismatch}$=2$ to identify the different branches, which correspond to sets of minima with different input sectors that can however be considered to represent the same ``theory'' at different truncations.
In this way we find that, among these minima, some are isolated and do not from any branch (orange diamonds in panel (b) of figure \ref{fig:FT3d-dsigdeps-oop}), while others form branches (green, pink and blue bullets). The only branch in which the action decreases as a function of  $\Delta^*$  is the one associated to the free theory, the blue branch in  panel  (b) of figure \ref{fig:FT3d-dsigdeps-oop}. 

In $d=4$ a similar  high-$T$ search with $\elmax = 6$ and free $\Delta_\sigma$ gives rise to the contour lines in the background of  panel (c) of figure \ref{fig:FT3d-dsigdeps-oop}. 
A search of the minima at different fixed values of $\Delta_\sigma=1+n/10$, $n=0,1,2,3$, with input sectors $1\_1\_1$, $1\_1\_1\_1$, $1\_1\_1\_1\_1$ and $1\_1\_1\_1\_1\_1$ with \texttt{relTol}$=0.1$ and \texttt{maxMismatch}$=2$ allows us to identify, also in this case, some local minima at fixed $\Delta_\sigma$. Like in the $d=3$ case,  some minima are isolated and do not from any branch (orange circles in panel (d) of figure \ref{fig:FT3d-dsigdeps-oop}), while others form branches (green, pink and blue circles). The only branch in which $S$  consistently decreases with $\Delta^*$  is the one associated to the FT, the blue branch in   figure \ref{fig:FT3d-dsigdeps-oop} (d).

As we will show in the following, the main structure of the $S$ landscape with two  elongated basins of attractions (valleys) will appear repeatedly in the analysis of more complex situations. Note the striking difference between the $d=3$ and $d=4$ cases: in the former the basins are two, in the latter the basin is only one. One could argue that this difference is due to the presence of another actual theory in $d=3$ close to the FT point, the Ising theory, 
and none in $4d$ within the above assumptions. 
Indeed, our results in sections \ref{subsec:1opspin} and \ref{subsec:Z2cft} further point to this explanation. 

We also find that if the unitarity bounds are relaxed (both on $\Delta$ and on the OPE coefficients squared), the aforementioned disjoint valleys in $d=3$ join in a non-unitary point of the parameter space. This 
is shown in figure \ref{fig:d3-nonUnit-OOPS}, where we compare the lowest-lying points in an unitary MC search to those of a non-unitary setting.

\begin{table}[t!]
    \centering
    \caption{Scaling dimensions of the first operators appearing in the OPE of $\sigma$ in the Free Scalar Theory. In these settings $\Delta_\sigma = (d-2)/2$ and $\Delta_T = d$. 
    }
    \begin{tabular}{|c|c|c|c|c|c|}
    \hline
      & $\Delta_0$ ($\Delta_\epsilon$) & $\Delta_4$ & $\Delta_6$ & $\Delta_8$ & $\Delta_{10}$\\
    \hline
     
         $d=3$ & $ 1.0000 $ & $ 5.0000 $ & $ 6.9999 $ & $ 8.9999 $ & $ 11.0178 $   \\
         $d=4$ & $ 1.9999 $ & $ 6.0000 $ & $ 7.9999 $ & $ 9.9993 $ & $ 12.0308 $   \\
         \hline
    \end{tabular}
    \label{tab:FT}
\end{table}
\begin{table}[t!]
    \centering
    \caption{Scaling dimensions of $\sigma$ and of the first operators appearing in the OPE of $\sigma$ in the Free Scalar Theory. Besides unitarity the only constraint imposed is $\Delta_T = d$.}
    \begin{tabular}{|c|c|c|c|c|c|c|}
    \hline
    & $\Delta_\sigma  $
      & $\Delta_0$ ($\Delta_\epsilon$) & $\Delta_4$ & $\Delta_6$ & $\Delta_8$ & $\Delta_{10}$\\
    \hline
         $d=3$ & $ 0.50009 $ & $ 0.9997 $  & $ 5.0009 $ & $ 6.9974 $ & $ 8.9973 $ & $ 11.1505 $   \\
         $d=4$ & $ 1.00000 $ & $ 2.0035 $  & $ 6.0000 $ & $ 7.9995 $ & $ 9.9997 $ & $ 12.0368 $ \\
         \hline
    \end{tabular}
    \label{tab:FT-freedsig}
\end{table}
\begin{table}[t!]
    \centering
    \caption{Comparison between the exact and the numerically determined values of the OPE coefficients squared for the first operators in the Free Scalar Theory. Besides unitarity the only constraint imposed is $\Delta_T = d$.}
    \begin{tabular}{|c|c|c|c|c|c|c|}
    \hline
      &  $\lambda^2_\epsilon$ & $\lambda^2_T$ & $\lambda^2_4$& $\lambda^2_6$ & $\lambda^2_8$ & $\lambda^2_{10}$\\
    \hline
         $d=3$ & $ 2.0022 $ & $ 0.093796 $ & $ 0.0042723 $ & $ 0.00022128 $ & $ 1.1604\times 10^{-5} $ & $ 8.3384 \times 10^{-7} $  \\
         Exact & $ 2.0 $ & $ 0.09375 $ & $ 0.0042725 $ & $ 0.0002203 $ & $ 1.1986\times 10^{-5} $ & $ 6.7214\times 10^{-7} $   \\
         \hline
         $d=4$ & $ 1.9927 $ & $ 0.33343 $ & $ 0.02853 $ & $ 0.0021857 $ & $ 0.00014554 $ & $ 1.4538\times 10^{-5} $     \\
         Exact & $ 2.0 $ & $ 0.33333 $ & $ 0.028571 $ & $ 0.0021645 $ & $ 0.0001554 $ & $ 1.0825\times 10^{-5} $      \\
         \hline
    \end{tabular}
    \label{tab:FT-OPEs}
\end{table}

The main take-away of this test is that the evolution with respect to $\Delta^*$ of the action in branches formed by akin minima seems to be a discriminant factor between spurious minima and physically meaningful ones (the FT in this case); namely, the spurious minima do not create long branches that go down into small values of the action $S$, while the FT is present for every truncation with very high consistency. 
We find that in the FT minimum high spin operators are determined quite accurately ($\sim 1\%$ error) except for the highest$-\Delta$ at each truncation, as can be seen in table \ref{tab:FT}. 

Although it is clear that the minima at $\Delta_\sigma = (d-2)/2$ are deeper than those away from it, we performed a MC search at $\elmax=10$ without any constraint on $\Delta_\sigma$ besides unitarity. This allows us to understand the precise shape of the landscape around the FT point. 
After $2\times 10^8$ steps of an extensive search ($T\sim 1$), we chose the global minimum and performed a $T = 10^{-4}$ minimization in order to refine the precise location of this minimum. 
In table \ref{tab:FT-freedsig} we report these values. For completeness we also show in table \ref{tab:FT-OPEs} the OPE coefficients determined in this case.

We finally note that for higher $\Delta^*$, namely higher $N_{{\rm Ops}}$, the FT minimum is much deeper than the other minima. This means that the latter are very poor solutions to the crossing equations.  In other words, we did not find other consistent approximate solutions to the crossing equations, besides the FT, where \eqref{eq:nl1} is satisfied up to $\elmax=10$.

\subsection{More operators per spin} 
\label{subsec:Z2cft}

We now use our protocol to study CFTs with more than one operator per spin
with the assumptions discussed at the beginning of the section (presence of an effective $\mathbf{Z}_2$ symmetry and only one relevant $\mathbf{Z}_2-$even scalar). 

We here consider viable solutions to the crossing equations the end-point of branches\footnote{Our parameters for determining the connection between minima are \texttt{relTol}$=0.2$ and \texttt{maxMismatch}$=2$.} that arrive up to a given $\ell_{max}$ and include at least 2 different minima with strictly decreasing $S$.  For brevity, we will refer to these end points  simply as ``end-minima''. For example, in the case of only one operator per spin, the ``end-minima'' are the blue, pink and green circles at the end of the branches in the right panels of figure \ref{fig:FT3d-dsigdeps-oop}.
The rationale behind concentrating on end-minima is that they summarize the information from minima at higher $\elmax$ and thus allow for a cleaner visualization.  
Since this approach might exclude  possible branches that start only at the largest $\Delta^*$ here considered, 
 we will also include in our discussion isolated minima at the largest $\elmax$ considered. They will be also called end-minima in a slight abuse of notation. 
 
Our whole search space includes spectra with $\elmax = 4, 6, 8, 10$. For  $d=2$ we limit the analysis to $\elmax = 8$, since at $\elmax = 10$ the  end-minima that we find are so many that they obscure the visualization.
 Here we only observe that, qualitatively, all the minima that we find at $\elmax = 10$ differ only by the least relevant operators.

The role of the stress-energy tensor deserves some consideration. For the study of local theories it is necessary to impose the presence of a spin$-2$ operator with $\Delta_T=d$. In practice, we have also found necessary to impose a gap on the scaling dimension $\Delta_{T'}$ of the second spin$-2$ operator so that it does not come too close to $d$ and create the effect of a non-local theory by ``supplanting'' the stress energy tensor. In section \ref{sec:DT} we will study the effect of allowing for non-local theories but hereafter it should be understood that unless specified otherwise we fix $\Delta_T=d$ and impose the gap $\Delta_{T^\prime}\geq d+1$.

We report the end-minima found in the $(\Delta_\sigma,\Delta_\epsilon)$-plane in figures \ref{fig:d2-overview}, \ref{fig:d3-overview} and \ref{fig:d4-overview} for the $d=2,3,4$ cases, respectively. We discuss these results separately in the sections below, but a few considerations apply to all cases. The contour lines of the lowest-lying configuration sampled by an high$-T$ MC
is shown in the background as a visual aid. The orange dashed line corresponds to rigorous bootstrap bounds, while the green continuous line is the line of GFTs. We use the radius of the points to represent the action as before (see \eqref{eq:riSmax}) and use transparent markers so that coincident minima appear darker. To understand the fact that several minima are superimposed in this representation  it is important to keep in mind that there can be up to $10$ other dimensions not shown in the plot. Indeed, in many cases there are end-minima which have almost identical values of $\Delta_\sigma$ and $\Delta_\epsilon $ but differ in the scaling dimensions of the other operators. 

\subsubsection{$d=2$}
\label{subsec:d2}

One can wonder if our method can work in $2d$ given that the landscape of CFTs is notoriously very rich. On the other hand, $2d$ is the ideal playground to test the method, since entire classes of CFTs are exactly soluble.

We have considered various MCs for 8 different fixed values of $
\Delta_\sigma$: 
\begin{equation}
  \Delta_\sigma = \frac{1}{8}, \, \frac{3}{20}, \, \frac{7}{40}, \,\frac{1}{5}, \,\frac{1}{4}, \,\frac{2}{7}, \, \frac{3}{10}\,,  \frac{\sqrt{5}}{10}\,.
  \label{eq:MM0}
\end{equation}
When $\Delta_\sigma$ is too small the numerical analysis becomes quite noisy. For this reason, we did not explore values of $\Delta_\sigma<1/8$. 
7 of the values in \eqref{eq:MM0} have been chosen to be ``small" rational numbers, while one was chosen irrational, e.g. numerically equivalent to a ``large" rational number. Rational scaling dimensions can be associated to exactly known rational conformal field theories (RCFT). 
As we will see, in order to limit the number of allowed theories and make the analysis feasible, the assumption of having only one relevant $\mathbf{Z}_2$-even scalar turns out to be particularly important in $2d$. Before presenting our results, it is useful to report some (by no means exhaustive) instances of known CFTs admitting scalars with the above scaling dimensions.

\paragraph{A glimpse at the $2d$ CFT Landscape}

The landscape of the known CFTs in $2d$ is amazingly rich. In contrast to the situation for $d>2$, in particular, $2d$ CFTs can have a continuum spectrum (e.g. Liouville field theory). It is clear that we cannot detect such theories.\footnote{Unless the continuum starts beyond $\Delta^*$ in all spin channels, in which case the CFT will simply appear discrete.}
Exactly marginal deformations are ubiquitous in $2d$. These theories will not give rise to minima of $S$, but to a valley of minima, each corresponding to a CFT as we vary the marginal couplings. Since we look for isolated minima in our search, such theories will also be difficult to detect.
The CFTs we can hope to find should have a discrete, relatively sparse, spectrum and should be isolated. Their central charge $c$ is expected to be of order one.
Prototypical theories of this kind are RCFTs. Even if we focus on these theories only, we need to further restrict our search space
to avoid a proliferation of theories. We will do so by assuming, as mentioned, that only one relevant scalar $\epsilon$ can appear in the $\sigma\times \sigma$ OPE.
The importance of this assumption can be appreciated by focusing on specific exactly soluble CFTs. We will briefly consider in what follows the space of diagonal unitary minimal models ($c<1$), the $S^1/\mathbf{Z}_2$ orbifold $(c=1)$ and $SU(2)_k$ Wess-Zumino-Witten  (WZW) models ($c>1$). 

Recall that the scaling dimensions of the (Virasoro) primary fields $\phi_{r,s}$ in diagonal unitary minimal models ${\cal M}(m+1,m)$ is (conventions as in \cite{di_francesco_philippe_conformal_1997}):\footnote{In the $2d$ CFT literature, Virasoro and $SL(2,\mathbb{C})$ primaries are usually denoted primaries and quasi-primaries (respectively). In numerical bootstrap analysis we usually care about $SL(2,\mathbb{C})$ primaries, which are the analogues of the primary fields in $d>2$ CFTs. Most of the considerations that follow can be made at the level of Virasoro primaries with no need to decompose them in $SL(2,\mathbb{C})$ primaries.}
\begin{equation}
    \Delta_{r,s} = h_{r,s} + \bar h_{r,s}\,, 
\end{equation}
where 
\begin{equation}
h_{r,s} = \bar h_{r,s} = \frac{((m+1)r - ms)^2-1}{4m(m+1)}\,, \;\; 1\leq r \leq m-1\,, 1\leq s \leq m\,, \; m \in \mathbf{Z}_{\geq 3}.
\label{eq:MM2}
\end{equation}
The range in $s$ in \eqref{eq:MM2} can be restricted to $1\leq s\leq r$
using the field equivalence
\begin{equation}
    \phi_{r,s} = \phi_{m-r,m+1-s}\,.
    \label{eq:MM3}
\end{equation}
Minimal models feature a faithful discrete $\mathbf{Z}_2$ symmetry for any $m$ (see e.g. \cite{Ginsparg:1988ui} for a pedagogical exposition).\footnote{The presence of such a $\mathbf{Z}_2$ symmetry is clear from the equivalence of minimal models with Wess-Zumino-Witten cosets of the form [$SU(2)_{m-2}\times SU(2)_1]/SU(2)_{m-1}$, where $\mathbf{Z}_2$ is the combination of the two $\mathbf{Z}_2$ center symmetries of $SU(2)_{m-2}$ and $SU(2)_1$ which is not gauged by the center of $SU(2)_{m-1}$.} Using the equivalence class \eqref{eq:MM3}, for any $r$ and $s$ we can choose the representative field $\phi_{r,s}$ with $r+s$ an even integer. On these representatives, 
$\mathbf{Z}_2$-even fields are those with $r$ and $s$ odd, while $\mathbf{Z}_2$-odd fields have $r$ and $s$ even (the identity operator is $\phi_{1,1}$).
It is a straightforward exercise to show that there are one or more unitary minimal models which contain a $\mathbf{Z}_2$-odd field $\sigma$ with scaling dimension equal to each of the first 7 values of $\Delta_\sigma$ reported in \eqref{eq:MM0}.\footnote{Of course, minimal models associated to the irrational value of $\Delta_\sigma$ reported in \eqref{eq:MM0} can also be found provided we take $m$ large enough and approximate $\sqrt{5}/10$ with an appropriately chosen rational number.} For instance, for $\Delta_\sigma = 3/20 $ within minimal models with $m\leq 100$ we find two which contain a Virasoro primary with that dimension: $\phi_{8,8} \subset {\cal M}(15,14)$ and 
$ \phi_{14,14} \subset {\cal M}(26,25)$. The fusion rules $\sigma \times \sigma$ contain respectively 42 and 132 Virasoro primaries. Aside from the identity operator, 11 and 21 of these $\mathbf{Z}_2$-even fields have $\Delta<2$, respectively. The smallest Virasono primaries in the two cases are $\phi_{5,5}$ and $\phi_{3,3}$, with scaling dimensions $\Delta_\epsilon = 2/105, 2/325$, respectively. 
Models in such kind of field correlators cannot be detected by our method, because i) their OPE contain too many light operators and ii) operators with a too small scaling dimension makes the numerical analysis unfeasible. 
Imposing that we have only one relevant $\mathbf{Z}_2$-even scalar removes most of these field identifications. The only identification left, for any $m\geq 3$, is the one where $\sigma=\phi_{m-1,m-1}(\sim \phi_{1,2})$ and  $\epsilon=\phi_{1,3}$, already considered in \cite{Rychkov:2009ij}, which have
\begin{equation}
    \Delta_\sigma = \frac 12 - \frac{3}{2(m+1)} \,, \qquad \Delta_\epsilon=2-\frac{4}{m+1}\,,
    \label{eq:DsDe}
\end{equation}
and the simple fusion rules
\begin{equation}
    \sigma \times \sigma = 1 + \epsilon\,.
\label{eq:MMFusion}
\end{equation}
Only 4 among the 7 rational values of $\Delta_\sigma$ in \eqref{eq:MM0} equal the scaling dimensions of $\phi_{m-1,m-1}$: ordinary $(m=3, \Delta_\sigma=1/8)$, tricritical $(m=4, \Delta_\sigma=1/5)$, tetracritical $(m=5, \Delta_\sigma=1/4)$ and pentacritical $(m=6, \Delta_\sigma=2/7)$ Ising theories. Note that for even $m$ we have $\mathbf{Z}_2(\sigma) = +\sigma$. However, as far as we consider the four-point function of $\sigma$ only, given the fusion rules \eqref{eq:MMFusion}, we can effectively assign an unfaithful $\mathbf{Z}^\prime_2$ charge to  $\sigma$ such that $\mathbf{Z}_2^\prime(\sigma) = -\sigma$ and identify such $\mathbf{Z}_2^\prime$ as the global discrete symmetry of the four-point correlator.

Another notable example which shows the importance of restricting to one relevant scalar is given by the theory of a free compact scalar on a $S^1/\mathbf{Z}_2$ orbifold. 
This CFT has two Virasoro primaries with $\Delta_{\sigma} = 1/8$, which are the twisted ground states of the orbifold. They are odd under the quantum $\mathbf{Z}_2$ global symmetry that emerges after the gauging and hence we can consider any of them as our field $\sigma$.
As well-known, the $S^1/\mathbf{Z}_2$ orbifold is a $c=1$ theory admitting an exactly marginal deformation, given by the radius $R$ of $S^1$. 
At $R^2=1$ the theory is equivalent to a decoupled pair of Ising theories.\footnote{In our normalizations, like those of \cite{di_francesco_philippe_conformal_1997}, the self-dual radius is $R^2=2$.} Deforming the radius gives rise to a family of theories related to the Ashkin-Teller model, namely the theory obtained by coupling two Ising models \cite{zamolodchikov_two-dimensional_1986}. The OPE of two twist fields give rise to an infinite number of Kaluza-Klein and winding vertex operators (Virasoro primaries) with scaling dimensions \cite{Dixon:1986qv}
\begin{equation}
    \Delta_{m,n} = \frac{m^2}{R^2} + \frac{n^2 R^2}{4}\,,\qquad  m\in \mathbf{Z}\,, n\in 2\mathbf{Z}\,.
    \label{eq:orbifold}
\end{equation}
At $R^2=1$, the two Ising theories are decoupled and this theory is indistinguishable from a single Ising copy. When $R\neq 1$ the two couples to each other and we get a continuum of $c=1$ CFTs parametrized by $R$.
For any $R$ such theories contain more than 1 relevant $\mathbf{Z}_2$-even scalar, as we see from \eqref{eq:orbifold} by setting either $m$ or $n$ to zero. In this case, restricting to only one relevant $\mathbf{Z}_2$-even scalar is crucial to avoid a continuum of theories, which would be a challenge for our numerical methods.

\begin{figure}[t!]
    \centering
    \includegraphics[width=0.8 \linewidth]{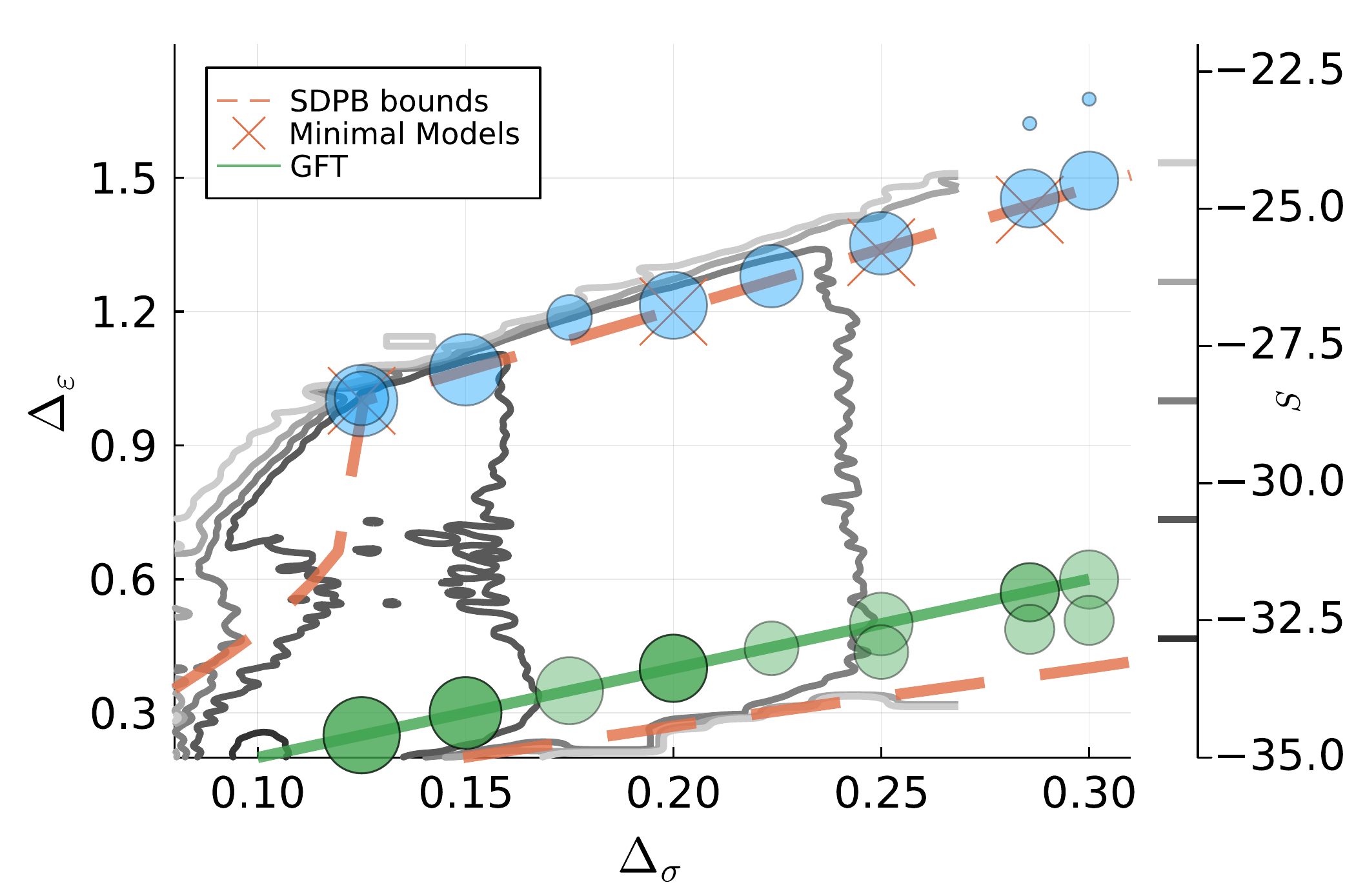}
\caption{$d=2$. Location of the end-minima with $\elmax \leq 8$ in the ($\Delta_\sigma,\Delta_\epsilon)$-plane. The end-minima belong to different sectors all of them contained in $ 4 \_4 \_3 \_2 \_1$.  The dashed orange line corresponds to the bounds obtained in \cite{Behan:2017rca} assuming only one $\mathbf{Z}_2$-even relevant scalar.
All the end minima correspond to local theories $\Delta_T = 2$, $\Delta_T' \geq 3$.}
    \label{fig:d2-overview}
\end{figure}

The external operator $\sigma$ could also be part of a CFT with a continuous global symmetry $G$. CFTs with continuous global symmetries should of course be analyzed in a covariant way exploiting the symmetry, see e.g. \cite{Rattazzi:2010yc}. However, if $G\supset \mathbf{Z}_2$ the field $\sigma$ could be considered some (real) component of a multiplet in a representation of $G$ for which $\mathbf{Z}_2(\sigma) = - \sigma$. For example, $\sigma$ could be the real component of a spin $j$ multiplet, with half-integer $j$, of a $SU(2)_k$ WZW model. More specifically, we could have
\begin{equation}
    \sigma \sim \phi_{j,m} (z) \bar\phi_{j,-m} (\bar z) +\phi_{j,-m} (z) \bar\phi_{j,m} (\bar z) \,,
    \label{eq:WZWspinj}
\end{equation} 
where $\phi_{j,m}(z)$ and $\bar \phi_{j,\bar m}(\bar z)$ are the holomorphic and  anti-holomorphic highest-weight state components of the spin $j$ field. 
The symmetry $\mathbf{Z}_2$ could be identified with the center of either the holomorphic or of the anti-holomorphic $SU(2)$ gauge group. With the identification \eqref{eq:WZWspinj}, we have
\begin{equation}
    \Delta_\sigma = \frac{2j(j+1)}{k+2}\,,\qquad  j\in \frac{\mathbf{Z}}{2} \,.
    \label{eq:WZW0}
\end{equation}
At $\Delta_\sigma=1/5, 2/7$ there are no solutions with integer $k$ and half-integer $j$ of \eqref{eq:WZW0}. For the remaining rational values in \eqref{eq:MM0} we have an infinite number of solutions with increasing values of $j$ and $k$. Solutions with $j>1/2$ have more than one relevant $\mathbf{Z}_2$-even scalar exchanged in the $\sigma\times \sigma$ OPE. On the other hand, for $j=1/2$ we get only one relevant $\mathbf{Z}_2$-even scalar with\footnote{We also have marginal scalars with $\Delta=2$ associated to the $J\bar J$ operators arising from the identity character.}
\begin{equation}
    \Delta_\epsilon = \frac{4}{k+2}\,.
\end{equation} 
This occurs for 4 of the values reported in \eqref{eq:WZW0}: 
\begin{equation}
    \Big(k=3,\Delta_\sigma =\frac{3}{10}\Big), \quad \Big(k=4, \Delta_\sigma =\frac{1}{4}\Big), \quad \Big(k=8, \Delta_\sigma =\frac{3}{20}\Big), \quad \Big(k=10,\Delta_\sigma =\frac{1}{8}\Big).
    \label{eq:possWZW}
\end{equation}

This quick detour on some well-known $2d$ CFTs shows the importance of reducing the space of viable theories by demanding only one relevant scalar.
Needless to say, the space of CFTs is vastly larger than the three classes we discussed, so it is likely that other CFTs with $\sigma$ operators with scaling dimensions as in \eqref{eq:MM0} and only one relevant $\mathbf{Z}_2$-even scalar exist. 

\begin{figure}[t!]
    \centering
    \includegraphics[width=0.75 \linewidth]{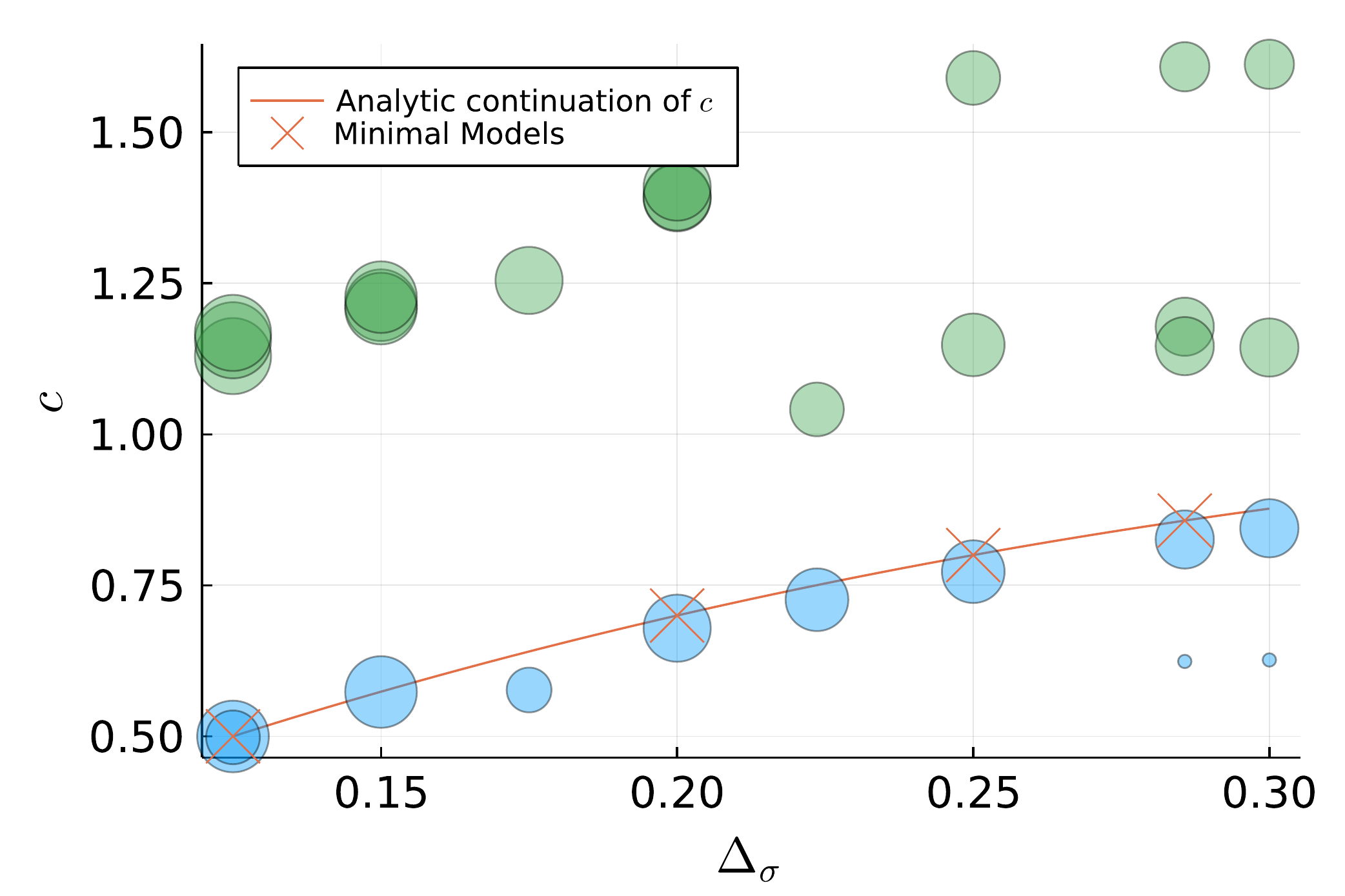}
\caption{Central charge $c$ as a function of $\Delta_\sigma$ for end-minima with $\elmax \leq 8$ in $d=2$. The end-minima belong to different sectors all of them contained in $ 4 \_4 \_3 \_2 \_1$. Comparison of our results to the analytic expectation \eqref{eq:cGMM} (red line) for the generalized minimal models.
All the end minima correspond to local theories $\Delta_T = 2$, $\Delta_T' \geq 3$.}
    \label{fig:d2-overview2}
\end{figure}

\paragraph{Results in $d=2$}

We report in figures \ref{fig:d2-overview} and \ref{fig:d2-overview2} the
location of the end-minima found with $\elmax \leq 8$ in the $(\Delta_\sigma,\Delta_\epsilon)$ and $(\Delta_\sigma,c$)-planes, respectively. The end-minima belong to different sectors, all of them contained in $4 \_4 \_3 \_2 \_1$. 

Most of the end-minima are aligned along two approximately straight trajectories in the $(\Delta_\sigma,\Delta_\epsilon$)-plane. The blue end-minima are along the upper extremal bootstrap bounds, while the green ones are aligned along the GFT line, despite the constraint $\Delta_T=2$. 

Let us first discuss the blue end-minima. Two less significant blue end-minima appear above the extremal line in the disallowed region, indication of being fake end-minima not associated to CFTs. The rest are in good agreement with the bootstrap bounds.
\begin{figure}[t!]
    \centering
    \includegraphics[width=0.75 \linewidth]{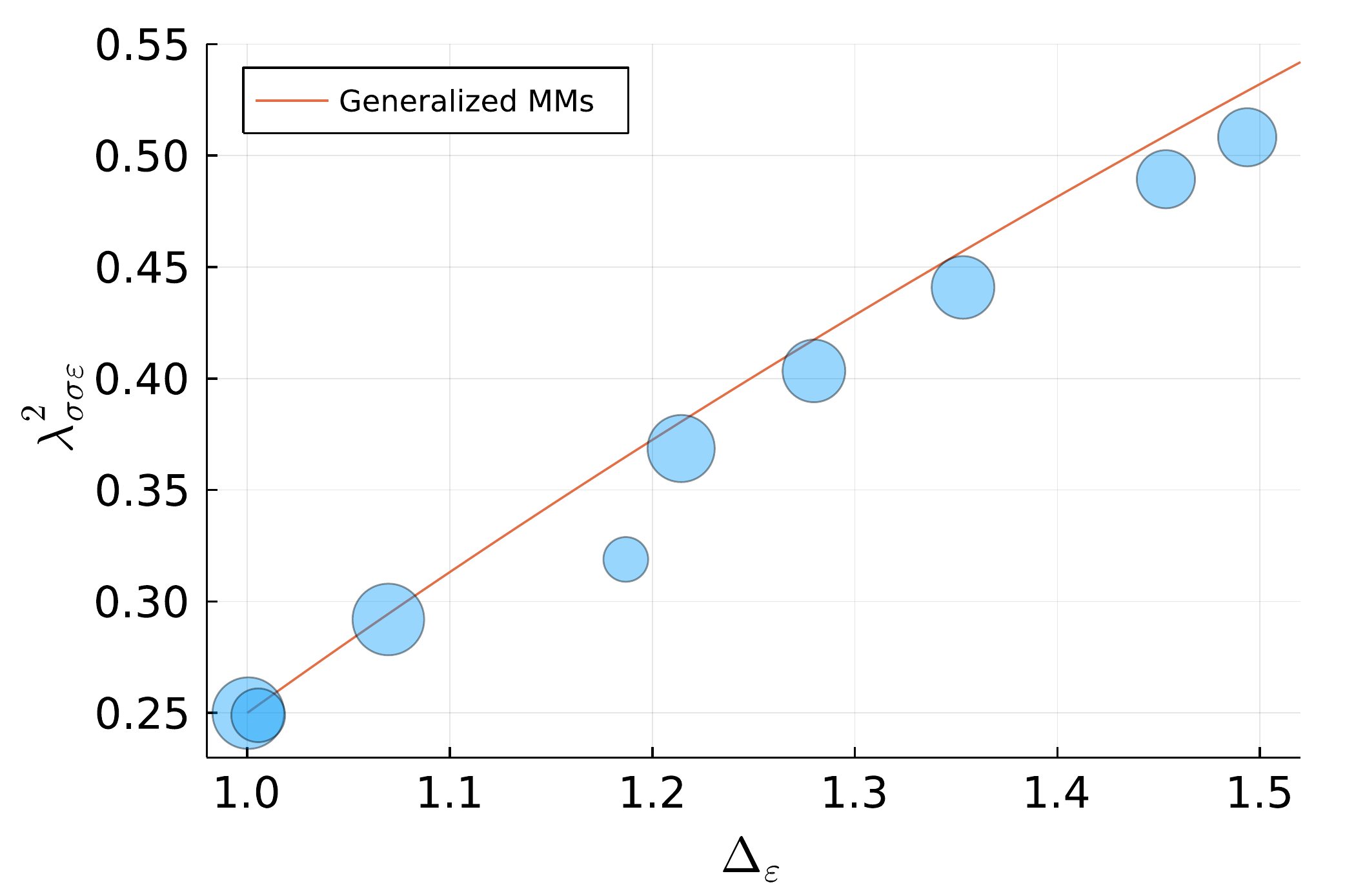}
\caption{$d=2$. Comparison between the OPE squared coefficient $\lambda^2_{\sigma\sigma\epsilon}$ numerically determined from the blue end-minima in figure \ref{fig:d2-overview} and its analytic expression for generalized minimal models (red line) \cite{Liendo:2012hy}, as a function of $\Delta_\epsilon$.}
    \label{fig:epsOPE}
\end{figure}

It is reassuring that, among such end-minima, we reproduce the first 4 minimal models (red crosses in figure \ref{fig:d2-overview}), with the field identification as in \eqref{eq:DsDe} discussed before. The remaining four points along the line correspond instead to the so called generalized minimal models \cite{Liendo:2012hy}, i.e. non-unitary minimal models obtained by analytically continuing the CFT data appearing in the 4-point correlation function from integer values of $m$ (minimal models) to real values of $m$ (generalized minimal models). Such theories are visible in the conformal bootstrap, despite the enforcement of unitarity, because the OPE coefficients squared of the exchanged quasi-primaries have been conjectured in \cite{Liendo:2012hy} (see also \cite{El-Showk:2014dwa}) and then proved in \cite{Behan:2017rca}, to be positive for any $m$. 

Given \eqref{eq:DsDe}, the blue end-minima should be aligned along the line
\begin{equation}
    \Delta_\epsilon = \frac 23 + \frac 83 \Delta_\sigma\,.
\end{equation}
In terms of the central charge, we have
\begin{equation}
    c = \frac{9\Delta_\sigma}{\Delta_\sigma+1}-4 \Delta_\sigma \,.
    \label{eq:cGMM}
\end{equation}
We compare in figure \ref{fig:d2-overview2} the central charge \eqref{eq:cGMM} with the one extracted from the OPE coefficient $\lambda_{\sigma\sigma T}$ by means of the relation
\begin{equation}
c = \frac{d}{d-1} \frac{\Delta_\sigma^2}{4 \lambda_{\sigma\sigma T}^2}\,,
    \label{eq:cc}
\end{equation}
valid for general $d$ dimensions. We see that the agreement is very good, despite the severity of our truncation, and confirm also the identification of such end-minima with generalized minimal models.

As a further check we compare in figure \ref{fig:epsOPE} the OPE coefficient squared $\lambda_{\sigma\sigma\epsilon}^2$ with the one analytically found for such models
in appendix B of \cite{Liendo:2012hy}.
The good agreement found is yet another indication of the accuracy of our determination and of the nature of the blue end-minima.
The only exception is the minimum at $\Delta_\sigma = 7/40$, which falls slightly above the SDPB bounds in figure \ref{fig:d2-overview} and its values of $c$ and $\lambda^2_{\sigma\sigma\epsilon}$ in figures \ref{fig:d2-overview2} and \ref{fig:epsOPE} are slightly misaligned with those of the generalized Minimal Models. We expect this to be due to some numerical artifact and that a more thorough study will find a lower action minimum that better agrees with the theoretical values.
 
\begin{figure}[t!]
    \centering
    \includegraphics[width=0.48 \linewidth]{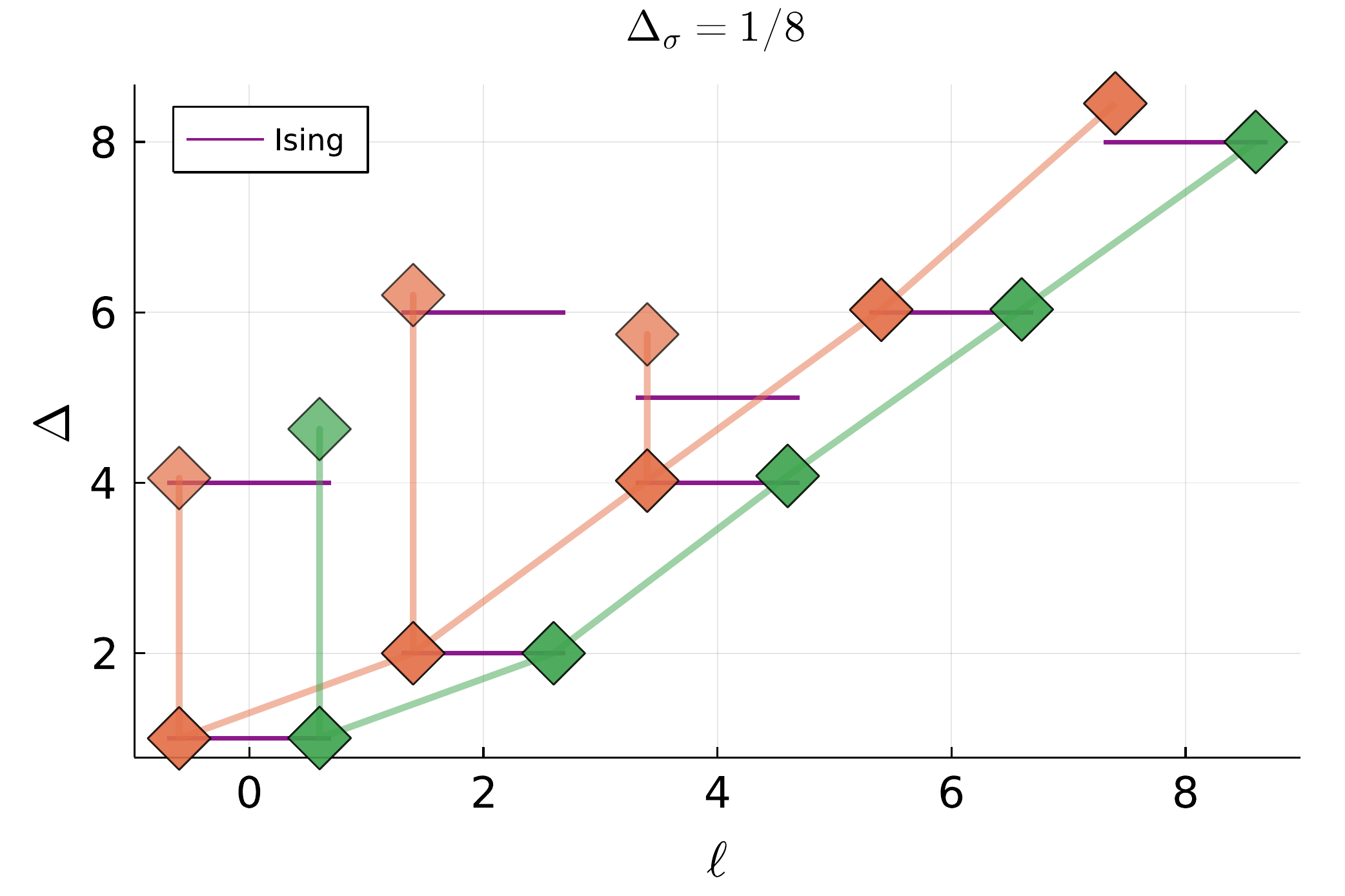}
    \includegraphics[width=0.48 \linewidth]{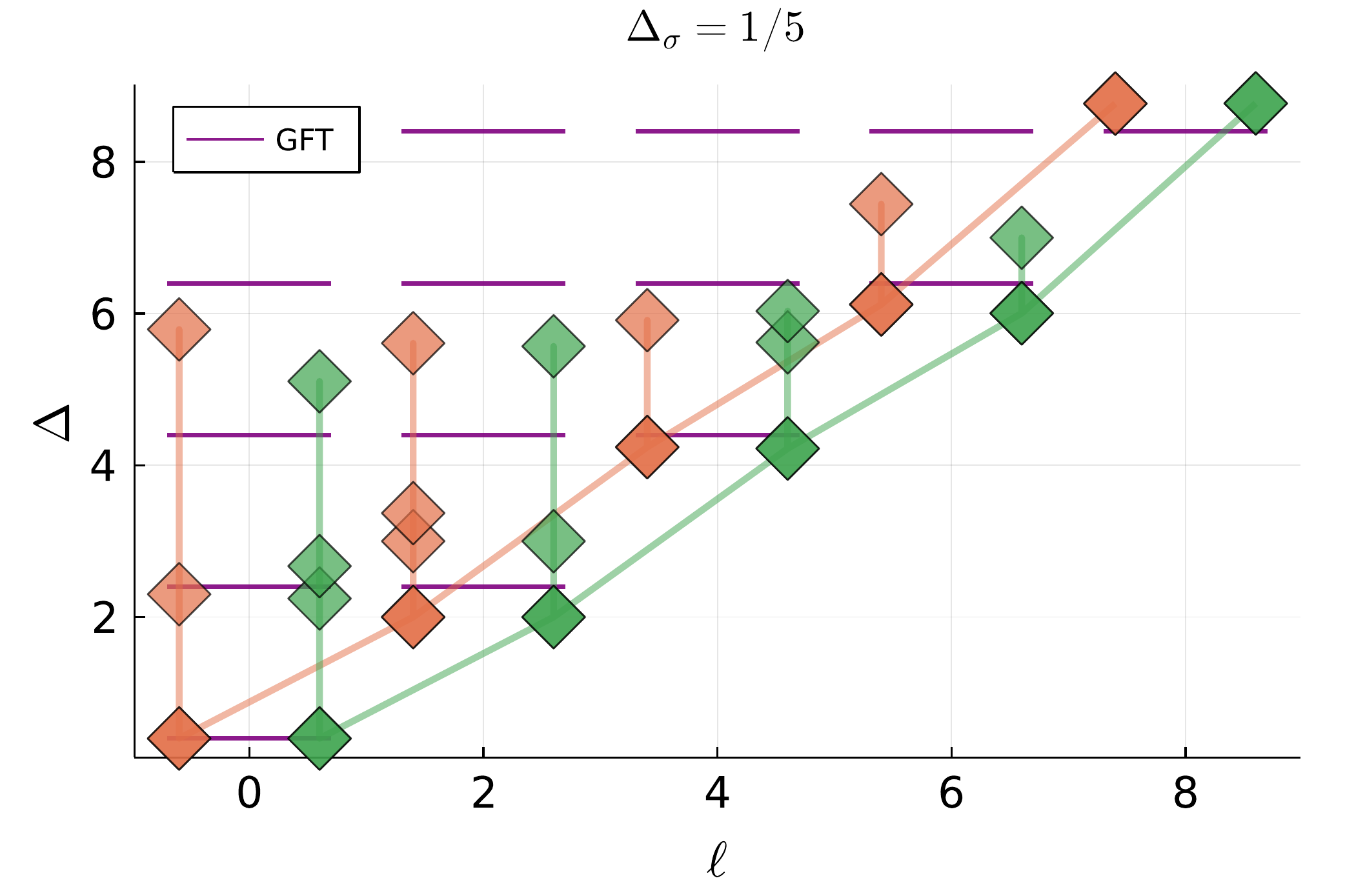}
\caption{Full spectrum of selected end-minima  in $d=2$
with $\Delta_T=2$ and $\Delta_{T^\prime}\geq 3$. 
    \emph{Left:} Ising-like end-minima.
    \emph{Right:} GFT-like end-minima.
The colors and the lines connect operators within the same end-minimum.  For comparison we also report the exact scaling dimension of the 2d Ising and GFT with purple lines.}
    \label{fig:dsig125Minima}
\end{figure}

Let us now discuss the green end-minima of figure \ref{fig:d2-overview}. 
In contrast to the blue ones, we do not have a convincing explanation for the nature of these end-minima as local CFTs, assuming they are not numerical artifacts. Most of them lie on the GFT line $\Delta_\epsilon = 2\Delta_\sigma$. 
A notable class of local CFTs that have in their spectrum operators with scaling dimensions related in this way are $\mathcal{N}=2$ CFTs. 
It would be interesting to understand if the green points along the GFT line (or some of them) can be identified as $\mathcal{N}=2$ Supersymmetric Minimal Models, provided an appropriate identification of $\sigma$ is made. 
This analysis would probably require also to understand whether generalized $\mathcal{N}=2$ minimal models, in the spirit of \cite{Liendo:2012hy}, exist and can be constructed.  

\begin{figure}[t!]
    \centering
    \includegraphics[width=0.85 \linewidth]{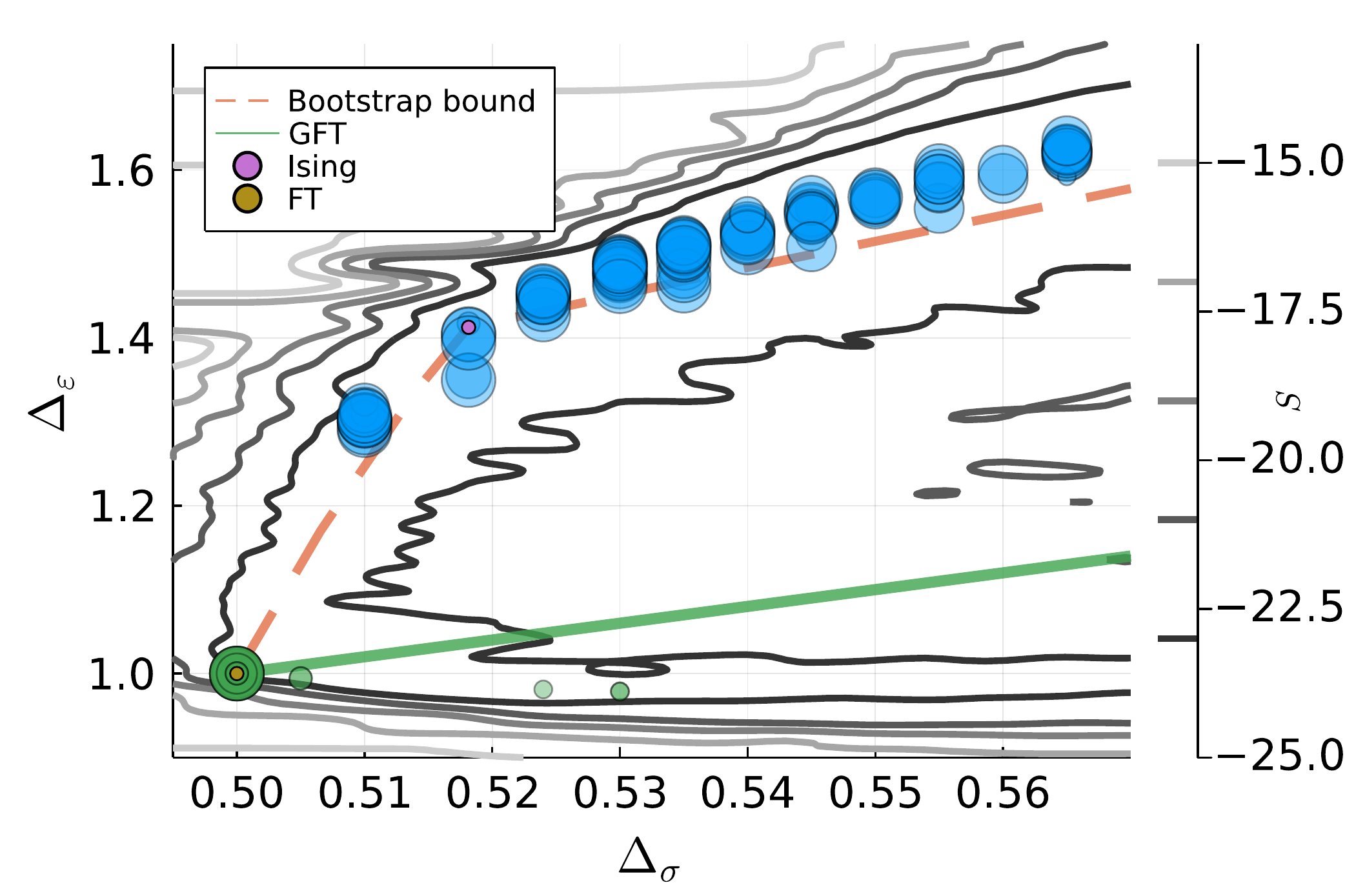}
\caption{$d=3$. Location of the end-minima with $\elmax \leq 10$ in the $(\Delta_\sigma, \Delta_\epsilon)$-plane.  The end-minima belong to different sectors all of them contained in $ 4 \_4 \_4 \_3 \_2 \_1$. The dashed line is the upper bound for $\Delta_\epsilon$  from \cite{El-Showk:2012cjh}.}
    \label{fig:d3-overview}
\end{figure}

Despite we enforced the presence of an energy momentum tensor, the green end-minima along the GFT line could simply be an approximation of GFTs themselves. 
This observation is supported by the fact that if we remove the gap assumption $\Delta_{T^\prime}\geq 3$ most of the green minima are unstable. In particular, $\lambda^2_{\sigma \sigma T}\sim 0$ and $\Delta_{T^\prime}\approx  2\Delta_\sigma+2$ effectively replaces $T$ as the first spin-2 operator.
Similarly, if we do not assume the existence of an energy momentum tensor, 
most minima turn out to be GFTs. As we will see in the next subsections, the analogue of the green points in figure \ref{fig:d2-overview} will not occur in $d>2$ when we enforce
the presence of an energy-momentum tensor. 

It is useful to also report the spectrum of $SL(2,\mathbb{C})$-primary operators which appear at
 the end-minimum at a given truncation. We report as an example in figure  \ref{fig:dsig125Minima} the 
 full spectrum of some end-minima found at $\Delta_\sigma = 1/8, 1/5$ (two in each case). 
 The operators associated to the same end-minimum are connected with lines. 
In the Ising case, where the identification of the end-minima with the Ising model is quite convincing, we see how the agreement of the whole low-lying spectrum is quite accurate, with small deviations occurring for higher-spin operators in the first Regge trajectory (the spin-8 orange diamond) or for operators in the second Regge trajectory (the second spin-0 green diamond, the second spin-4 orange diamond).   

In the GFT-like case, as already mentioned, we do not have a convincing explanation for the nature of such theories. Despite having enforced the presence of a spin 2 operator at $\Delta=2$, we can see from the right panel of figure  \ref{fig:dsig125Minima} that the spectrum resembles roughly the one of GFTs. The presence of more operators with respect to the Ising case makes the accuracy of the numerics less accurate. In particular, it is not possible to disentangle if the significant deviations from the GFT spectrum at higher Regge trajectories are, in addition to numerical noise, due trivially to the fact that we have imposed the presence of a non-existent operator (the stress-tensor), or because the theory under question is in fact distinct from a GFT. See appendix \ref{app:min-endmin} for some further details, in particular for examples of selected branches, the endpoints of which correspond to the blue and green end-minima discussed above.

It is well possible that other theories with a relatively sparse spectrum satisfying our assumptions (or effectively doing it, like the generalized minimal models)  might have not been found by our search protocol, given also the limitations imposed by machine precision computations and the small values of $\Delta^*$ and $N_{\rm{Ops}}$ we sampled. Possible theories of this kind are for instance the $SU(2)_k$ WZW models \eqref{eq:possWZW} with the field identification \eqref{eq:WZWspinj}.

\subsubsection{$d=3,4$}

\label{subsec:d34}

\begin{figure}[t!]
    \centering
    \includegraphics[width=0.75 \linewidth]{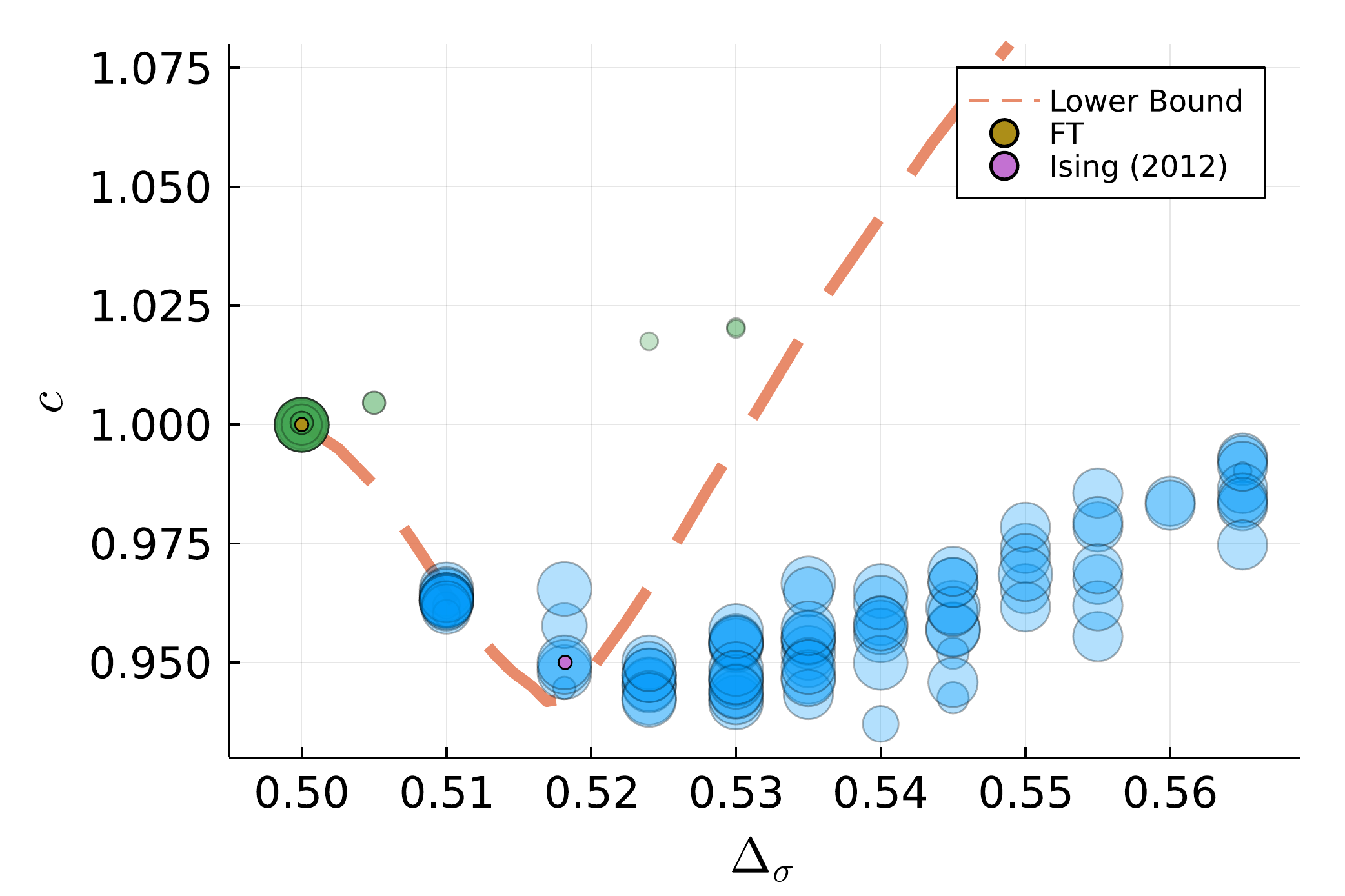}
\caption{Central charge $c$ as a function of $\Delta_\sigma$ for end-minima with $\elmax \leq 10$. The end-minima belong to different sectors all of them contained in $ 4 \_4 \_4 \_3 \_2 \_1$.  The rigorous lower bounds are taken from \cite{El-Showk:2014dwa}. The color code of the end-minima is the same as in figure  \ref{fig:d3-overview}.}
    \label{fig:d3-cvsdsig-comparison}
\end{figure}

In $d=3$ we run our protocol for the following values of $\Delta_\sigma$:
\begin{equation}
    \Delta_\sigma = 0.5 , \,0.505 , \,0.51  , \,0.5181489  , \,0.524 , 
     \,0.53 , \,0.535 , \,0.54 , \,0.545 , \,0.55 , \,0.555 , \,0.56 , \,0.565 .
     \label{eq:dsigma3d}
\end{equation}
We plot in figure \ref{fig:d3-overview} the end-minima found in the $(\Delta_\sigma,\Delta_\epsilon)$-plane, with $\Delta_T=3$, $\Delta_{T^\prime}\geq 4$, $\Delta_{\epsilon^\prime}\geq 3$. All the end-minima lie at the extremality bound, like in the $d=2$ case. In contrast to the latter, no end-minima along the GFT line appear, while we find a couple of faint end-minima with $\Delta_\epsilon < 1$, below the GFT line.\footnote{At low $\ell_{max}$ we get an entire line of minima, 
but these do not create consistent branches. As we will see later when discussing non-local theories, end-minima along the GFT line are reassuringly found when we fix $\Delta_T = 2\Delta_\sigma + 2$ (the GFT value), as expected.}

Reassuringly, we get end-minima in correspondence of the 3d Ising model, which sits at the kink of the dashed orange lines. End-minima are found for all the $\Delta_\sigma$ in \eqref{eq:dsigma3d} with the exception of $\Delta_\sigma = 0.505$. The lack of an end-minimum at such value is due to the imposed gap $\Delta_{\epsilon^\prime}\geq 3$. Indeed, by repeating the MC analysis of section \ref{subsec:sigmafree} with $\Delta_\sigma$ unconstrained assuming $\Delta_{\epsilon^\prime}\geq 3$, the ``flow" towards the FT passing through the valley connecting the Ising point with the FT in figure  \ref{fig:lowTMCevolution} gets interrupted around $\Delta_\sigma = 0.510$. This is an indication that possible extremal CFTs with $0.5 \leq \Delta_\sigma \leq 0.510$ do have more than one $\mathbf{Z}_2$-even scalar.\footnote{Note that this region is devoid of not only end-minima, but of minima at all.}

\begin{figure}[t!]
    \centering
    \includegraphics[width=0.49 \linewidth]{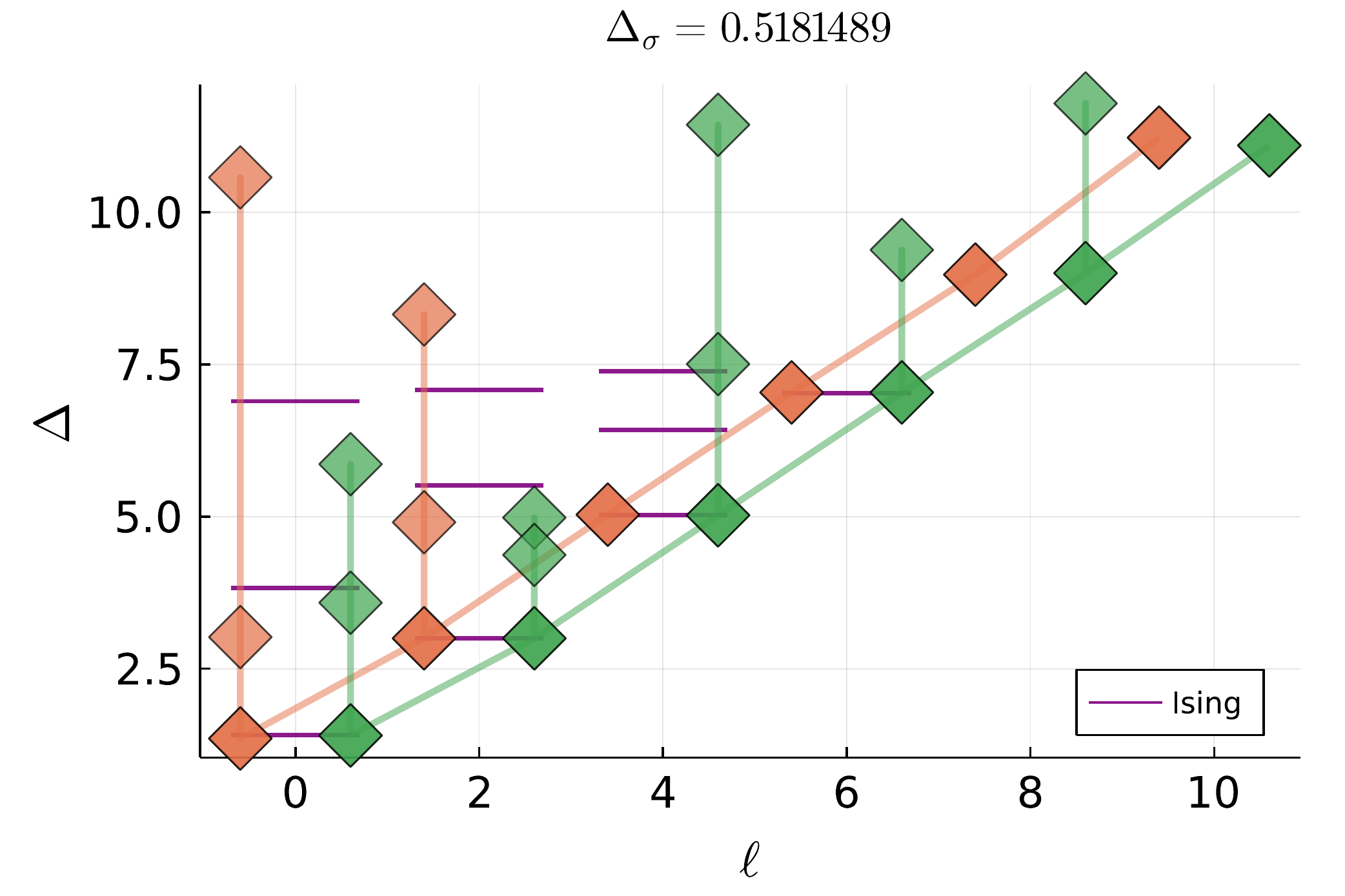}
    \includegraphics[width=0.49 \linewidth]{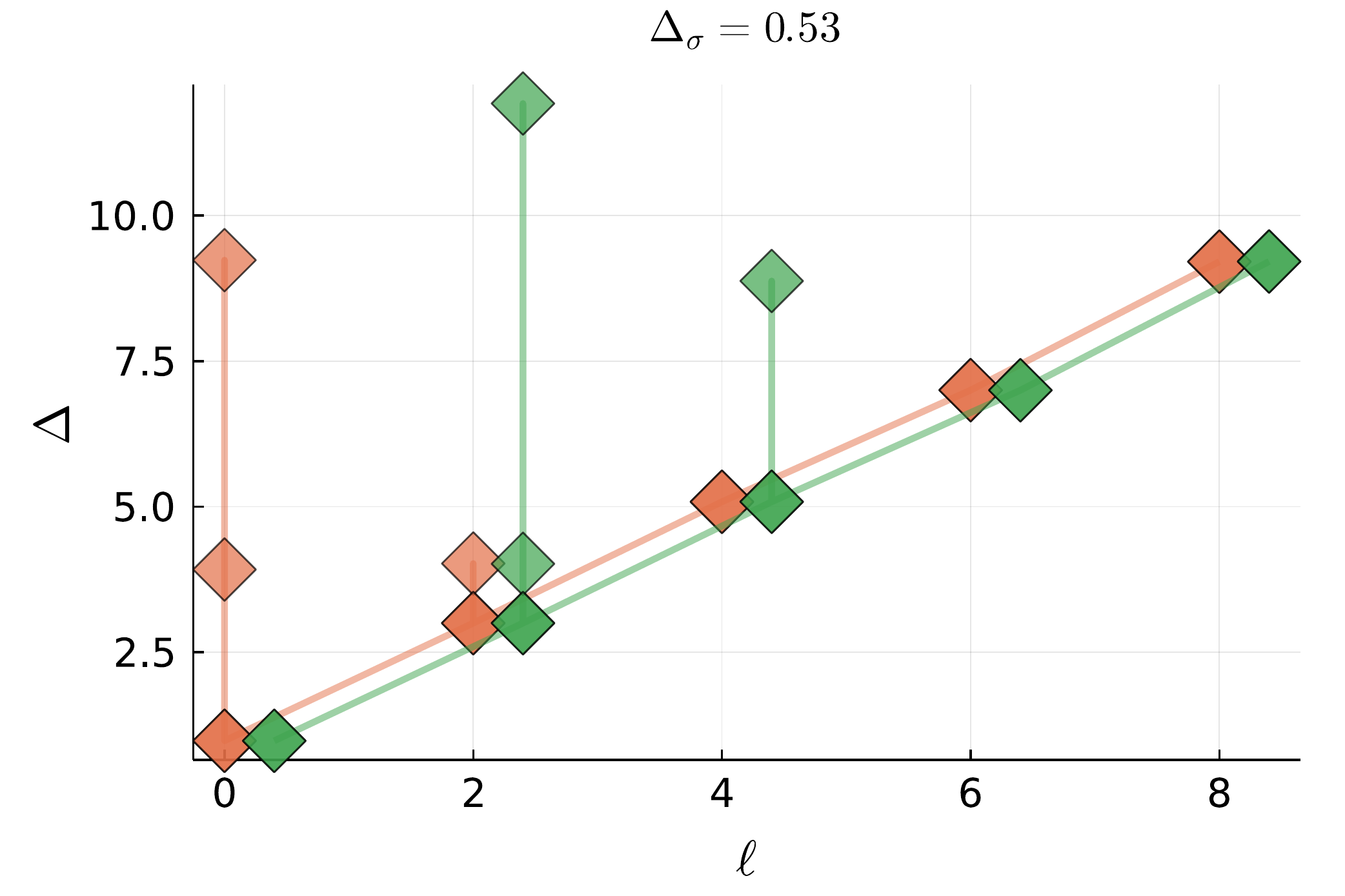}
\caption{\emph{Left:} Full spectrum of the minima found at $\Delta_\sigma = 0.5181489$, $\Delta_T=3$ that resemble the critical Ising model in $d=3$.
Ising reference values taken from \cite{Simmons-Duffin:2016wlq}.
\emph{Right:} Full spectrum of the minima found at $\Delta_\sigma = 0.53$, $\Delta_T=3$ that have $\Delta_\epsilon$ below the GFT line. The colors and the lines connect operators within the same end-minimum.}
    \label{fig:3dspecs}
\end{figure}

We plot in figure \ref{fig:d3-cvsdsig-comparison} the central charge of the end-minima, computed using \eqref{eq:cc}, as a function of $\Delta_\sigma$. The dashed orange lines
delineate the rigorous lower bound of \cite{El-Showk:2014dwa}. 
Most of the end-minima in figures \ref{fig:d3-overview} and \ref{fig:d3-cvsdsig-comparison} lie in the forbidden region, slightly in the former case and more evidently in the latter. This difference should not come as a surprise, since it is known that the bounds on the central charge are more sensitive than those on $\Delta_\epsilon$ to higher dimensional operators. Compare e.g. figures  3 and 4 of \cite{CastedoEcheverri:2016fxt}
to appreciate the different convergent properties of the two bounds as the degree of truncation is varied.

The spectra associated to the branches with $\Delta_\sigma =  0.5181489,0.53$ is reported in figure \ref{fig:3dspecs}. 
In the left panel we also report as reference values the Ising spectrum found in \cite{Simmons-Duffin:2016wlq} using the Extremal Functional Method (purple lines).  The operators associated to the same end-minimum are connected with lines. 
In the Ising case we see how the accuracy is significantly worse with respect to the $2d$ case. We have a reasonably good agreement of the low-lying spectrum in the first Regge trajectory, but the remaining operators have a significant indeterminacy. 
Similarly, in the right panel of figure \ref{fig:3dspecs} the operators in the first Regge trajectory agree among the two orange and green branches, but significantly differ at higher levels.

\begin{figure}[t!]
    \centering
    \includegraphics[width=0.99 \linewidth]{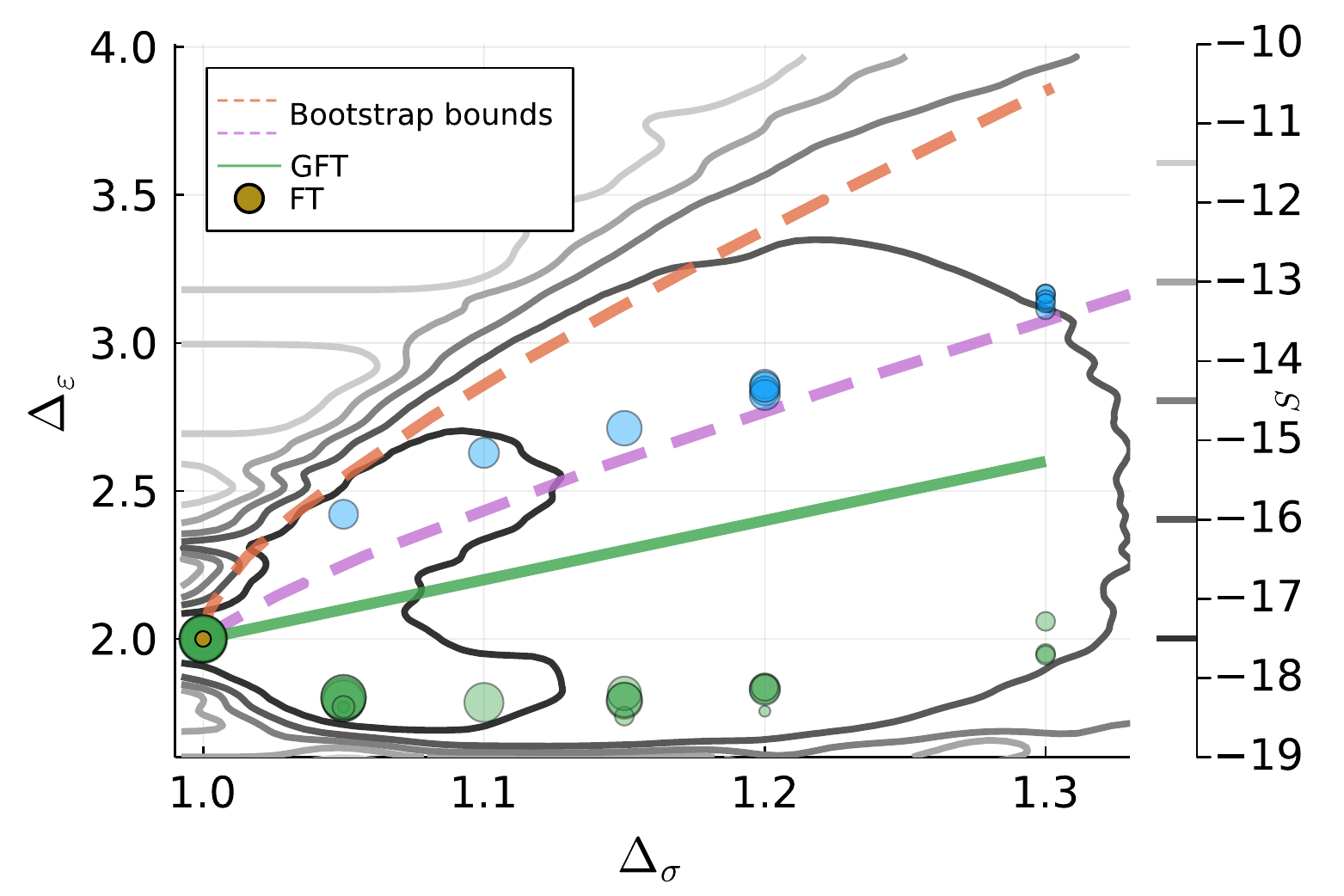}
\caption{
$d=4$.
Location of the end-minima with $\elmax \leq 10$ in the $(\Delta_\sigma, \Delta_\epsilon)$-plane. The end-minima belong to different sectors all of them contained in $ 4 \_4 \_4 \_3 \_2 \_1$.  The dashed orange and purple lines correspond to the bounds obtained in \cite{Rattazzi:2008pe} and \cite{Poland:2011ey}, respectively.}
    \label{fig:d4-overview}
\end{figure}

In $d=4$ we run our protocol for the following values of $\Delta_\sigma$:
\begin{equation}
\Delta_\sigma = 1, \, 1.05, \, 1.1 , \, 1.15, \, 1.2, \, 1.3.
\end{equation}
Due to the limited numerical accuracy of our protocol in $d=4$ we just report in figure  \ref{fig:d4-overview} the end-minima found in the $(\Delta_\sigma,\Delta_\epsilon)$-plane with $\Delta_T=4$, $\Delta_{T^\prime}\geq 5$, $\Delta_{\epsilon^\prime}\geq 4$, with $\elmax \leq 10$. Once again we have end-minima along the extremality bound but it is interesting to see how our minima satisfy the older bounds (with less derivatives) of \cite{Rattazzi:2008pe} but are slightly excluded by the bounds in \cite{Poland:2011ey}.
Interestingly enough, for each value of $\Delta_\sigma$ sampled we also find end-minima below the GFT line. 

\subsection{Non-local theories} 
\label{sec:DT}

In this subsection we analyze the structure of the minima found when relaxing the condition $\Delta_T=d$. 
The most notable and exactly calculable non-local theory is the GFT.  
At a given $\Delta^*$ the number of operators in the GFT spectrum is larger than the maximal truncation we considered in this paper. Although in principle this rules out the possibility of finding exactly the GFT as end-minima in our approach, the limitation is only theoretical, considering that our numerical accuracy does not anyhow allow to precisely determine operators beyond the first Regge trajectories, as we have seen. 
In fact, we will find end-minima which seem quite compatible with the leading operators of the GFT at different values of $\Delta_\sigma$.

\begin{figure}[t!]
    \centering
    \includegraphics[width=0.49 \linewidth]{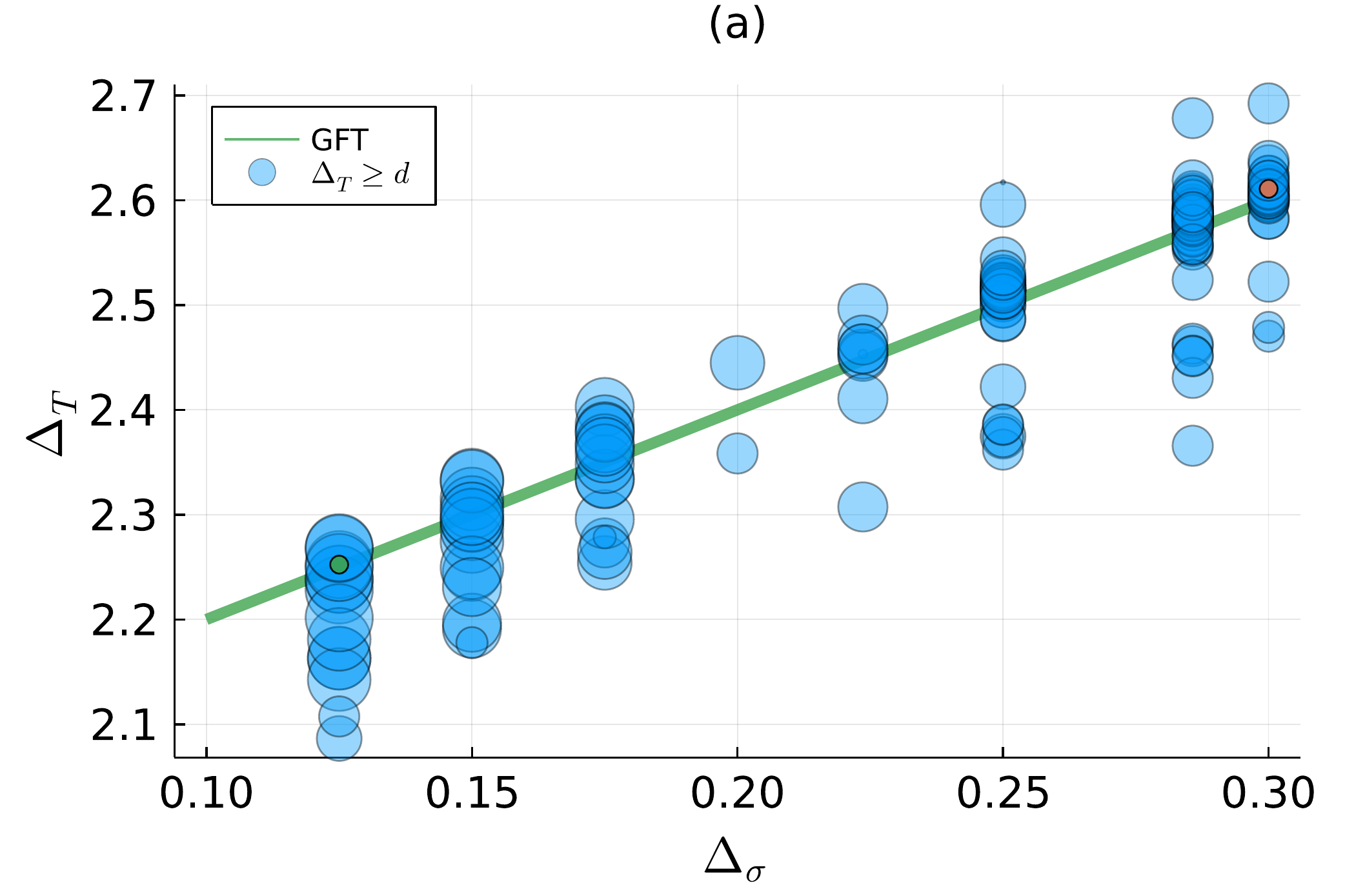}
    \includegraphics[width=0.49 \linewidth]{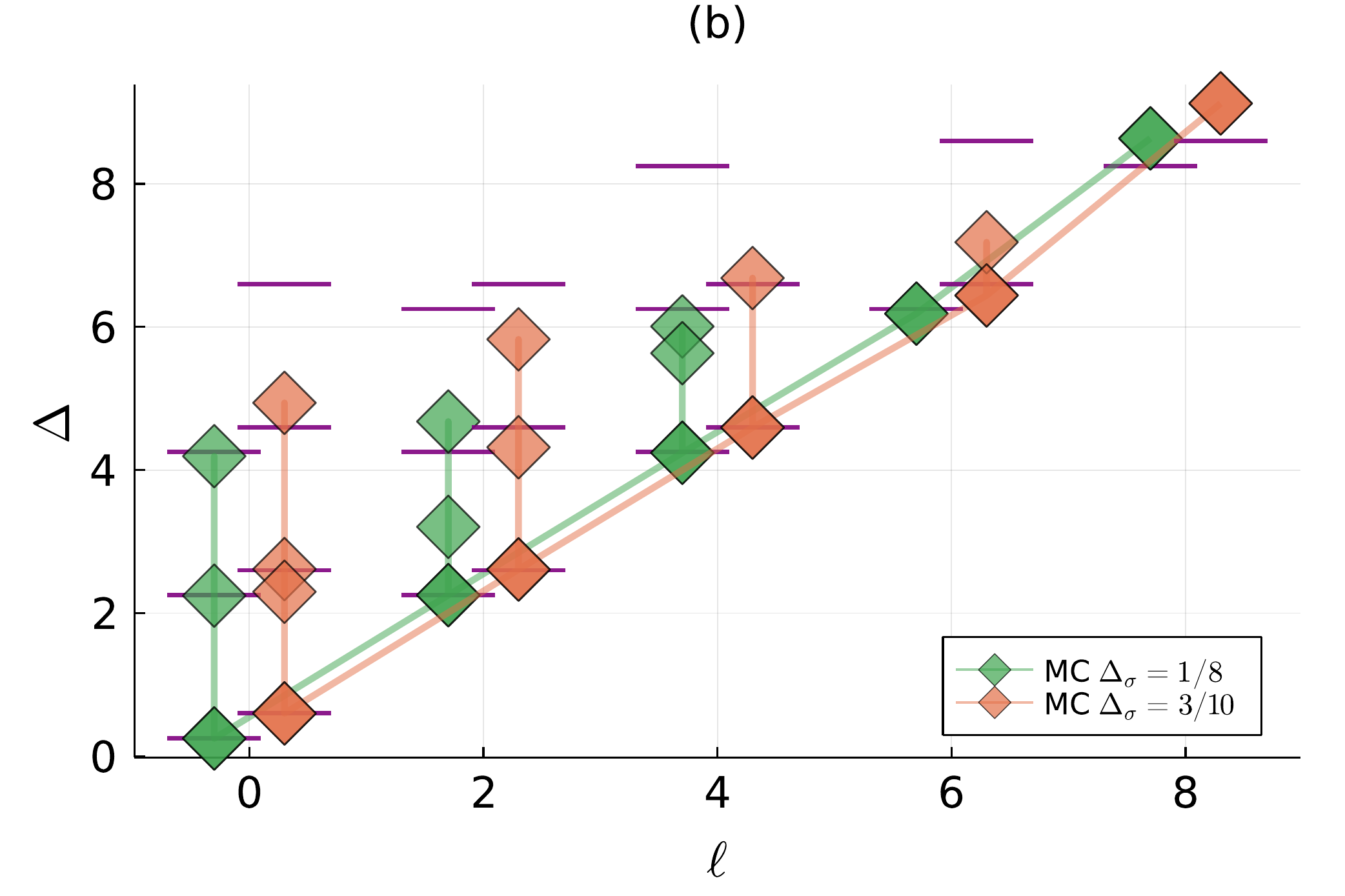}
\caption{(a): End-minima in $d=2$ with no condition on $\Delta_T$ besides unitarity.
 (b): Spectra for the closest match to the GFT for $\Delta_\sigma = 1/8, 3/10$. The colors and the lines connect operators within the same end-minimum. Purple lines represent the theoretical values of the GFT at that $\Delta_\sigma$.
}
    \label{fig:dtvsdsig}
    
\end{figure}

The landscape of end-minima is now much richer and cover most of the allowed region. Interestingly enough, along the $\Delta_T$ direction we find that there is a clear tendency of the minima to cluster around $\Delta_T \approx 2\Delta_\sigma +2$, see figure \ref{fig:dtvsdsig} (a) for $d=2$ (a similar phenomenon occurs in $d=3$ and $d=4$),
where we report the distribution of the end-minima in the $(\Delta_\sigma, \Delta_T)$ plane.
As can be seen, we have many more end-minima with respect to those found when  
$\Delta_T=d$ and $\Delta_{T^\prime}>d+1$ were imposed.
Interestingly enough, most end-minima have $d\leq \Delta_T \leq d+2\Delta_\sigma$.
Note that only for a limited range in $\Delta_\sigma$, end-minima with $\Delta_T=d$ are found. This is particularly evident in $d=2$, where an end-minimum at $\Delta_T=2$ is found only at $\Delta_\sigma = 1/8$. This signals the fact that the extremal end-minima associated to the (generalized) minimal models with $m>3$ found in subsection \ref{subsec:d2} become unstable when we relax $\Delta_T$ and move towards the GFT region. In panel (a) of figure \ref{fig:dtvsdsig} two specific end-minima points have been  singled out at $\Delta_\sigma = 1/8$ (green circle) and at $\Delta_\sigma=3/10$ (red circle). They have been selected as those resembling most closely the GFT conformal data using \eqref{eq:quality} as criterion. The truncated spectrum at those points is reported in panel (b) of figure \ref{fig:dtvsdsig}. Note the good agreement beyond the first Regge trajectory, in particular in the scalar sector where the first three scalars are well reproduced in both theories.

\begin{figure}[t!]
    \centering
    \includegraphics[width=0.85 \linewidth]{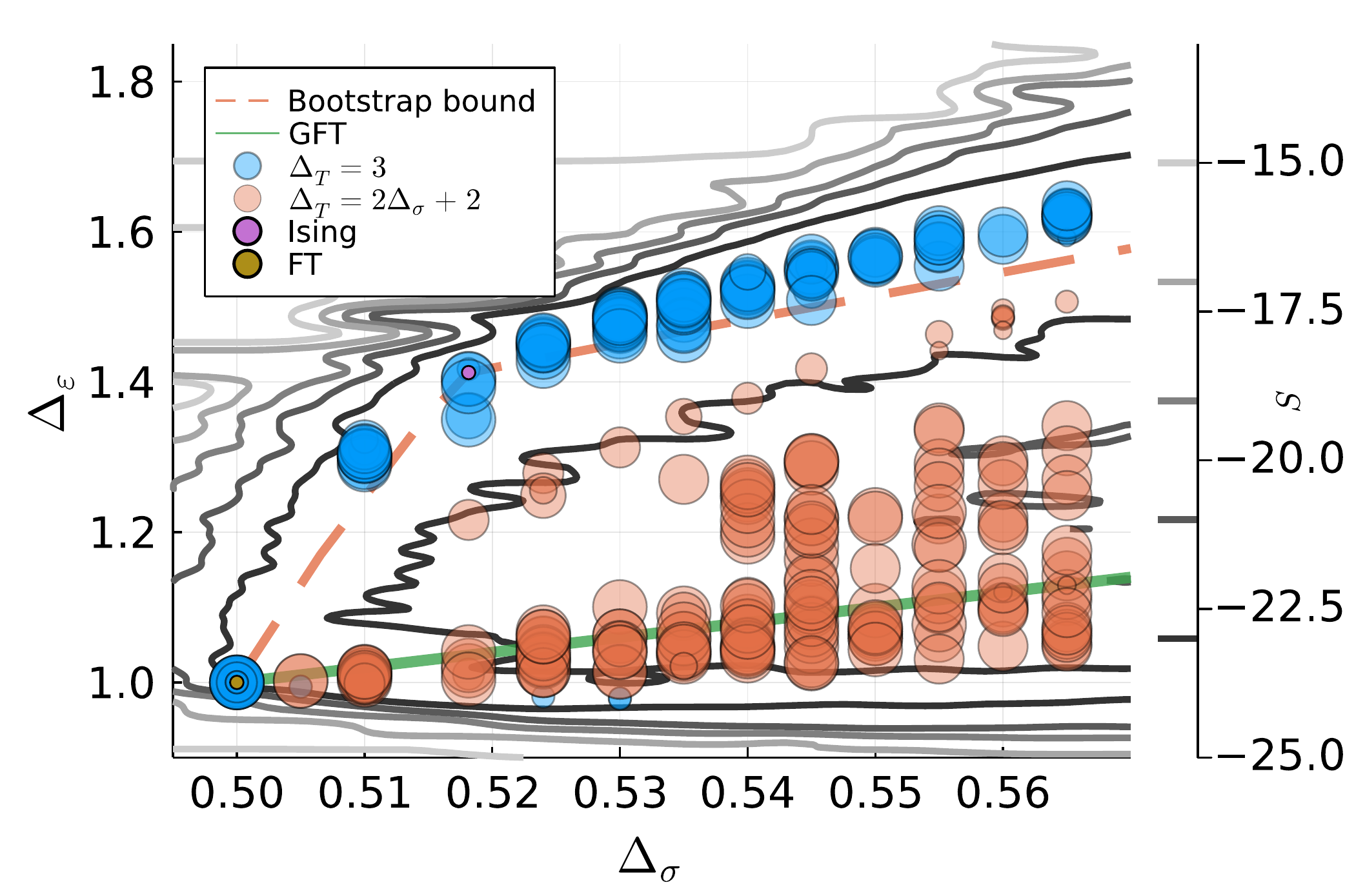}
\caption{$d=3$. Same as figure \ref{fig:d3-overview} but we also show with red circles the minima found by fixing $\Delta_T$ to the GFT value. Note that we show here in blue all the end-minima that in figure \ref{fig:d3-overview} were reported in blue and green.}
    \label{fig:d3-overview-GFT}
\end{figure}

We have also explored the effect of fixing the scaling dimension of the first spin 2 operator at the exact GFT value, still assuming $\Delta_{T^\prime}\geq d+1$. This is useful to understand whether minima at fixed $\Delta_T$ are ``harder'' to reach than similar points when $\Delta_T$ is allowed to vary.
We show these results in figure  \ref{fig:d3-overview-GFT} for the $d=3$ case, where the blue circles are obtained at $\Delta_T=3$ and are the same as in figure \ref{fig:d3-overview}, while the red ones are those arising at  $\Delta_T=2\Delta_\sigma+2$. The red end-minima align along the GFT line, as expected, but they spread a wider area, which however does not include the extremality region. Similar considerations apply also in $d=2$ and $d=4$.

We saw in figure \ref{fig:dtvsdsig} (a) that some of the minima found by our protocol at $\Delta_\sigma = 1/8$ do resemble local theories, although they are clearly disfavoured in front of the GFT-like ones. We would like to determine more precisely how well we reproduce local theories when no constraint on $\Delta_T$ is imposed.
This can be attained by taking the 2d Ising theory and the GFT as reference points and defining a distance measure between our minima and these exactly known CFTs.
We will here take the scaling dimensions of the first scalar and first spin-$2$ operators and their associated OPEs coefficients, and compute the average relative error with respect to those of the known theory:
\begin{equation}
Q(\mathbf{v}_{min},\mathbf{v}_{exact})  =\frac{1}{4}
\sqrt{\sum_{i=1}^{4} \left( \frac{{v}_{min}^i - {v}_{exact}^i}{{v}_{exact}^i} \right)^2 } .
    \label{eq:quality}
\end{equation}
where $\mathbf{v} = (\Delta_\epsilon,\Delta_T,\lambda_{\sigma\sigma\epsilon}^2,\lambda_{\sigma\sigma T}^2 )$.

\begin{figure}[t!]
    \centering
    \includegraphics[width=0.85 \linewidth]{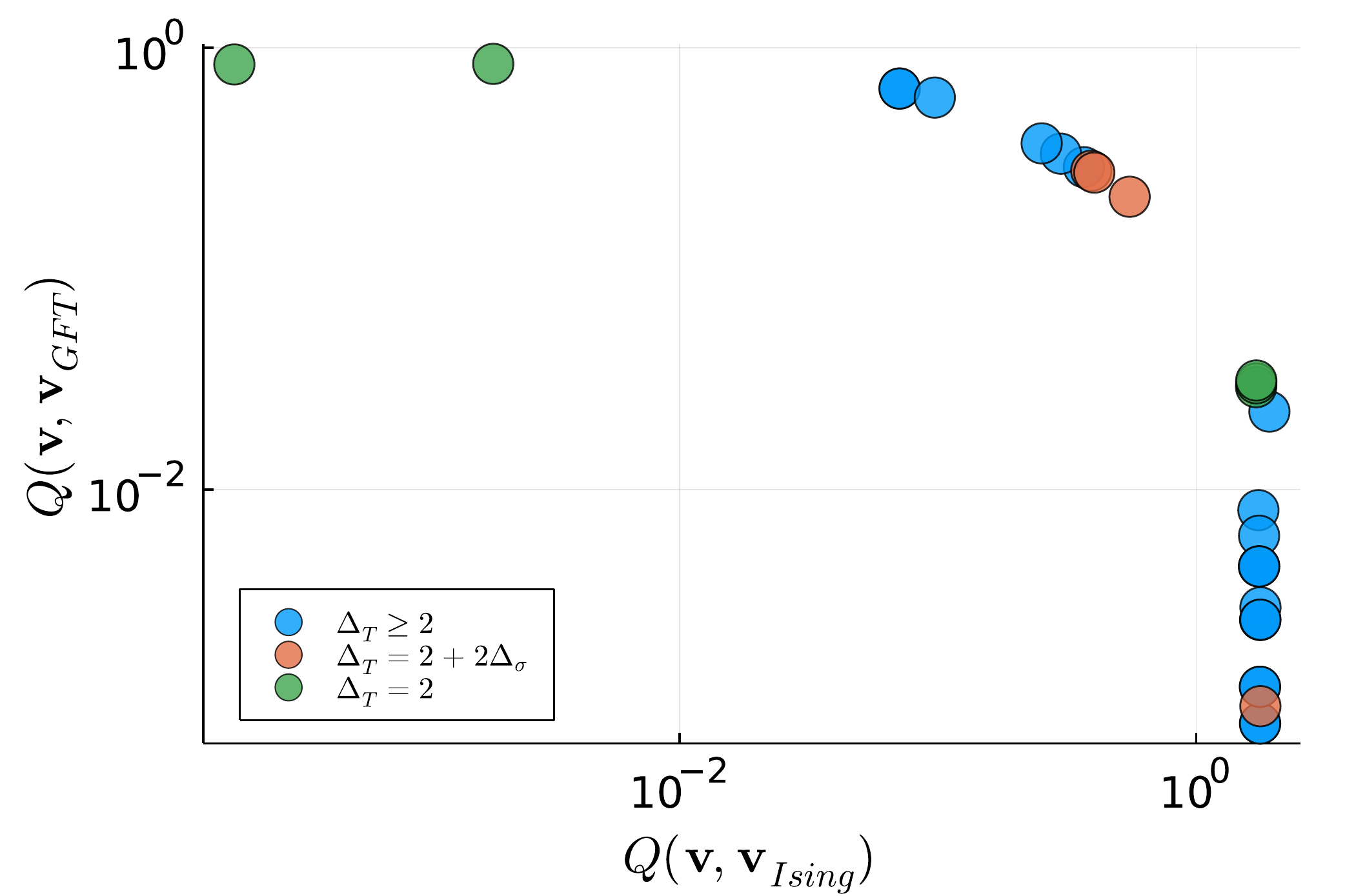}
\caption{Distance of different end-minima to the $d=2$ Ising and the GFT with $\Delta_\sigma = 1/8$. The quantity $Q$ is computed according to \eqref{eq:quality} for end-minima with $\Delta_T = 2, 2 + 2\Delta_\sigma$ and those in which $\Delta_T$ is determined dynamically.}
    \label{fig:distances}
\end{figure}

In figure \ref{fig:distances} we show how the blue circles reach configurations that are not too far away from the 2d Ising theory (roughly $10\%$ mean relative error)  but also cover configurations which reproduce the leading conformal data of the GFT even better than the MC which had $\Delta_T = 2 + 2\Delta_\sigma$ to begin with. Finally, figure \ref{fig:distances} supports the interpretation of the green minima in figure  \ref{fig:d2-overview} to be ``images'' of the GFT, since we can appreciate that they correspond to solutions which reproduce the Conformal Data of this theory with an error of less than $10 \%$.

    \subsection{A taste of non-unitarity: The 2d Yang-Lee Model}
    \label{subsec:2dYL}

The MC search protocol of section \ref{sec:protocol} does not rely on unitarity. Although we mostly focus on unitary models in this paper, we would like to briefly show here a proof of concept of its use in the non-unitary realm, by checking that the $d=2$ Yang-Lee (YL) model can easily be found using our method. Non-unitary theories have received less attention from the bootstrap community due to their non tractability with linear or positive semidefinite programming methods. The best approach so far to solve non-unitary theories with the conformal bootstrap is given by \cite{Gliozzi:2013ysa,Gliozzi:2014jsa}, which also severely truncates the crossing equations (as we do). 
As well-known, the $d=2$ YL model can be described by the non-unitary minimal model ${\cal M} (5,2)$, which contain only two Virasoro primaries related by the following fusion rule:
\begin{equation}
    \Phi \times \Phi \sim 1 + \Phi.
\end{equation}
This means in practice that the external dimension $\Delta_\phi$ will also be the scaling dimension of the first exchanged scalar and that we must not impose positivity of the OPE coefficients. 

\begin{figure}[t!]
    \centering
    \includegraphics[width=0.85 \linewidth]{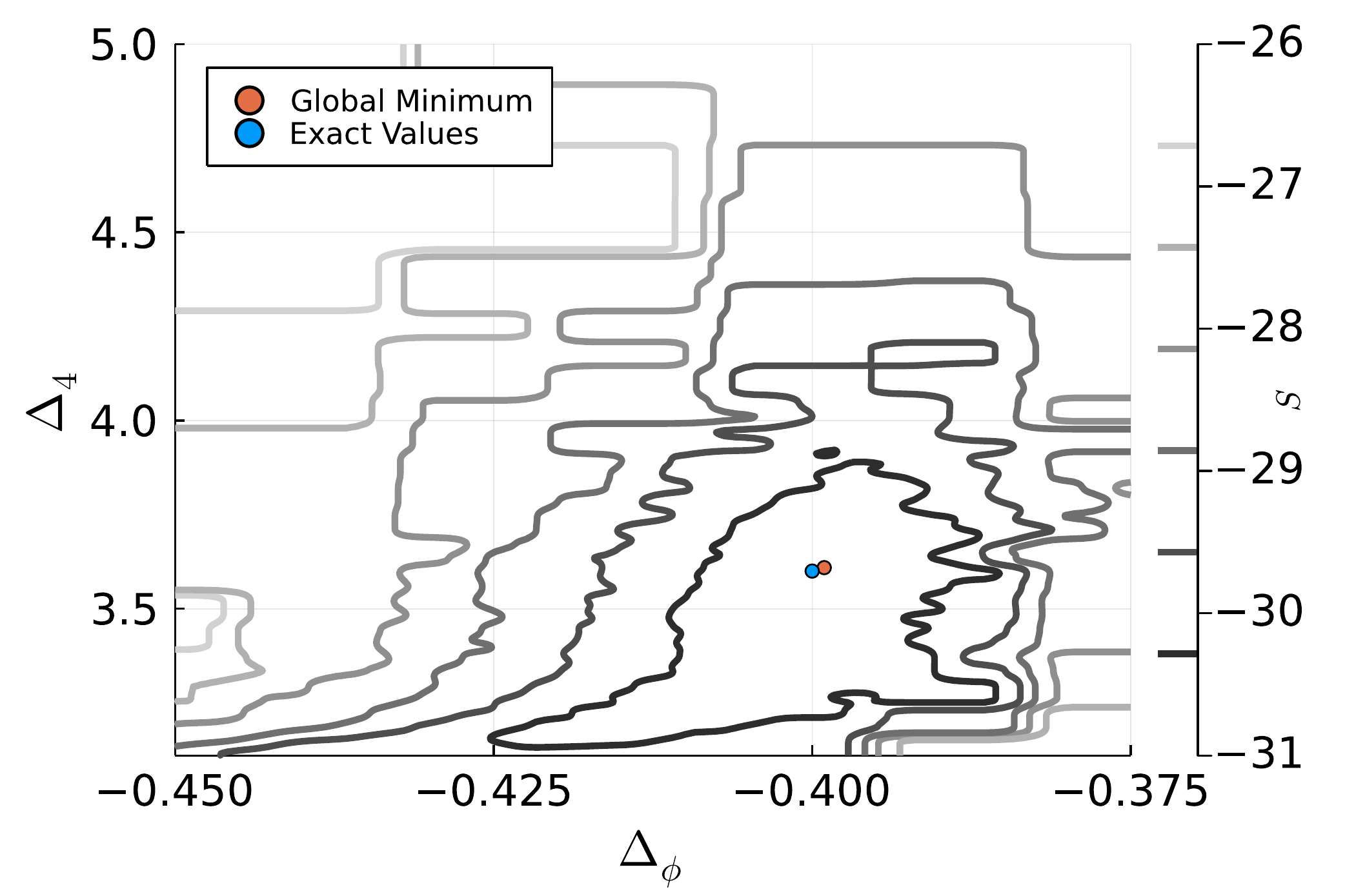}
\caption{Projection of the value of the action $S$ on the $(\Delta_\phi,\Delta_4)$- plane. The lowest-action point visited by the MC at $T=0.5$ and the exact values are shown as red and blue circles, respectively.}
    \label{fig:YL}
\end{figure}

We launched a Metropolis Monte Carlo search with $T=0.5$ and a spectrum $3\_1\_2\_1$ ($\Delta^* = 9$) to sample a parameter space close to the one containing the exact CFT data of the $2d$ YL model. We find that the global minimum is compatible with the exact values, as can be seen in figure   \ref{fig:YL} for $\Delta_\phi$ and $\Delta_4$ (the first spin 4 operator). At the global minimum the other operators show also similar agreements with the exact values.

We note in passing that in $d=3$, if we relax the positivity constraint on the OPE coefficients, a search with a $1\_1\_1\_1$ spectrum shows another basin of attraction besides those discussed in section 
\ref{subsec:1opspin}.
The basin is in a region well below the unitarity bounds for $\Delta_\sigma$ and lies roughly along the line $\Delta_\sigma = \Delta_\epsilon$. Besides being an interesting feature of the landscape in itself, we consider this to be an  indication that the $d=3$ YL model should also appear as a minimum if investigated with our method.

\section{Outlook}

\label{sec:outlook}

Motivated by the pressing need of a tool to explore the inner part of the allowed space of CFTs,  we have introduced in this paper a method based on a stochastic minimization via a Metropolis algorithm to find approximate numerical solutions with a sufficiently sparse spectrum of operators to truncated bootstrap equations for CFTs in arbitrary dimensions. 
Given an initial value of scaling dimensions for a reduced set of primary operators up to some scaling dimension $\Delta^*$, the action \eqref{eq:DefAction} is minimized and an approximate set of CFT data found.

While for all physical CFTs we should eventually have $F=e^S\rightarrow 0$ in the limit when all operators are taken into account, at fixed operator truncation the rate in which this limit is reached does matter.
For instance, we have seen that for arbitrary severe truncations of the operator spectrum, the free theory minimum dominates over all and no other theory can be found, see e.g. figure \ref{fig:lowTMCevolution}. In order to find other theories, extra conditions, such as fixing the external operator dimension, have to be imposed. 
It is clear that, at fixed truncation, not all CFTs are equally accessible, even assuming the same sparsity of spectrum. We expect that richer set of solutions can be found by generalizing our method to multiple correlators and by imposing further constraints that would halt the MC evolution towards specific attractors
(e.g. the extremality line or the GFT line in $d=2$). 
Though we mostly focused on unitary theories, the method does not rely on it and can efficiently be used in the non-unitary realm. This is particularly important considering that for non-unitary theories we still do not have rigorous numerical protocols. It would be interesting to see how our numerical method performs with respect to Gliozzi's method \cite{Gliozzi:2013ysa,Gliozzi:2014jsa} and its subsequent versions \cite{Esterlis:2016psv,Li:2017ukc}. 
The aim of this work is very similar to that of \cite{Kantor:2021kbx,Kantor:2021jpz}, where approximate solutions to crossing have been achieved
using a similar logic but a different technique (reinforcement learning instead of MC techniques). 
It would be interesting to systematically compare the two approaches.

In this work the bootstrap equations have been evaluated in a given sample of points in cross-ratio space. This choice has been primarily dictated by the flexibility, when developing the algorithm, of changing the number of points and the possibility (used in earlier versions of the code) of sampling stochastically these points. The actual performance of the code sensitively depends on how points in cross-ratio space are chosen. It would be interesting to further improve on this aspect. In particular, an improvement could result by choosing derivatives evaluated at the crossing symmetric point $z=\bar z = 1/2$, as currently done in functional bootstrap studies.

The \texttt{FORTRAN}-code used in this work is limited to machine-like precision.
While this allowed us to perform extensive MC searches at little cost, 
limited precision does not allow us to systematically increase the value of $\Delta^*$ and $N_{\rm Ops}$. For example we do not have enough accuracy to understand the nature of the green end-minima we found in $2d$ (figure \ref{fig:d2-overview}),  whether they  are unidentified local CFTs of some kind or some sort of numerical artifacts resembling GFT theories. For $d>2$ the accuracy is even less and the nature of the found end-minima is yet to be understood. Despite the limited accuracy of the algorithm, we think the results of this paper have shown that the method works. 
Upgrading the current code to arbitrary precision will be the key for a fully working program and is the most important thing to do in the near future. We believe this is feasible, given the ample room for improvement of the code performance in terms of speed.

In summary, we hope that the non-rigorous approach initiated in this work, properly extended and improved, might be a useful  tool to navigate in the space of allowed CFT and guide subsequent analysis performed using more rigorous numerical algorithms.

\section*{Acknowledgments}

ULV and MS thank the organizers of the Bootstrap 2019 conference held in PI for the hospitality and all the participants for interesting discussions. 
MS also thanks F. Benini, S. Cecotti, L. Eberhardt and G. Mussardo for discussions on $2d$ CFTs. Work partially supported by INFN Iniziativa Specifica ST\&FI.

    \appendix
    
   \section{Newton Rhapson minimization}
   \label{sec:NR}
The Newton-Rhapson algorithm is a way of finding the local minima  of $\mathcal{C}^2$-functions with a positive definite Hessian $H$.
We will now discuss the basic idea behind this method and the implementation used in this work. For a detailed explanation the reader is suggested to consult \cite{num_rec}.
 
In short,  the method uses the hessian matrix to approximate the function to be minimized up to quadratic order and solves iteratively for the minimum of this hyper paraboloid. 
    In this work a modified version of this method is used (known as the Levenberg–Marquardt algorithm \cite{levenberg_method_1944,marquardt_algorithm_1963}), in which a damping factor $\beta$ interpolates between  gradient descent and the bare Newton-Rhapson method, avoiding the overshooting of the minimum when very close to it. 
    This is necessary because due to numerical noise some eigenvalues of the hessian might not be positive, 
    which violates the conditions for the convergence Newton-Rhapson method.
    
    More concretely, let  $f(x):\mathbb{R}^n \to \mathbb{R}$ be a twice-differentiable function. Given an initial guess $x_0$ we find the next point by using the following formula:
\begin{equation}
x_1 = x_0 - \left( H(x_0) + \beta \mathbb{1} \right)^{-1} \cdot \nabla f(x_0) \,.
    \label{eq:NR}
\end{equation}
At each step, if $f(x_1)<f(x_0)$ then  $\beta$ is decreased by a factor of $3$. Otherwise it is increased by a factor of $2$. 
This is known in the literature as ``Delayed gratification''\cite{transtrum_geometry_2011} and gives stable results in the minimization problems explored 
in this work.

The factor $\beta$ also gives the termination condition for the algorithm:
 when $\beta>10^6$ it is clear that a stationary point has been reached.

\section{Implementation details}
\label{sec:impDetails}

The protocol detailed in section
\ref{sec:protocol} was implemented in a series of automatized bash scripts that used \texttt{Julia} for the data processing and 
\texttt{FORTRAN90} (compiled with \texttt{ifort v2021.2}) for the numeric evaluation of the action 
 and the different minimization algorithms.
\texttt{LAPACK} was used for the linear algebra manipulations.
 They were run on \texttt{Intel Xeon}\textsuperscript{TM}  processors 
 in an institutional cluster. The CPU time to run $10^8$ steps was of roughly $12$ hours 
 for representative spin partitions.
The parameters used in the searches are shown in table \ref{tab:params}.

\begin{table}[t!]
    \centering
    \caption{Summary of the relevant parameters for the wide searches (first step of the protocol).}
    \label{tab:params}
    
    \begin{tabular}{|c|c|p{7cm}|}
   \hline
   Parameter       & Value(s)   & Description\\
   \hline
  \texttt{ nZ                }& $200     $  & Number of points taken in the $z-$plane.\\
   \hline
  \texttt{ lambda0           }& $0.42    $  & Size of the region in which the points are sampled.\\
   \hline
  \texttt{ NT                }& $ 2 \times 10^8$ &  Number of steps (MC time). \\
   \hline
  \texttt{ positiveOPEs      }& \texttt{true }&     Enforcement of real  OPE coefficients. \\
   \hline
  \texttt{ externalInOPE     }& \texttt{false} &    Fixing the first exchanged scalar to be equal to the external operator.\\
   \hline
  \texttt{ Temp              }& $ 0.3-0.9  $ &    Temperature ($T$) of the Metropolis Monte Carlo algorithm.\\
   \hline
  \texttt{ MCstep            }& $ 0.001    $ &    Overall scale factor for the MC-step.   \\
   \hline
  \texttt{ wall              }& $ 10^4   $ &    Constant for the quadratic penalty that enforces the bounds on $\Delta_i$. \\
   \hline
  \texttt{ frameRate  } & $ 1000     $ &  The MC saves one out of   \texttt{ frameRate  } steps as a frame.\\
   \hline
  \texttt{ nops              }& $ 3 - 20   $ &  Number of operators (besides the identity).\\    
   \hline
  \texttt{ boundaries              } & $ [\Delta_{unit.},\Delta^*]   $  &  Range to which each operator is constrained.\\    
   \hline
  \end{tabular}  
\end{table}

In order to provide the technical details  we will discuss in the following subsections
the different elements of our pipeline and their implementation.

\subsection{Numerical evaluation of $S$}
We start by discussing the approximation and numerical evaluation of \eqref{eq:DefAction}.
 We describe  each of the approximations used in order to render computationally feasible sampling $O(10^8)$ points in $\sim 10$ hours on a single CPU.

\paragraph{i) Discretization in $z$}
The grid is parametrized as 
$z = x_0 + an + y_0 + ibm$ with $m,n\in \mathbb{N}$ and $x_0=1/2 + 1/1000$, $y_0 = 1/1000$.
In this work we used $a=1/100$ and $b=1/50$ but in the computation of $S$ only a randomly chosen sample of \texttt{nZ} points was considered. This was done by ordering them with respect to their value of $\lambda$ and then sampling this list.

\paragraph{ii) Truncation of the spectrum }
This was implemented by passing to the \texttt{FORTRAN} routine the list of the spins of each operator in the truncation under study. It is in this step that we input important parameters such as the spin partition, $\Delta^*$ and $\elmax$.

\paragraph{iii) Evaluation of the conformal blocks}
The conformal blocks were tabulated using 160 digits of internal precision in \texttt{Mathematica} and stored as plain text files for each $\ell$. In $d=2,4$ we used the  closed expressions in terms of hypergeometric functions \cite{Dolan:2000ut}. For $d=3$ the recursion relations implemented as in \cite{Kos:2013tga} were used.
For each point in $z$ we computed  the CBs at $5000$ points evenly spaced in $[\Delta_{unit}(\ell) -99/100, \Delta*]$ with  $\Delta^* = d + 15$. We note in passing that this gives a lower bound on the error in our determination of the scaling dimensions 
($\Delta$) of the exchanged operators.
In the \texttt{FORTRAN} code, the interpolation is done with cubic splines. 

\begin{figure}[t!]
    \centering
    \includegraphics[width=0.8 \linewidth]{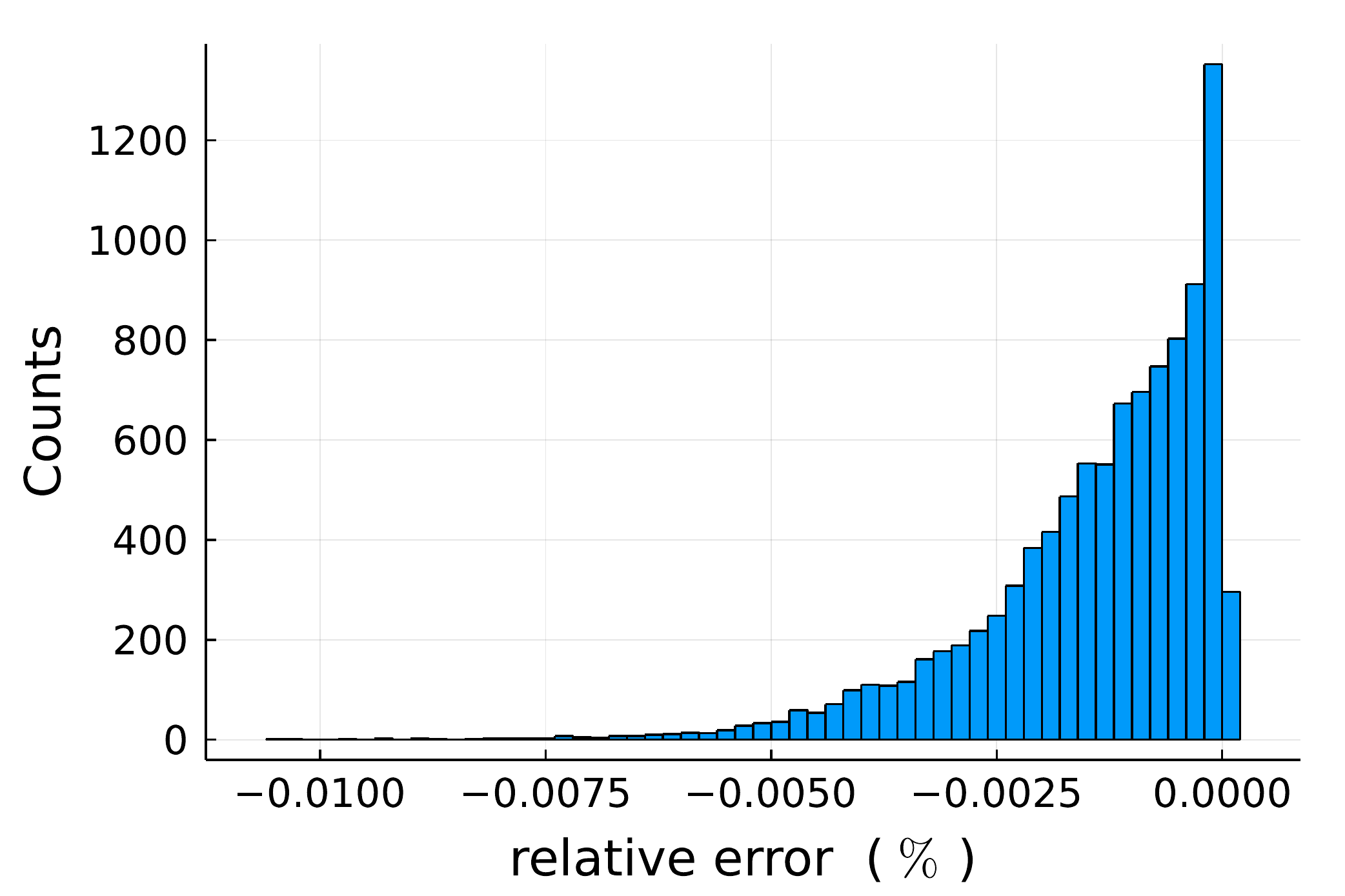}
    
\caption{Distribution of the relative error defined as 
$S_{\mathtt{Mathematica}} - S_{\mathtt{FORTRAN}}$. The action was evaluated at $10^4$ points normally distributed around an apparent local minimum  in $d=3$ with spin sector $2\_2\_1\_1\_1$, $\lambda_0 = 0.42$ and $\texttt{nZ} = 200$. 
Note the long tail for negative values.}
    \label{fig:hist_err}
\end{figure}

\paragraph{iv) Determination of the OPE coefficients}
Once the matrix of the conformal blocks for each $z$ and operator $(\Delta,\ell)$ has been computed,
the determination of the OPE coefficients is then a weighted least squares problem which we solve using \texttt{LAPACK} (see \ref{eq:error}).

If \texttt{positiveOPEs} is true, this is not the end of the story because the OPE coefficients that solve the quadratic problem might as well be negative. To enforce unitarity, we iteratively decouple the offending operators until every OPE coefficient is positive. It is important to underline that this is done at each point evaluated in our algorithm, and thus operators do not decouple for the whole test but only for those particular configurations.

\paragraph{v) Computation of $S$} As described in section \ref{sec:method}, we obtain $S$ by taking the logarithm of the sum of the squared residuals. If any operator has a $\Delta$ outside its allowed boundaries we impose a quadratic penalty that is then added to the action. This guarantees the analyticity of our potential and thus avoids noxious boundary effects. 

More concretely, suppose an operator bound to be in the interval $[\Delta_0,\Delta_1]$ has $\Delta <\Delta_0$. In that case, the effective action  will be
\[
S_{eff} = S + \mathtt{wall} (\Delta-\Delta_0)^2,
\]
where the constant \texttt{wall} is set to $10^4$.

\subsection{Numerical error}
\label{sec:error}

After all the approximations discussed above, reasonable doubt could remain about the accuracy of $S$ computed with our method. Thus, we consider it necessary to 
quantify the magnitude of the numerical errors in our implementation. We accomplish this by computing  $S$ at representative points with our framework and by comparing them to the values of $S$ obtained in \texttt{Mathematica} with arbritrary precision. 

From this analysis we conclude that the action computed with \texttt{Mathematica} is systematically lower than the approximated one using our \texttt{FORTRAN} implementation. However, the discrepancy is such that at the values of $\Delta^*$ studied in this paper we can still trust the results. Naturally, the errors become larger for bigger $N_{\rm Ops}$ and $\Delta^*$, but even in the cases with the worst agreement the relative errors are smaller than $10^{-3}$.
Moreover, the largest discrepancy sets a natural lower bound for $T$, since we must allow for fluctuations of this height in order not to wrongly identify artifacts as minima. 
For example, in $d=3$ the minimum temperature for $\elmax = 6$ is $T \sim 10^{-4}$, whereas for $\elmax = 8$ is $T \sim 10^{-3}$. 
We note that this resolution is likely enough for any physical minimum, given that in the $\Delta^* \to \infty$ limit any physical theory should have $S \to \infty$.

As an example, we show in figure  \ref{fig:hist_err} the relative error computed for $10^4$ gaussian perturbations of a 
 local minimum in sector $2\_2\_1\_1\_1$ for $d=3$, $\lambda_0 = 0.42$ and $\texttt{nZ}  = 200$.  
The fact that the error is mostly negative can be easily understood from the delicate numerics of the bootstrap equations. It is known that in order to solve \eqref{eq:bootstrap} to a high degree of precision, very fine-tuned cancellations between the terms must occur. When using our double-precision interpolation of the blocks, there will be points where these errors will spoil the cancellations and thus the value of $S$ can only be bigger than the ``exact'' one.

Another point that deserves discussion is the determination of the OPE coefficients ($\rho$ in eq. \eqref{eq:DefAction}). 
Solving for $\rho$ amounts to finding the minimum of a quadratic form. We find that the curvature of this form is very anisotropic, with hierarchies of as much as 10 orders of magnitude. This is in fact one of the main bottlenecks to adding too many operators: the curvature of the subleading operators at some points becomes too small to be resolved with double precision and 
the OPE coefficient of these operators cannot be reliably determined.

\begin{figure}[t!]
    \centering
    \includegraphics[width=0.48 \linewidth]{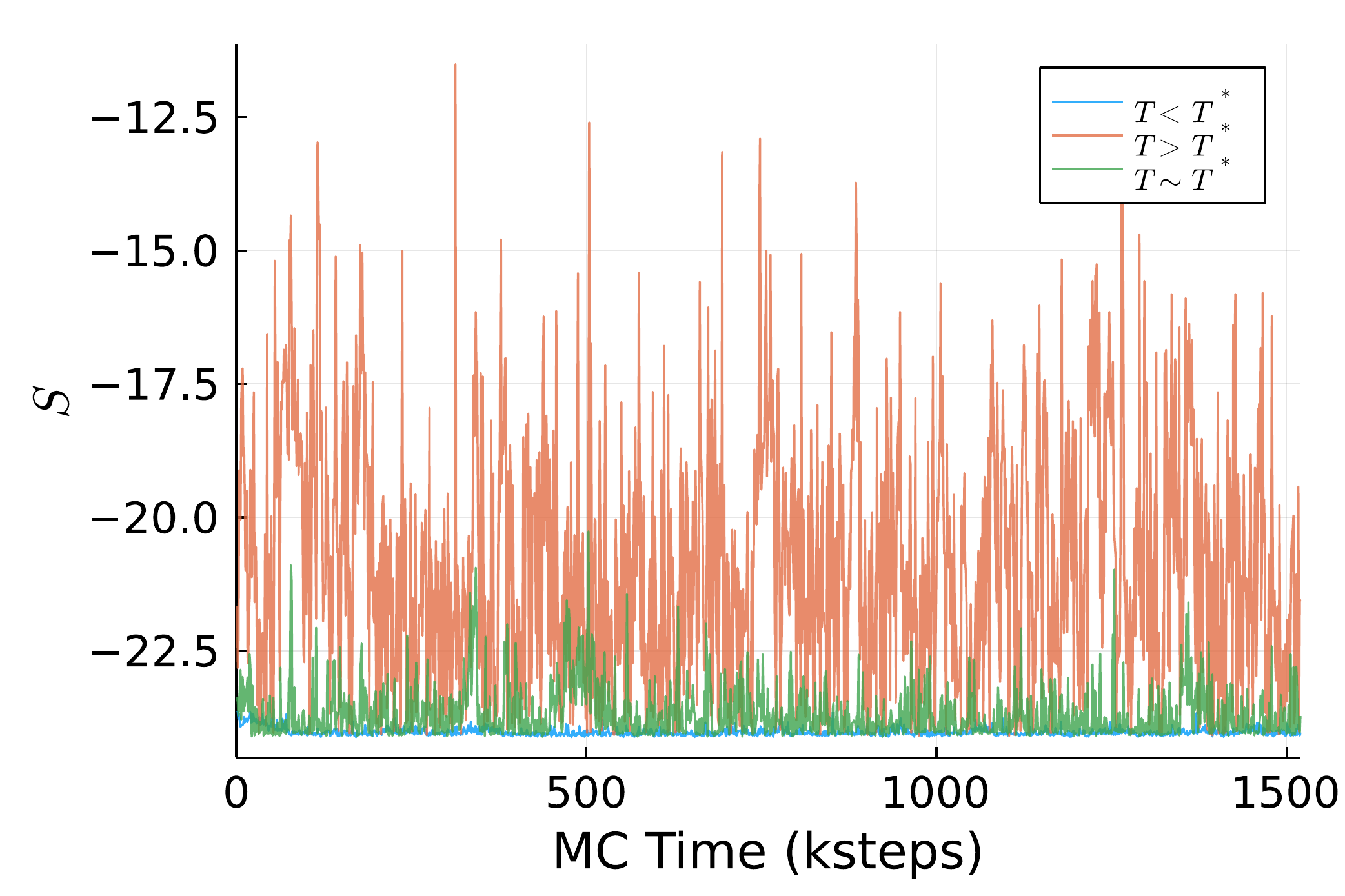}
    \includegraphics[width=0.48 \linewidth]{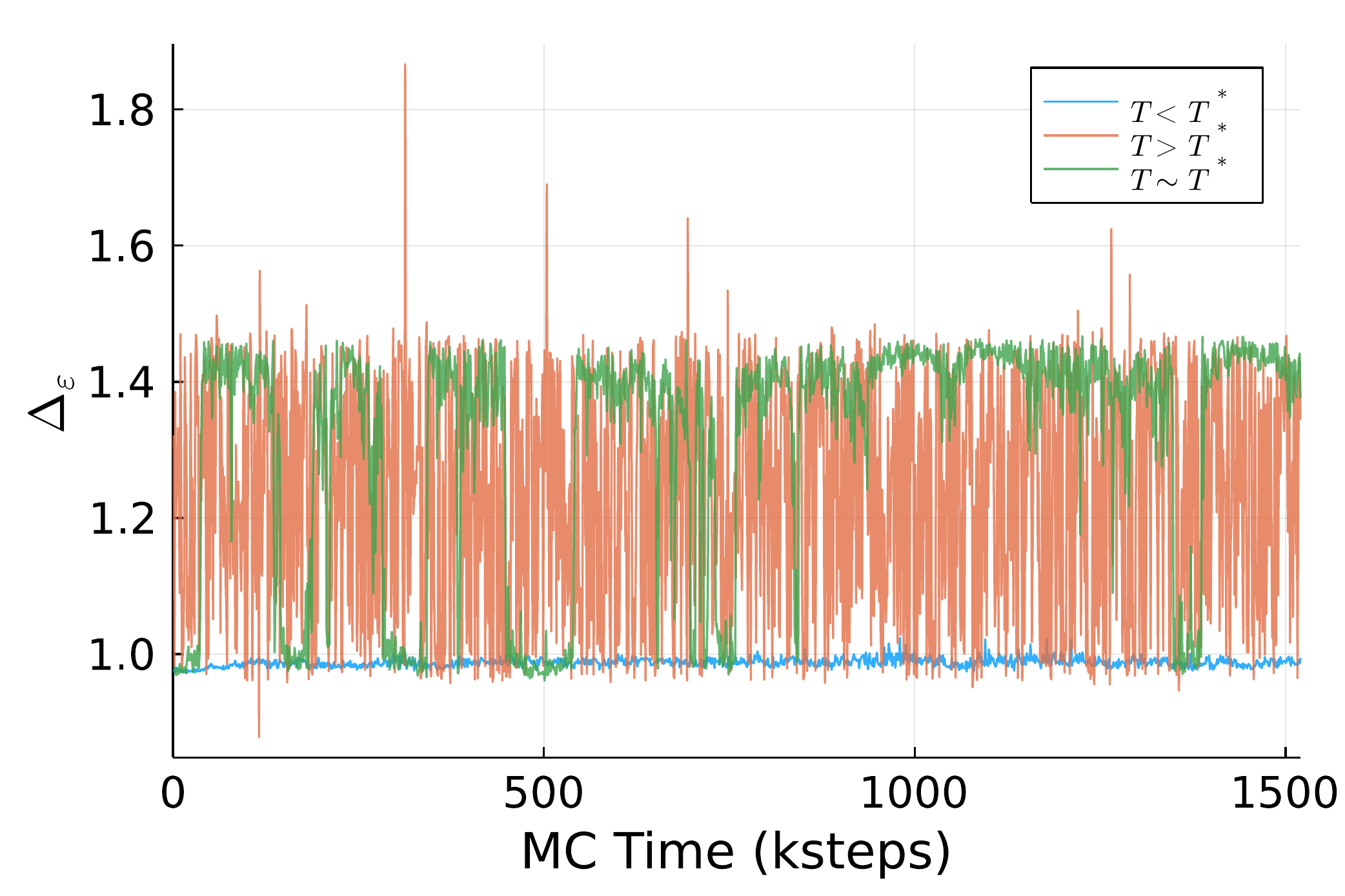}
\caption{Sample trajectories for MC runs at different temperatures. 
\emph{Left:} Value of the action $S$ of the points visited. 
\emph{Right:} Scaling dimension of the lightest exchanged scalar operator. 
This particular test was done in $d=3$ by taking $\Delta_\sigma = 0.519$ fixed, with a spin partition $3\_ 2 \_ 2 \_ 1$ and $\Delta^* = 10$. The trajectories are shown for $T=0.05,0.3,0.9$. In this case, $T^* = 0.3$.}
    \label{fig:trajectories}
\end{figure}

\subsection{Ideal Temperature}
\label{sec:Temperature}

As mentioned earlier in section \ref{sec:method}, the key step of our protocol is the wide search performed using the Metropolis Monte Carlo algorithm. In order for this step to be efficient the temperature must be chosen wisely. 
To this end we performed test searches for different values of $d$, $N_{\rm Ops}$ and $\Delta^*$, where several different temperatures were used. Then,  we determined by inspection the ideal one in each 
setting, according to the criterion described in section \ref{sec:protocol}.

In figure \ref{fig:trajectories}
we show a concrete example of how to identify the ideal temperature. We show  ``cold'',  ``hot'' and just right temperatures in blue, red and green (respectively). We can see clearly that the blue line ``freezes'' into the first local minimum found by the MC and then stays there for the whole test, whereas the red trajectory travels back and forth randomly. We consider the green trajectory  
to be representative of an efficient search because it oscillates around local minima for some time before making a transition into another one. While this could mislead the reader into thinking that the time spent oscillating around a local minimum is in some sense ``wasted'' one must bear in mind that during those steps, the MC effectively samples the rest of the scaling dimensions, thus refining the solution to crossing in that neighborhood.

\section{Minima and End-Minima}
\label{app:min-endmin}

 \begin{figure}[t!]
    \centering
    \includegraphics[width=0.37\linewidth]{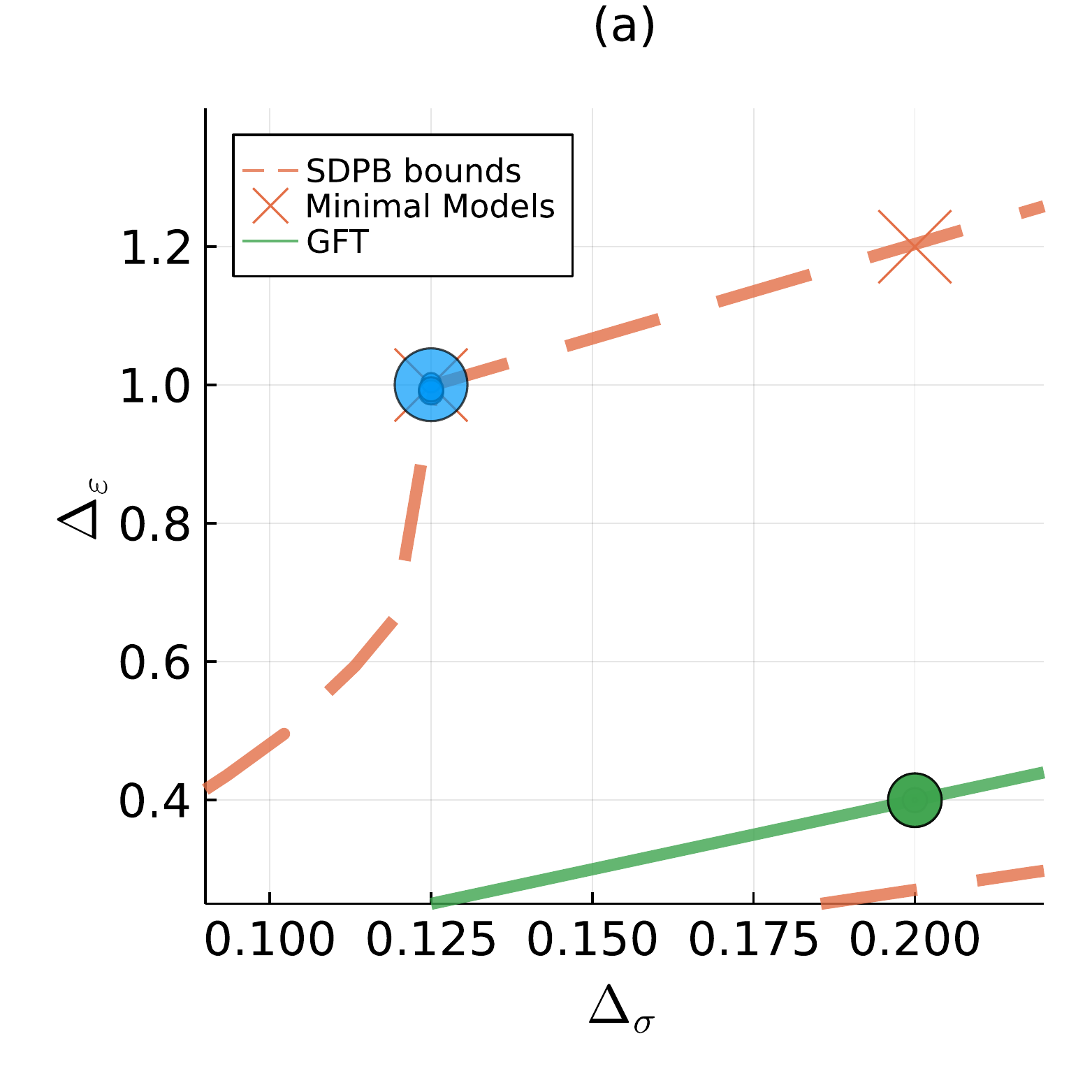}
    \includegraphics[width=0.62 \linewidth]{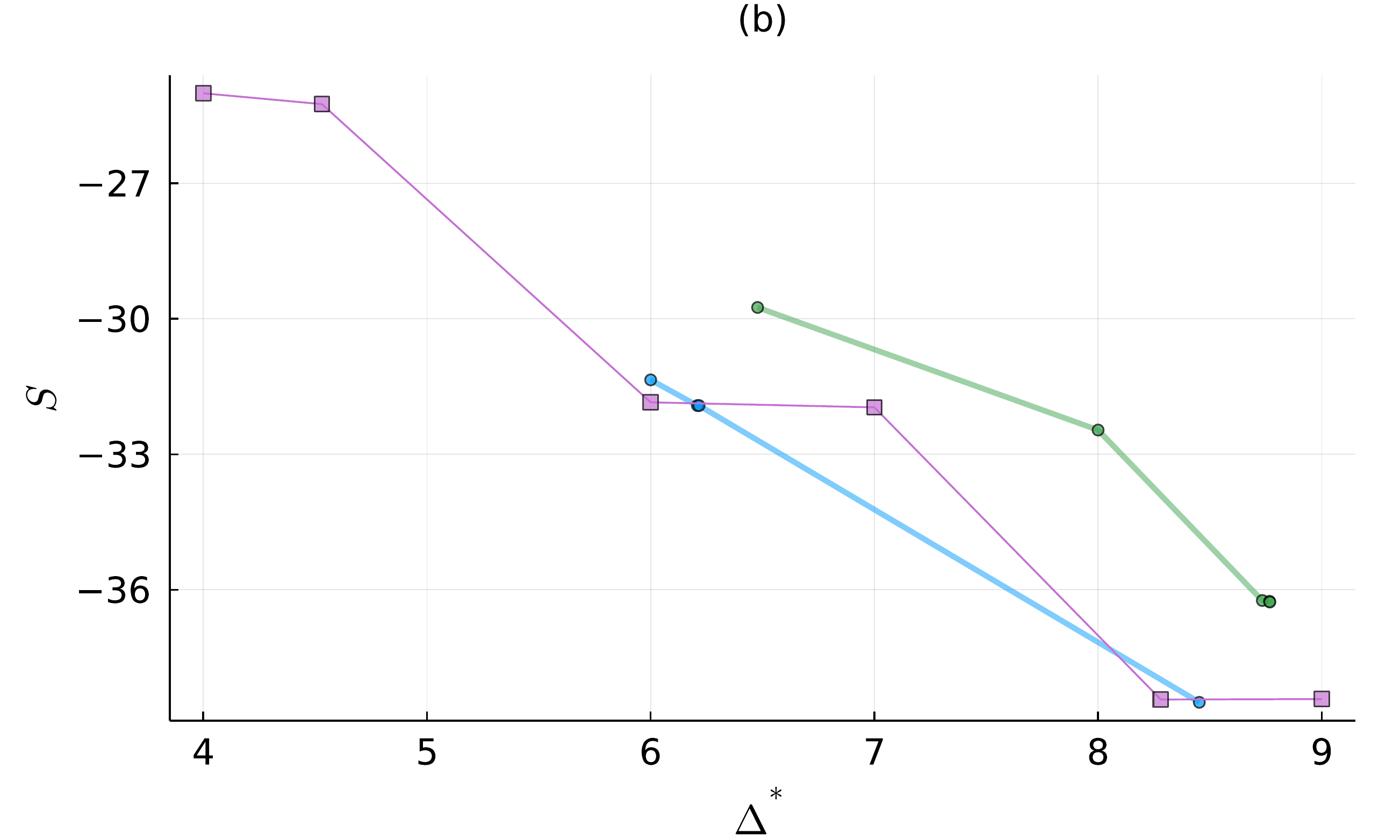}
\caption{Representative branches in $d=2$ local CFTs.
    (a): Location of individual minima (colored circles) with $\elmax \leq 8$ in the $(\Delta_\sigma,\Delta_\epsilon)$-plane. The minima belong to different sectors all of them contained in $ 4 \_4 \_3 \_2 \_1$.  The dashed orange line corresponds to the bounds obtained in \cite{Behan:2017rca} assuming only one $\mathbf{Z}_2$-even relevant scalar.
        (b) Value of the action as a function of $\Delta^*$ for each branch. 
        Color codes are the same in the two panels. Namely, each minimum in (a) corresponds to a related small circle
        in (b) of the same color. The purple Ising ref. branch is the one obtained by our protocol when starting from the exact scaling dimensions for the 2d Ising model.}
    \label{fig:d2-BranchSummary}
\end{figure}

In this appendix we illustrate the relation between minima and end-minima.
As discussed in the main text, end-minima are defined as those special minima which 
are the end-points of branches. The latter are a set of akin minima, supposedly associated to the same CFT.

We report in the left panel of figure  \ref{fig:d2-BranchSummary} the individual minima found in $d=2$ for some selected branches at $\Delta_\sigma = 1/8$ (Ising) and at $\Delta_\sigma= 1/5$ (GFT-like). 
Note the difference with respect to figure \ref{fig:d2-overview}. There we reported only the end-minima 
associated to the end-points of {\it all} the branches. Here, we pick up a given branch and show all the minima contained in it. Since most of these minima overlap in figure \ref{fig:d2-BranchSummary} and similarly several branches overlap in figure \ref{fig:d2-overview} (detectable from the darker color of the circles),
the difference between the two figures can be overlooked at first glance. In the right panel of figure \ref{fig:d2-BranchSummary} we report the value of the action $S$ for each of these minima making the branches manifest. We also note in passing that the spectrum at the end-minimum for each value of $\Delta_\sigma$ is shown in figure \ref{fig:dsig125Minima} in orange (the green spectrum corresponds to a similar spectrum not belonging to the same branch).

In order to have an estimate of how the action is supposed to decrease with $\Delta^*$, we also report the branch obtained by starting low $T$-MCs from several truncations of the exact Ising scaling dimensions (purple squares in the right panel of figure \ref{fig:d2-BranchSummary}). 
This implies that there are several minima in the vicinity of the exact Ising values and that our protocol is able to find at least a subset of them. It is also reassuring to see that the decrease in $S$ of the branches found and of the Ising benchmark branch are fully compatible. The good behaviour of the GFT-like branches is the main reason why we believe such theories are not numerical artifacts.

\begin{figure}[t!]
    \centering
    \includegraphics[width=0.37 \linewidth]{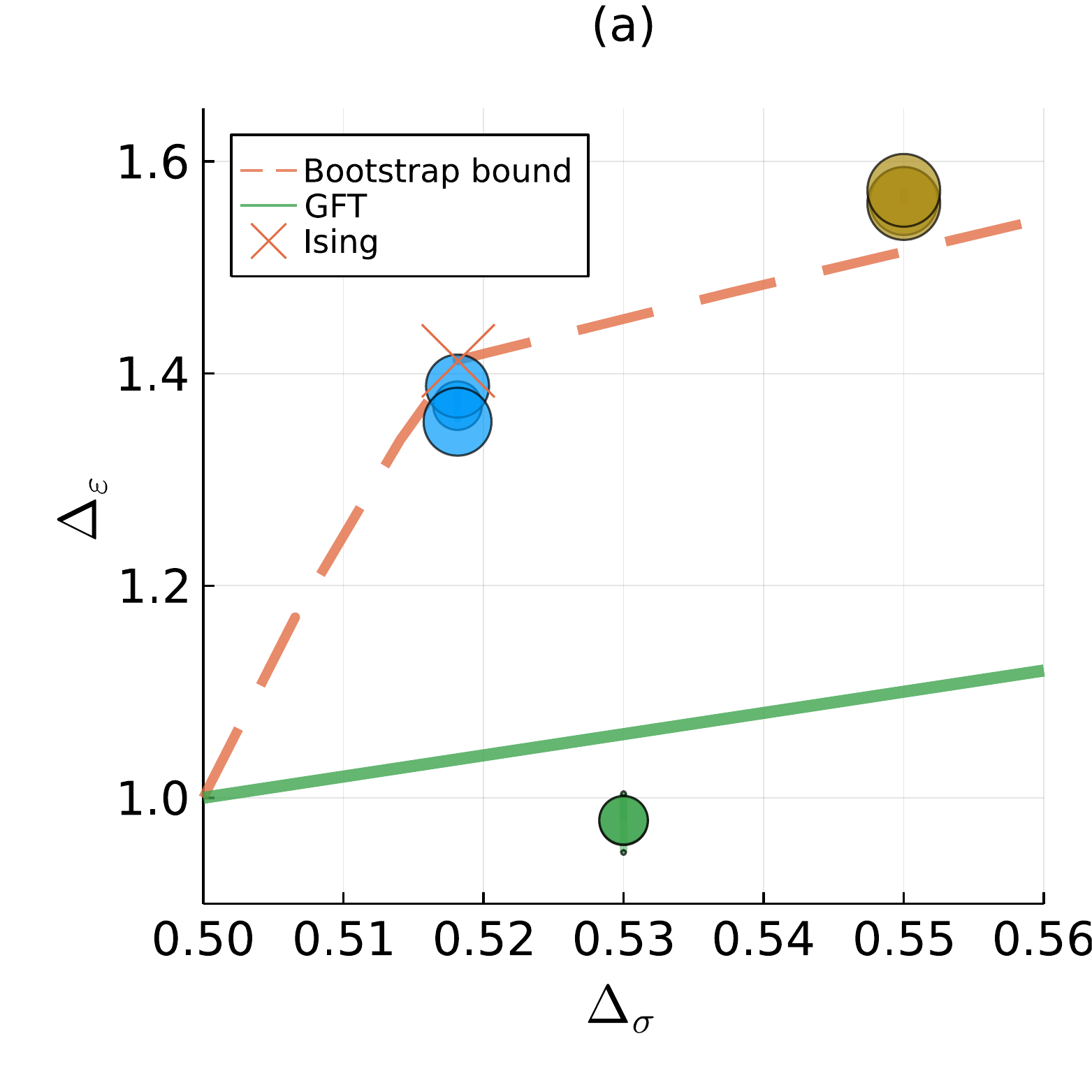}
    \includegraphics[width=0.62 \linewidth]{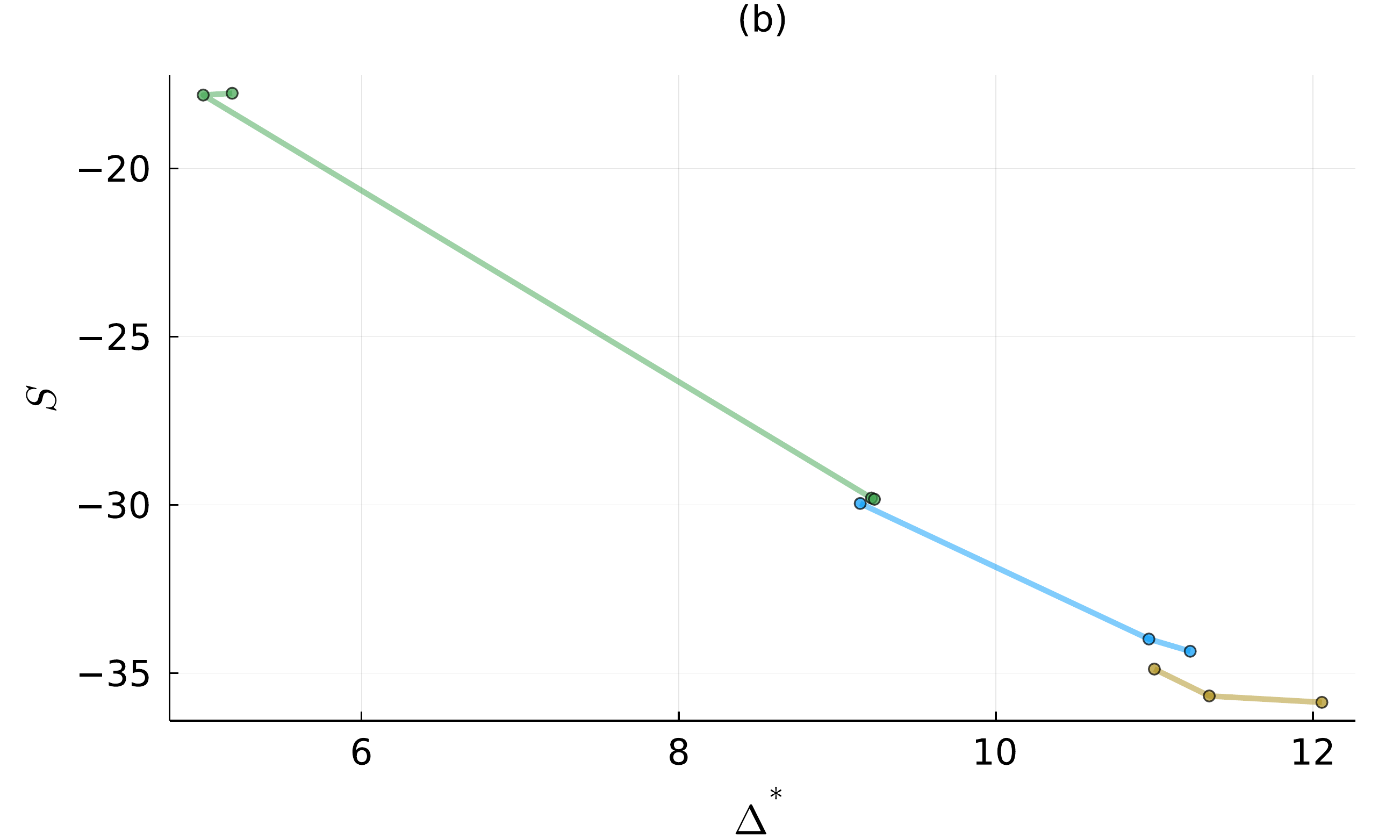}
\caption{Representative branches in $d=3$, $\Delta_T=3$.
    (a): Location of individual minima (colored circles) with $\elmax \leq 10$ in the $(\Delta_\sigma,\Delta_\epsilon)$-plane. The minima belong to different sectors all of them contained in $ 4 \_4 \_4 \_3 \_2 \_1$. Several minima overlap and appear as a single one. The dashed orange line is the bound reported in \cite{El-Showk:2014dwa}.
        (b) Value of the action as a function of $\Delta^*$ for each branch. Color codes are the same in the two panels. Each minimum in (a) corresponds to a related small circle in (b) of the same color.
}
    \label{fig:d3-BranchSummary}
\end{figure}

As further example we show in the left panel of figure  \ref{fig:d3-BranchSummary} the individual minima for three selected end-minima in $d=3$. The blue one corresponds to the Ising model, the green one to the end-minimum below the GFT green line and the olive one to one of the end-minima at $\Delta_\sigma = 0.55$. In the right panel we show the value of the action $S$ for each of these minima (the color code indicates the different branches). 

We point out in passing that a consistent branch should not necessarily be monotonically decreasing as $\Delta^*$ increases, because it can happen that minima with less operators have a  slightly higher $\Delta^*$ than a related minima with a denser spectrum. For instance, the green branch in figure \ref{fig:d3-BranchSummary} (b) would seem to ``turn around'' at the beginning, but the important fact is that we make sure that minima with more operators have strictly smaller $S$.

\bibliographystyle{JHEP}
\bibliography{references.bib}

\end{document}